\documentclass[a4paper,11pt]{article}
\usepackage{pos}

\newcommand{\bea}{\begin{eqnarray}}
	\newcommand{\eea}{\end{eqnarray}}
\newcommand{\bee}{\begin{eqnarray*}}
	\newcommand{\eee}{\end{eqnarray*}}
\newcommand{\al}{\begin{align*}}
	\newcommand{\eal}{\end{align*}}
\newcommand{\be}{\begin{equation}}
	\newcommand{\ee}{\end{equation}}

\newcommand{\bem}{\begin{pmatrix}}
	\newcommand{\eem}{\end{pmatrix}}

\def\l{\lambda}

\newcommand{\tl}[1]{\tilde{#1}}

\newcommand{\comment}[1]{}

\newcommand{\RR}{{\mathbb R}}%Reals
\newcommand{\CC}{{\mathbb C}}%Complex
\newcommand{\PP}{{\mathbb P}}%Projective
\newcommand{\ZZ}{{\mathbb Z}}%Integers
\newcommand{\str}{\operatorname{{str}}}

\theoremstyle{definition}

\theoremstyle{remark}

\numberwithin{equation}{section}
\title{TASI Lectures on the Mathematics of String Dualities}
\author[a]{Niklas Garner}
\author*[a]{Natalie M. Paquette}

\affiliation[a]{Department of Physics, University of Washington, \\ Seattle, WA, 98195, USA}

\emailAdd{nkgarner@uw.edu}
\emailAdd{npaquett@uw.edu}

\abstract{In these lecture proceedings, we describe some of the fundamental mathematical concepts that underlie supersymmetric string theory and field theory, and their role in describing and testing dualities. In particular, we provide a pedagogical introduction to topological and holomorphic twisting, descent, and higher algebraic structures. Our primary examples are worldsheet theories of topological strings, namely the A- and B-models, which we briefly review. These proceedings are based on lectures given by the second author at TASI 2021.}

\FullConference{%
	Theoretical Advanced Study Institute 2021 (TASI 2021)
	June 7 - July 2, 2021\\
	University of Colorado, Boulder CO 80309 USA}

\begin{document}
\maketitle

\section{Introductory remarks}
These proceedings are based on lectures given at TASI 2021, the aim of which was to provide an overview of basic dualities in string theory and the mathematical techniques used to test and explore these dualities. In particular, these proceedings focus on the content of the final two lectures in the series, which served to highlight certain mathematical structures and frameworks that are ubiquitous in contemporary studies of physical mathematics and supersymmetric string theory. The techniques we introduce have proven essential for both formalizing and conceptualizing these theories, and have enjoyed wide use in testing putative dualities. Furthermore, they have led to a host of intellectual inquiries in their own right. This review aims to be self-contained, while still providing a succinct pedagogical introduction to the selected topics; as such, the selection of topics (and references) is very limited and, necessarily, incomplete. The first two lectures in the series, which we will not review in these proceedings, provided a lightning introduction to string theory and a summary of basic supersymmetric string dualities on various backgrounds. We omit them because these topics are already discussed thoroughly elsewhere, e.g. \cite{V97}, as well as in the foundational string theory texts.

Before diving into the subject at hand, we sketch some brief general philosophy, following \cite{Polchinski:2014mva}. Dualities are equivalences: a (typically very complicated) change of variables that leaves the underlying physics, i.e. observables, invariant. The reason dualities, unlike more pedestrian basis changes, are nontrivial is that they are typically non-manifest in any weak coupling description of the theory. Rather, they are \textit{exact} ``symmetries'' of the theory.\footnote{In these lecture notes we will focus on exact, rather than infrared, dualities.} Symmetry is in quotes here because a duality often involves, in addition to a transformation of the field variables, a nontrivial change of  background parameters such as the coupling constant.\footnote{Note that in string theory, however, these parameters are set by vacuum expectation values of moduli fields, which should be contrasted with the situation in field theory.} In many string dualities, the strongly coupled limit of one string theory is related by a duality to (i.e., is equivalent to) a weakly coupled description of another, in general different, string theory. 

Morally, one can think of such dualities as akin to Fourier transforms. Recall that we can write the classical action of a weakly coupled, say, string theory as
\[
e^{i S/ (g^2 \hbar)}
\]
where we have made a rescaling of the fields so that the coupling constant appears as an overall factor of ${1 \over g^2}$ multiplying the action. Since the coupling constant then appears only in the combination $g^2 \hbar$, we see that small $g^2$ is equivalent to small $\hbar$, and large $g^2$ is equivalent to large $\hbar$. A weak-strong duality transforms the coupling constant as $g \sim 1/g$, so that as we tune $g$ from weak coupling to strong coupling, we can trade the original description of the theory for that of the dual theory whose coupling $g' \sim 1/g$ is becoming increasingly weak. One can think equivalently about the duality as trading a description in which quantum fluctuations are becoming increasingly large for one where quantum fluctuations are becoming increasingly suppressed. This is similar to a Fourier transform, in which a well-localized quantity in position space is spread out in momentum space, and vice versa. In a weak-strong duality, however, the map between variables involves highly nonlinear transformation on an infinite-dimensional field space. In general, we do not know how to map good field variables in the weakly coupled description to their counterparts on the dual strongly coupled side. This statement is already true in field theories, and in string theories the situation may be even more complicated.

We will focus on theories in which supersymmetry is preserved. In these cases, though not only these cases, certain aspects of dualities can be related to mathematically well-defined quantities once more. For example, as we will see, path integrals can localize to finite dimensional integrals, and protected quantities can enjoy invariance under deformation of the coupling constant (or more generally under deformation of the parameters that transform under the duality of interest). Protected quantities may be computable in any duality frame, although they may have very different descriptions in the different frames, and provide useful probes of nonperturbative physics. Their proposed invariance under dualities also leads to highly nontrivial (conjectural) mathematical equivalences. We will explain how to isolate the physics of such protected quantities by means of \textit{twisting}, and discuss some of the algebraic structures that govern their operator products. In fact, when focusing on the mathematical avatars of such protected quantities, aspects of some string dualities can indeed be seen to reduce to Fourier transforms on the nose. More precisely, natural mathematical generalizations of Fourier transformations, such as the Fourier-Mukai transform or even Koszul duality\footnote{See \cite{P21} for an exposition on this point of view for a non-expert audience.}, are ubiquitous. Although we will not rely on this perspective in the remainder of the notes, we advise the reader to keep this analogy in mind when they embark on a (supersymmetric) duality chase.

Our ur-example of string dualities, which will play a starring role throughout these notes, is mirror symmetry. Mirror symmetry is a perturbative string duality; it is not a strong-weak duality in the string coupling constant $g_s$. A useful analogue to keep in mind is T-duality on a circle of radius $R$ which behaves as a ``strong-weak duality” in the stringy, or worldsheet loop expansion parameter $\alpha'/R^2$,\footnote{Indeed, although we will not discuss it further in these notes, mirror symmetry can be conjecturally understood as a composition of T-dualities \cite{SYZ} on the fiber of a Calabi-Yau, when the latter is viewed as a special Lagrangian fibration. We also remark that the action of T-duality on D-branes, when the latter is modeled by coherent sheaves (as we review in \ref{sec:Bbranes}) can be formalized as a Fourier-Mukai transform.} but not in the string coupling $g_s$, which can remain small in both duality frames. Other dualities will be genuinely strong-weak in the way they act on $g_s$ and are hence nonperturbative dualities. Although some of these dualities may be less well-understood than perturbative dualities like mirror symmetry\footnote{It is, however, worth emphasizing that string theory may blur apparently sharp distinctions between these cases: for example, a non-perturbative S-duality in a fundamental string is perturbative T-duality for the dual solitonic string \cite{MDuff}.}, the mathematical concepts and techniques presented here have broad applicability to these situations. 

In the remainder of these notes, we will emphasize the natural role of \textit{cohomology} in studying mathematical aspects of string theory and dualities, in the context of twisting supersymmetric theories. We will then discuss \textit{homotopical algebras}, also called \textit{higher algebras}, which provide a more refined lens to the physics of twisted theories. 

\section{Twisting \& mirror symmetry}
Although dualities need not follow from the presence of supersymmetry (indeed, many very interesting dualities do not! Recall the Kramers-Wannier duality, and for additional modern examples, see for example \cite{KTT19, SSWW16, MN21}), supersymmetry is a particularly convenient way to motivate and check dualities because of the ubiquity of BPS quantities, which we can compute in both putative duality frames. One can isolate the protected parts of these full physical dualities expediently using a procedure known as \textit{twisting}. Twisting localizes mathematically ill-defined quantities to well-defined objects that nonetheless retain a great deal of physical and mathematical richness. We begin by providing an introduction to twisting, which has applications to supersymmetric field theories as well as superstring theories. We will focus, in particular, on mirror symmetry as a string duality with close connections to mathematics and whose study has flourished due to the precision afforded by this twisting procedure. For more details, the reader is encouraged to consult the two thorough textbooks \cite{HKKPTVVZ03, ABCDKMGSSW09}, whose presentation we will follow at various points.

\subsection{A modern introduction to twisting}
\label{sec:twistinto}
A twist is an operation we can perform on a supersymmetric field theory to, roughly speaking, restrict the subspace of physical observables under consideration to a simpler, more manageable (BPS) subset. The discussion in this section applies very generally to supersymmetric field theories, though we will shortly be interested in applying it to 2d supersymmetric field theories that describe some superstring worldsheets. Performing a twist directly on a spacetime theory with gravitational dynamics is more subtle and an active area of research that goes beyond the scope of these proceedings, but see for instance \cite{CL16, LT20, dWMR18, BR18}. We will primarily work in Euclidean signature, where twisting is best understood and well-defined. 

Twisting begins with making a choice of nilpotent supercharge, i.e. a supercharge $Q$ in the supersymmetry algebra of interest that squares to zero $[Q,Q] = 0$.%
\footnote{We use conventions that $[-,-]$ is the graded commutator, i.e. $[a,b] = a b - (-1)^{F(a) F(b)} b a$ is a commutator if one of $a$ or $b$ is bosonic (at least one of $F(a)$ and $F(b)$ is even) and an anti-commutator otherwise (both $F(a)$ and $F(b)$ are odd). It is graded antisymmetric with respect to fermion parity $[a, b] = -(-1)^{F(a)F(b)}[b,a]$ and satisfies a graded version of the Jacobi identity $[a, [b,c]] = [[a,b],c] +(-1)^{F(a)F(b)}[b,[a,c]]$ that says the linear map $[a,-]$ is a derivation (with the same statistics as $a$) of the bracket $[-,-]$.} %
This is a fermionic symmetry much like a BRST symmetry, and behaves the same way.  We consider $Q$-invariants, so that operators that survive the twist are \textit{$Q$-closed}: $[Q, O] = 0$. If we further assume that the supersymmetry $Q$ isn't spontaneously broken, so $Q$ annihilates the vacuum, it follows that correlation functions of $Q$-closed $O$ with other $Q$-closed local operators are invariant under $O \to O + [Q, \Lambda]$:
\be
\langle (O + [Q, \Lambda]) \ldots \rangle =  \langle O \ldots \rangle + \langle [Q, \Lambda \ldots] \rangle = \langle O \ldots \rangle.
\ee
This implies the equivalence relation $O \sim O + [Q, \Lambda]$, where operators of the form $[Q, \Lambda]$ are called \textit{$Q$-exact}. Therefore, insofar as a QFT is defined by the data of its correlation functions, local operators in the twisted theory are $Q$-closed local operators, modulo the addition of $Q$-exact local operators: they are labeled by elements of \textit{$Q$-cohomology}.

Imagine that we are quantizing a theory with some gauge symmetry in the BRST formalism, so that we are taking cohomology with respect to some BRST differential $d$. One simple way to formulate twisting in this language is to simply augment the BRST differential $d$ by this new supercharge $Q$, and take cohomology with respect to the deformed differential $d_Q = d +  Q$: operators in the twisted theory are given by $d_Q$-closed operators, modulo $d_Q$-exact operators. To a first approximation, operators in the twisted theory can be though of as gauge/BRST-invariant ($d$-closed) operators that are simultaneously $Q$-invariant ($Q$-closed). This can't be quite right: this answer must be corrected to account for operators that are neither $d$-closed nor $Q$-closed but are nonetheless $d_Q$-closed. There is a way to compute these corrections in a cohomological manner, i.e. where at each step we compute a suitable cohomology, called a \textit{spectral sequence}. Computing the intersection of the $d$ and $Q$-cohomologies can be a useful approximation to the complete answer, but in general we caution the reader that the twisted theory does not always capture a simple subset of observables in the original theory.

\subsubsection{The nilpotence variety}
\label{sec:nilpot}
For a given supersymmetry algebra, there are often many nilpotent supercharges $Q$ and hence possible twists. Suppose there are $N$ supercharges $Q_i$, $i = 1, ..., N$ ($i$ is a combination of the spinorial and $R$-symmetry indices), with brackets $[Q_i, Q_j] = \Gamma_{ij}^\mu P_\mu$ for $P_\mu$ the spacetime momentum operators; if $Q = q^i Q_i$ then the nilpotence of $Q$, i.e. $[Q, Q] = 0$, translates to a collection of quadratic equations for the $q^i$: $\Gamma_{ij}^\mu q^i q^j = 0$ for $\mu = 1,\ldots,d$. We require that at least one $q^i$ is nonzero (otherwise $Q = 0$) and note that if $Q$ solves this equation then so does $\lambda Q$ for any $\lambda \in \CC^*$; thus, the moduli space of nilpotent supercharges in a given supersymmetry algebra, sometimes called the \textit{nilpotence variety} of said algebra, is naturally a closed subvariety of $(N-1)$-dimensional projective space $\PP^{N-1}$ \cite{ESW20, ESW21}.

The nilpotence variety is typically singular, but has natural actions of (the complexification of) Spin$(d)$ and (the complexification of) the supersymmetry algebra's $R$-symmetry group $G_R$ rotating the $i$ index of the homogeneous coordinates $q^i$. Additionally, it has a natural stratification by the rank of the $d \times N$ matrix $\Gamma_{ij}^\mu q^i$, i.e. by the number of translations $P_\mu$ that belong to the image of the map $[Q, -]$. The translations $P_\mu$ in the image of $[Q, -]$, being $Q$-exact by definition, are necessarily trivial in the twisted theory: if we have operators $Q_\mu$ such that $[Q, Q_\mu] = iP_\mu \sim \partial_\mu$, then correlation functions of $Q$-closed observables \textit{are locally constant in $x^\mu$}. Explicitly (focusing on a single insertion at $x$): 
\be
\begin{aligned}
	\partial_\mu \langle O(x) \ldots \rangle &\sim  \langle [Q, [Q_\mu, O(x)]] \ldots \rangle \\
	&\sim \langle [Q, [Q_\mu, O(x)]  \ldots] \rangle = 0
\end{aligned}
\ee
The operators $Q_\mu$ will reappear in Section \ref{sec:homotopy} and underpin much of the homotopy-algebraic structures of observables in twisted QFTs.

The structure of the $\Gamma$ matrices in various supersymmetry algebras implies that the translations trivialized for a given $Q$ organize themselves into two types: for a suitable choice of coordinates, we find that $P_\mu$ is trivialized for, say, $\mu = 1, ..., n \leq d$ but only the complex combination $P_{\bar{a}} = \tfrac{1}{2} (P_{n+2a-1} + i P_{n+2a})$ for $a = 1, ..., m$ with $d = n + 2m$. It follows that the cohomology of $Q$ is invariant under arbitrary translations along $x^\mu$ as well as anti-holomorphic translations along $\bar{z}^{\bar{a}}:= x^{n+2a-1} - i x^{n+2a}$. The simplest and most commonly studied nilpotent supercharges $Q$ are where $n = d$, these are called a \textit{topological supercharges} and the corresponding twists \textit{topological twists} \cite{W88}. More generally, $Q$ is called a \textit{holomorphic-topological} or \textit{mixed supercharge} and, when $d$ is even with $d = 2m$, $Q$ is called a \textit{holomorphic supercharge} \cite{Johansen:1994aw, nekrasov1996four}, with the twists named correspondingly.

Topologically twisted theories often have the property that all components of the stress tensor, which govern variations of the metric, are $Q$-exact:
\be
T_{\mu \nu} = {\delta S \over \delta h^{\mu \nu}} =  [Q,  G_{\mu \nu}].
\ee
A common way to realize this condition in practice is that the entire Lagrangian can itself be written as a $Q$-commutator. From this condition, metric-independence of correlators in the theory, ${\delta \langle O_1 \ldots O_n \rangle \over \delta h^{\mu \nu}} = 0$, is immediate, because one brings down a factor of $T_{\mu \nu}$ from the variation of the action, which is $Q$-exact and hence zero in correlation functions. Mixed holomorphic-topological theories have similar properties. Heuristically, one expects a TQFT in the topological directions and a holomorphic QFT in the holomorphic directions; the latter has the structure of a vertex algebra or chiral algebra in two dimensions, as is familiar from CFT, and in higher dimensions it is given by a ``higher'' analogue of a chiral algebra. We will elaborate more on higher structures in the sequel.

As a running example, consider the 2d $\mathcal{N}=(2,2)$ supersymmetry algebra. We work in Euclidean signature, and choose complex coordinates $z, \bar{z}$ on the worldsheet. The supersymmetry algebra is generated by $N=4$ supercharges $Q_\pm, \bar{Q}_\pm$ with non-vanishing brackets (in the absence of central charges):
\be
[Q_+, \bar{Q}_+] = 2 P_{\bar{z}} \qquad [Q_-, \bar{Q}_-] = -2 P_z
\ee
for $P_z, P_{\bar{z}}$ the holomorphic/anti-holomorphic momenta.%
\footnote{In Lorentzian signature, we have the left/right moving momenta $P_\pm$ with anti-commutators $[Q_\pm, \bar{Q}_\pm] = 2 P_\pm$. The Lorentzian left/right moving coordinates $x^\pm = x^0 \pm x^1$ are Wick rotated to the Euclidean holomorphic coordinates as $x^+ = \bar{z}$ and $x^{-} = -z$.} %
This superalgebra is invariant under $U(1)_V$ vector $R$-symmetry rotations (generated by a charge $R_V$) acting on the supercharges as
\be
[R_V, Q_\pm] = Q_\pm, \qquad [R_V, \bar{Q}_\pm] = - \bar{Q}_\pm,
\ee
and $U(1)_A$ axial $R$-symmetry rotation (generated by $R_A$) acting as
\be
[R_A, Q_\pm] = \pm Q_\pm, \qquad [R_A, \bar{Q}_\pm] = \mp \bar{Q}_\pm.
\ee

\begin{figure}[h!]
	\centering
	\includegraphics{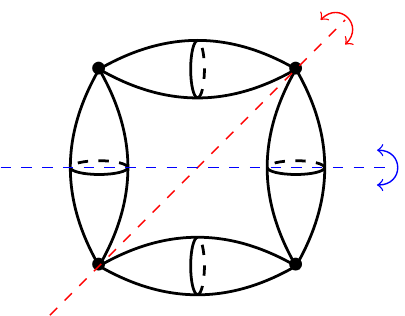}
	\caption{The nilpotence variety for the 2d $\mathcal{N}=(2,2)$ supersymmetry algebra. The poles of each $\PP^1$ correspond to rank 1 (holomorphic) supercharges with the remaining points being rank 2 (topological) supercharges. The red symmetry axis corresponds to the parity transformation $(Q_+, \bar{Q}_+, P_{\bar{z}}) \leftrightarrow (Q_-, \bar{Q}_-, P_z)$ and the blue symmetry axis corresponds to charge conjugation $Q_\pm \leftrightarrow \bar{Q}_\pm$.}
	\label{fig:22nilpotence}
\end{figure}

The supercharge $Q = q^\alpha Q_\alpha + \bar{q}^\alpha \bar{Q}_\alpha$ is nilpotent if and only if
\be
q^+ \bar{q}^+ = 0 \qquad q^- \bar{q}^- = 0.
\ee
Thus, the nilpotence variety for this supersymmetry algebra can be identified with four copies of $\PP^1$ touching at their poles; see Figure \ref{fig:22nilpotence}. Up to symmetries of the algebra, e.g. overall scaling, parity $+ \leftrightarrow -$, or charge conjugation $Q_\pm \leftrightarrow \bar{Q}_\pm$, we can always choose $\bar{q}^+ = 1$ (and hence $q^+ = 0$). There are then three possibilities: 1) $\bar{q}^- = 0 = q^-$, 2) $\bar{q}^- = 0$ and $q^- \neq 0$, or 3) $\bar{q}^- \neq 0$ and $q^- = 0$. It is easy to see that the first case $Q = Q_{H} = \bar{Q}_+$ is a rank 1, holomorphic supercharge. The corresponding twist was historically called the \textit{half twist} \cite{W98, K05, W07}. For the second (resp. third) case, we can always use (complexified) $R$-symmetry rotations and rescaling to choose $q^- = 1$ (resp. $\bar{q}^- = 1$) to see there is essentially a single choice $Q_A = \bar{Q}_+ + Q_-$ (resp. $Q_B = \bar{Q}_+ + \bar{Q}_-$) and that it is a rank 2, topological supercharge. The resulting topological twists are correspondingly called the \textit{$A$ twist} and \textit{$B$ twist}. 

\subsubsection{Twisting homomorphisms}
\label{sec:twistinghomoms}
Depending on the type of supercharge we are considering, we can attempt to put the twisted theory on non-trivial spacetimes/backgrounds compatible with the twist. For a mixed holomorphic-topological supercharge, with $n$ topological directions and $m$ holomorphic directions, we can at best expect to put the twisted theory on a spacetimes that locally look like $\RR^n \times \CC^m$ with transition functions that are holomorphic on $\CC^m$, i.e. $(x, z, \bar{z}) \to (y(x, z, \bar{z}), w(z), \bar{w}(\bar{z}))$. Manifolds of this form are said to have a \textit{transverse holomorphic foliation} (THF). For example, we can hope to put a holomorphically-twisted theory ($d = 2m$) on a general complex manifold or a topologically-twisted theory ($d=n$) on any manifold.

In our Euclidean setting, the original physical theory will have an action of the Euclidean spin group Spin$(d)$ in $d$-dimensional flat space. When considering a twisted theory, the supercharge $Q$ isn't compatible with spacetime rotations: since $Q$ transforms as a spinor, the action of $Q$ doesn't commute with spacetime rotations. For the moment, consider a topological supercharge $Q$. The typical resolution of this problem is to choose an injective group homomorphism $\iota:$ Spin$(d) \rightarrow G_R$, where $G_R$ is the $R$-symmetry group of the physical QFT. If the homomorphism $\iota$ is such that $Q$ transforms trivially under the action of Spin$(d)$ via the map $\textrm{id} \times \iota:$ Spin$(d) \rightarrow$ Spin$(d) \times G_R$, then we can use this inclusion to define a modified action of Spin$(d)$ compatible with the action of $Q$. We call this the \textit{twisted spin} and call such a homomorphism $\iota:$ Spin$(d) \rightarrow G_R$ a \textit{twisting homomorphism}.%
\footnote{It is also possible and often useful to include other symmetries in twisting homomorphisms. For example, if the supersymmetric theory has an internal symmetry $H$ (necessarily commuting with $G_R$ and Spin$(d)$) then we can consider more general homomorphisms $\eta: \textrm{Spin}(d) \to G_R \times H$. To ensure that the modified action of Spin$(d)$ is compatible with the twist, we need that $\iota = \pi_1 \circ \eta$ is a twisting homomorphism in the usual sense.} %

Similarly, if $Q$ is a general mixed holomorphic-topological supercharge then it suffices to redefine rotations compatible with the THF: we can restrict to the subgroup Spin$(n;m)$, which we define to be the subgroup of Spin$(d)$ that preserves the splitting $\RR^n \times \CC^m$, and then modify the action of said rotations via a twisting homomorphism $\iota_{\textrm{mixed}}:$ Spin$(n;m) \to G_R$ as above. For example, in the simplest holomorphic example $n =0$, $m = 1$ there is no change: Spin$(0;1) \cong $ Spin$(2) \cong U(1)$ (which is a double cover of rotations $SO(2) \cong U(1)$ of $\CC$). In the simplest mixed example $n = 1$, $m = 1$ the 3d Euclidean spin group Spin$(3) \cong SU(2)$ gets reduced to Spin$(1;1) \cong U(1)$ (a double cover of $SO(2) \cong U(1)$ rotations around the $\RR$-axis of $\RR \times \CC$).

Consider again the running example of a 2d $\mathcal{N}=(2,2)$ theory. We work with conventions where the supercharges $Q_+, \bar{Q}_+$ (resp. $Q_-, \bar{Q}_-$) have spin $+\tfrac{1}{2}$ (resp. $-\tfrac{1}{2}$) under Spin$(2) \cong U(1)$ with generator $J$. The typical choice of twisting homomorphism for the $A$ twist (resp. $B$-twist) uses the vector $R$-symmetry $U(1)_V$ (resp. axial $R$-symmetry $U(1)_A$) with the twisted spin generator $J_A = J + \tfrac{1}{2} R_V$ (resp. $J_B = J + \tfrac{1}{2} R_A$) so that $[J_A, Q_A] = 0 = [J_B, Q_B]$.

Before moving on, we mention that there is one minor caveat: this construction requires that the underlying supersymmetric QFT has a non-anomalous action of the $R$-symmetry group $G_R$, or at least the subgroup used in the twisting homomorphism. For example, it may be that the symmetry isn't even realized classically: 2d $\mathcal{N}=(2,2)$ Landau-Ginzburg models only realize the vector $R$-symmetry if their superpotential has vector $R$-charge 2. Thus, it is \textit{not} possible to consider their $A$-twist on curved worldsheets. It may also happen that a classically realized $R$-symmetry suffers from a quantum anomaly: the classical axial $R$-symmetry of a 2d $\mathcal{N}=(2,2)$ sigma model with general K{\"a}hler target suffers from an anomaly unless it has vanishing first Chern class $c_1$, i.e. unless it is Calabi-Yau. Thus, the $B$-twist of a general K{\"a}hler sigma model is incompatible with a general worldsheet. This phenomenon is often related to the aforementioned Landau-Ginzburg examples via mirror symmetry.

The choice of the twisting morphism, however, is optional provided one is interested in studying the twisted theory on flat space.%
\footnote{In fact, holomorphically twisted theories can be defined on Calabi-Yau manifolds. This can be generalized to K{\"a}hler manifolds with a suitable twisting morphism.} %
To place the theory on more general spacetimes, a twisting homomorphism is required to ensure the action of $Q$ is compatible with changes of coordinates. Somewhat more precisely, the twisting supercharge, being a spacetime spinor, will transform non-trivially on spacetimes with non-trivial spin structures. To compensate for this, we introduce a background $R$-symmetry bundle whose transition functions exactly cancel those of the spin structure, at least for the twisting supercharge $Q$. The choice of twisting homomorphism concisely encodes what background to introduce: if $P_{\textrm{Spin}}$ is the spin bundle over spacetime, we take the Spin$(d) \times G_R$ bundle to have transition functions given by composing those of $P_{\textrm{Spin}}$ with the twisting homomorphism: i.e. on two patches $U, V$ we have $U \cap V \to \textrm{Spin}(d) \overset{\iota}{\hookrightarrow} \textrm{Spin}(d) \times G_R$.

More generally, we can ask that a background (e.g. metric, $R$-symmetry bundle, $\ldots$) preserves some amount of the supersymmetry. Note that the flat space supersymmetry algebra isn't compatible with a general spacetime -- a general metric isn't invariant under translations, generated by the coordinate vector fields $\partial_\mu$, let alone any supersymmetric extension thereof. Instead, such a manifold may admit isometries, generated by \textit{Killing vector fields} $K = K^\mu \partial_\mu$. Similarly, we can ask that a given background admits some number of (generalized) \textit{Killing spinor fields}, or simply (generalized) \textit{Killing spinors} $\xi$, satisfying $\nabla_\mu \xi \propto \Gamma_\mu \xi$, for $\nabla_\mu$ the full covariant derivative on our background Spin$(d) \times G_R$ bundle. Together with the Killing vector fields, these generalized Killing spinors generate some \textit{rigid supersymmetry algebra}. The algebra realized above from a twisting homomorphism has (at least) a single (generalized) Killing spinor corresponding to the twisting supercharge $Q$. 

The constraints imposed by supersymmetry on a given background are often easily extracted by promoting the background fields to full supermultiplets and then requiring that the supersymmetry variation of the background fermions vanish. In the context of working on non-trivial Riemannian manifolds, we couple the theory to (the rigid limit of) a supergravity multiplet and read off the constraints imposed by supersymmetry from the gravitini variations; this approach was first described by Festuccia and Seiberg in 4d $\mathcal{N}=1$ \cite{FS11}, but can be applied quite generally. See, e.g., \cite{P17} for explicit examples across various dimensions and Contribution 3 \cite{BLF17} of loc. cit. or \cite{CC14} for computations relevant to 2d $\mathcal{N}=(2,2)$ theories.

\subsubsection{Gradings}
\label{sec:gradings}
A supersymmetric QFT in BRST quantization has two natural \textit{gradings} or \textit{degrees}; in other words, prior to taking the cohomology, the complex modeling the Hilbert space of the theory is stratified according to (at least) two conserved charges: ghost number $\in \ZZ$ and fermion parity~$\in \mathbb{Z}/2\mathbb{Z}$. We work in conventions where fermion parity alone determines signs in algebraic manipulations. In terms of this data, the BRST differential $d$ has bidegree $(1,-)$ and the twisting supercharge $Q$ has bidegree $(0, -)$. At the end of the twist, the theory only retains $\mathbb{Z}/2\mathbb{Z}$ grading by fermion parity.

We can do a little bit better than this when there is extended supersymmetry. Choose a map $\gamma: U(1) \rightarrow G_R$ into the $R$-symmetry group $G_R$%
\footnote{The above caveat makes a minor appearance once again: we also need the underlying supersymmetric QFT to have a non-anomalous action of the $U(1)$ subgroup generated by $\gamma$. Unlike with the twisting homomorphism discussion above, the lack of these $R$-symmetries doesn't render the theory inconsistent, there is just less control over the twisted theory.} %
such that $Q$ has charge +1 under the $U(1)$ action and that the $U(1)$ grading by $\gamma$, also called \textit{$R$-charge}, coincides modulo 2 with the fermion parity grading, i.e. if the corresponding generator is denoted $R$ then $(-1)^R = (-1)^F$. Such a choice enables a full $\mathbb{Z}$-grading on the twisted theory (so long as all $R$-charges are integral) given by the sum of ghost number, measured by an operator $\textrm{gh}$, and $R$-charge, measured by an operator $R$, which we call the \textit{cohomological grading}: $C = \textrm{gh} + R$.%
\footnote{It is also possible to consider to cohomological gradings that includes the internal symmetry group $H$. Note the modified spacetime rotations should have cohomological degree 0, i.e. the twisted spin generators should commute with the generator $C$ that measures cohomological degree.} %
Moreover, since ghost number is correlated with parity for the BRST fields, e.g. the $c$ ghost has $(\textrm{gh}, R, (-1)^F)$ charge/degree $(1,0,-)$, in such situations the cohomological grading determines fermion parity.

In our main example, 2d $\mathcal{N}=(2,2)$ theories, the typical choice of cohomological grading in the $A$-twist (resp. $B$-twist) combines ghost number and the axial (resp. vector) $R$-charge $C_A = \textrm{gh} - R_A$ (resp. $C_B = \textrm{gh} - R_V$) so that $[C_A, Q_A] = Q_A$ and $[C_B, Q_B] = Q_B$.

It is worth noting that 2d $\mathcal{N}=(2,2)$ is somewhat exceptional because the full $U(1)_V \times U(1)_A$ $R$-symmetry group commutes with the $A$ and $B$ twisted spins $J_A, J_B$. In higher dimensions, the spin group is necessarily nonabelian and thus the $R$-symmetry group used in the twisting homomorphism of a topological twist no longer commutes with twisted spin and is lost to the twisted theory. Thus, so long as $U(1)_A \times U(1)_V$ are symmetries of the theory, the topological $A$ and $B$ twists both admit two natural $\ZZ$ gradings: one cohomological ($C_A$ and $C_B$) and one internal ($R_V$ and $R_A$).

\subsubsection{Equivariant cohomology}
\label{sec:equivcoho}
Another modification to this basic recipe is as follows. Instead of choosing a supercharge that squares to zero on the nose, one may study a supercharge that squares to some bosonic generator $Q^2 = J$. Even though $Q^2 \neq 0$ when acting on most operators, it will if we restrict our attention to $J$-invariant operators. See \cite{Cordes:1994fc} for a thorough introduction to this subject for physicists. We will only discuss the abelian case for simplicity. 

This idea is based on the Cartan model of \textit{equivariant cohomology}: on a smooth manifold $M$ equipped with an action of, say, $U(1)$ generated by a vector field $V$, there is a natural generalization of the de Rham differential $d_M$ acting on differential forms tensored with an algebra of polynomials $\mathbb{C}[\sigma] \otimes \Omega^\bullet(M)$%
\footnote{More generally, if we were not specializing to the abelian case, we would have $S(\mathfrak{g}^*) \otimes \Omega^\bullet(M)$, forms valued in the symmetric algebra of the dual of the Lie algebra.} %
given by $d_\sigma = d_M + \sigma \iota_V$, where $\iota_V$ denotes contraction with the vector field $V$ and $\sigma$ is a formal variable called the \textit{equivariant parameter}. Note also that $d_{\sigma}\sigma = 0$. Cartan's formula (the Lie derivative $\mathcal{L}_V$ with respect to $V$ can be expressed as $\mathcal{L}_V = d_M \iota_V + \iota_V d_M$) implies that $d_\sigma^2 = \sigma \mathcal{L}_V$. The equivariant cohomology in the abelian case is essentially the $d_\sigma$ cohomology of $U(1)$-invariant differential forms, which are forms $\omega \in \Omega^\bullet(M)$ with $\mathcal{L}_V \omega = 0$. More precisely, we adjoin to these differential forms our formal parameter $\sigma$, which gives a realization the $U(1)$-equivariant cohomology of $M$: 
\be
H^\bullet_{U(1)}(M) = H((\mathbb{C}[\sigma] \otimes \Omega^\bullet(M))^{U(1)}, d_\sigma) = H(\Omega^\bullet(M)^{U(1)}[\sigma], d_\sigma).
\ee
Equivariant cohomology admits a $\ZZ$ grading if we give the formal parameter $\sigma$ cohomological degree $2$, with the total degree being form degree plus twice the degree in $\sigma$. This modification comes up frequently when studying, e.g., twisted theories 1) with central charges, 2) in the presence of an Omega-background \cite{N03}, where $M$ is a suitable space of fields and the $U(1)$ symmetry arises as rotations around some axis in spacetime, or 3) in studying supersymmetric theories in certain supergravity backgrounds. A generalization of these considerations also underlies \textit{supersymmetric localization}, which we briefly sketch in Appendix \ref{sec:localization}. 

Finally, we briefly mention that one proposal for twisting supergravity theories directly, due to Costello and Li \cite{CL16}, modifies the twisting procedure we outlined in a concrete way in terms of classical supergravity data: a spacetime manifold $\Sigma$ and bundles $P_{Spin}, P_R$ over $\Sigma$ with connections. 
\begin{enumerate}
	\item Instead of choosing a nilpotent $Q$, we turn on the vacuum expectation value of a bosonic ghost field associated to supertranslations. \\
	\item The twisting homomorphism $\iota$ is replaced by a choice of $G$-bundle over spacetime $P_G \rightarrow \Sigma$, including a choice of connection such that the bundle $P_{Spin} \times P_R$ is induced via the homomorphism $G \rightarrow Spin(d) \times G_R$. \\
	\item If applicable, the analogue of the map $\gamma: U(1) \rightarrow G_R$ is a choice of trivial $U(1)$ subbundle in $G_R$ on which the connection restricts to zero. 
\end{enumerate}

\subsection{Mirror symmetry from the worldsheet}
\label{sec:worldsheetMS}
Mirror symmetry, and its enrichment including categories of branes, is the string duality with perhaps the most profound impact on mathematics. At minimum, it has been among the most prominent historical examples of the interaction between string theory and mathematics. The twisted version of mirror symmetry, including the brane categories, is often called \textit{homological mirror symmetry}, though we emphasize that mirror symmetry is a duality of the full physical (untwisted) theories.

Mirror symmetry is a physical equivalence between the IIA and IIB string theories on two distinct Calabi-Yau manifolds, which are called \textit{mirror manifolds}. We can study each side of the duality perturbatively, for instance using the worldsheet formalism. Furthermore, we can twist the superstring worldsheet to access a mathematically rich but much simpler subset of the full physical duality, where stringent tests of the duality can be performed and novel mathematical results can be produced.

Here, we illustrate some of the basic ingredients that feed into the study of mirror symmetry in string theory and, relatedly, 2d $\mathcal{N}=(2,2)$ QFTs in anticipation for some relatively basic statements in homological mirror symmetry in Section \ref{sec:worldsheettwist}. Although we focus on 2d $\mathcal{N}=(2,2)$ QFTs, we pay particular attention to the $U(1)_A\times U(1)_V$ $R$-symmetries, since the preservation of both $U(1)$s is a necessary condition to obtain superconformal invariance in the IR, and the latter is a necessary condition for a string worldsheet theory. As described in the previous section, this also implies that such theories admit two (at least $\ZZ$-graded) topological twists that will take center stage in Section \ref{sec:worldsheettwist}.

\subsubsection{Mirror symmetry of the 2d $\mathcal{N}=(2,2)$ superalgebra}
\label{sec:MSalgs}
Already, without going into details of any specific theories and Lagrangians, we can describe what mirror symmetry is. Just as in the case of T-duality, it arises from an innocuous-looking worldsheet isomorphism, with dramatic spacetime consequences, that we can see already at the level of the $\mathcal{N}=(2, 2)$ superalgebra. (This may not be so surprising, since we discussed earlier how mirror symmetry can be viewed as a certain sequence of T-dualities.)

We briefly mentioned the structure of the 2d $\mathcal{N}=(2,2)$ supersymmetry algebra in Section \ref{sec:nilpot}, we now describe it in a bit more detail. We denote by $J$ the generator of rotations of the worldsheet so that, e.g., $[J, P_z] = - P_z$. As described above, the 2d $\mathcal{N}=(2,2)$ supersymmetry algebra has $4$ supercharges $Q_\pm,\bar{Q}_\pm$, where the subscript $\pm$ denotes the spin/chirality of the supercharges
\be
[J, Q_\pm] = \pm \tfrac{1}{2} Q_\pm, \qquad [J, \bar{Q}_\pm] = \pm \tfrac{1}{2} \bar{Q}_\pm.
\ee
In addition to the anti-commutation relations presented above, we can introduce two types of complex central charges $Z, \tilde{Z}$ and their conjugates $Z^*, \tilde{Z}^*$; the 2d $\mathcal{N}=(2,2)$ supersymmetry algebra, now in the presence of central charges, is given by the following brackets:
\be
\begin{aligned}
	{}[Q_+, \bar{Q}_+ ] & = 2P_{\bar{z}} \qquad & [Q_-, \bar{Q}_-] & = -2P_z\\
	[\bar{Q}_+, \bar{Q}_-] & = Z \qquad & [Q_+, Q_-] & = Z^*\\
	[\bar{Q}_+, Q_-] & = \tilde{Z} \qquad & [Q_+, \bar{Q}_-] & = \tilde{Z}^*\\
\end{aligned}
\ee
However, because these central terms must commute with all other algebra elements, $Z, Z^*$ must be zero if $R_V$ is conserved, and $\tilde{Z}, \tilde{Z}^*$ must be zero if $R_A$ is conserved. Nonetheless, the introduction of a complex mass $Z$ (resp. twisted complex mass) deforms the nilpotence of of the $B$-type supercharge $Q_B$ (resp. $A$-type supercharge $Q_A$) as we saw in Section \ref{sec:equivcoho}.%
\footnote{It turns out that the comparison can be made rather precise: complex mass deformations of 2d $\mathcal{N}=(2,2)$ theories typically arise from turning on a scalar component $\sigma$ of background vector multiplet coupling to (a torus of) the flavor symmetry $F$, whereby $Z$ acts as $n \sigma$ on the charge $n$ superselection sector. Twisted complex mass deformations of $\mathcal{N}=(2,2)$ theories often arise from Fayet-Illiopoulos, complexified by the 2d $\theta$-angle.} %
These central charges provide the mass of BPS solitons in various theories of interest and so $Z$ (resp. $\tilde{Z}$) is often called a complex mass (resp. twisted complex mass), though we will not study them in these lectures.

The above algebra has several outer automorphisms, i.e. symmetries of the algebra that aren't induced by commutation with some fixed element. As usual, there is the $\ZZ_2$ parity transformation sending $Q_\pm \to Q_\mp$ and $\bar{Q}_\pm \to \bar{Q}_\mp$ and the $\ZZ_2$ charge conjugation that sends $Q_\pm \leftrightarrow \bar{Q}_\pm$. But there is a third $\ZZ_2$ \textit{mirror} automorphism that only exchanges the left-moving supercharges: $Q_-~\leftrightarrow~\bar{Q}_-$. To preserve the $R$-symmetry group, we see that mirror symmetry must exchange the axial $U(1)_A$ and vector $U(1)_V$ $R$-symmetries: $R_A \leftrightarrow R_V$. Similarly, mirror symmetry must exchange complex masses $Z$ and twisted complex masses $\tilde{Z}$. Finally, we note that the mirror automorphism preserves the holomorphic supercharge $Q_H = \bar{Q}_+$ but exchanges the topological supercharges $Q_A:= \bar{Q}_+ + Q_- \leftrightarrow Q_B = \bar{Q}_+ + \bar{Q}_-$.

We say that two $\mathcal{N}=(2, 2)$ theories $\mathcal{T}$ and $\tilde{\mathcal{T}}$ are \textit{mirror} to one another if there is an equivalence of these two theories (e.g. an identification of: states in their physical Hilbert spaces, partition functions, correlation functions of local and extended operators, $\ldots$) that intertwines the supersymmetry generators via the above mirror involution. In particular, since the mirror automorphism exchanges the two topological supercharges, it follows that the $A$-twist of $\mathcal{T}$ must be equivalent to the $B$-twist of $\tilde{\mathcal{T}}$, and vice versa: $\mathcal{T}^A \cong \tilde{\mathcal{T}}^B$ and $\mathcal{T}^B \cong \tilde{\mathcal{T}}^A$. As we are anticipating from this very general discussion, mirror symmetry is not a phenomenon restricted to Calabi-Yaus, even though it was discovered in that context \cite{Candelas:1990rm}. 

\subsubsection{Mirror symmetry of chiral and twisted chiral supermultiplets}
\label{sec:MSchirals}
We saw above that mirror symmetry can be interpreted as a certain outer automorphism of the 2d $\mathcal{N}=(2,2)$ supersymmetry algebra. In this subsection, we describe some simple representations of the $\mathcal{N}=(2, 2)$ superalgebra called \textit{chiral multiplets} and \textit{twisted chiral multiplets}. Although we will not touch upon them in these lectures, there are also \textit{vector multiplets} and \textit{twisted vector multiplets}, used in building supersymmetric gauge theories. There are useful classes of gauge theories called \textit{gauged linear sigma models} or \textit{GLSMs} that can be engineered, for example, to flow to nonlinear sigma models in the IR with Calabi-Yau target.

The representations of the 2d $\mathcal{N}=(2,2)$ SUSY algebra that we will be interested in can be packaged in terms of 2d $\mathcal{N}=(2,2)$ \textit{superspace}. In addition to the complex bosonic coordinates $z,\bar{z}$, we introduce four spinorial, fermionic coordinates $\theta^\pm, \bar{\theta}^{\pm}$. The 2d $\mathcal{N}=(2,2)$ supersymmetry algebra can be realized as translations on superspace: we introduce the fermionic vector fields
\be
\begin{aligned}
	Q_+ & = \partial_{\theta^+} + i \bar{\theta}^+ \partial_{\bar{z}} \qquad & \bar{Q}_+ &= -\partial_{\bar{\theta}^+} - i \theta^+ \partial_{\bar{z}}\\
	Q_- & = \partial_{\theta^-} - i \bar{\theta}^- \partial_{z} \qquad & \bar{Q}_- &= -\partial_{\bar{\theta}^-} + i \theta^- \partial_{z}
\end{aligned}
\ee
from which it follows that the desired anti-commutators, e.g. $[Q_+, \bar{Q}_+] = -2 i \partial_{\bar{z}} = 2 P_{\bar{z}}$. The axial $U(1)_A$ and vector $U(1)_V$ $R$-symmetries naturally arise on superspace as rotations of the fermionic coordinates:
\be
(e^{i \alpha}, e^{i\beta}) \in U(1)_A \times U(1)_V \rightsquigarrow (\theta^\pm, \bar{\theta}^\pm) \mapsto (e^{\mp i \alpha - i \beta}\theta^\pm, e^{\pm i \alpha + i \beta}\bar{\theta}^\pm).
\ee

This realization affords us several natural representations of the 2d $\mathcal{N}=(2,2)$ algebra in terms of functions (or sections of a more general bundle/sheaf over) superspace $S(z,\bar{z}; \theta^{\pm}, \bar{\theta}^{\pm})$ called a \textit{superfield}. The fermionic nature of the coordinates $\theta^\pm, \bar{\theta}^\pm$ implies that a Taylor expansion about $\theta^\pm = \bar{\theta}^\pm = 0$ is necessarily finite; indeed, a general superfield $S(z,\bar{z}; \theta^{\pm}, \bar{\theta}^{\pm})$ incorporates $2^4 = 16$ fields. Once we choose the superfield's intrinsic vector and axial $R$-charges $q_V, q_A$, i.e. the charges of the constant term in the fermionic Taylor expansion, the charges of the component field are uniquely determined via
\be
\begin{aligned}
	e^{i \alpha R_A}S(z,\bar{z}; \theta^{\pm}, \bar{\theta}^{\pm}) &= e^{i \alpha q_A}S(z,\bar{z}; e^{\mp i \alpha}\theta^{\pm}, e^{\pm i \alpha}\bar{\theta}^{\pm})\\
	e^{i \beta R_V}S(z,\bar{z}; \theta^{\pm}, \bar{\theta}^{\pm}) &= e^{i \beta q_V}S(z,\bar{z}; e^{-i \beta}\theta^{\pm}, e^{i \beta} \bar{\theta}^{\pm}) 
\end{aligned}
\ee
Similarly, the fermionic parity of the constituent fields is determined by the intrinsic fermionic parity of the superfield $S(z,\bar{z};\theta^\pm, \bar{\theta}^\pm)$. For example, if $S = a + \theta^+ \alpha + ...$ is a bosonic (resp. fermionic) superfield then $a$ is a boson (resp. fermion) and $\alpha$ is a fermion (resp. boson).

It turns out that a general superfield $S(z,\bar{z}; \theta^{\pm}, \bar{\theta}^{\pm})$ does not lead to an irreducible representation of the 2d $\mathcal{N}=(2,2)$ superalgebra. Superspace comes to our aid once again by providing natural differential operators $D_\pm, \bar{D}_\pm$, called \textit{superderivatives}, that anti-commute with the supersymmetry algebra:
\be
\begin{aligned}
	D_+ & = \partial_{\theta^+} - i \bar{\theta}^+ \partial_{\bar{z}} \qquad & \bar{D}_+ &= -\partial_{\bar{\theta}^+} + i \theta^+ \partial_{\bar{z}}\\
	D_- & = \partial_{\theta^-} + i \bar{\theta}^- \partial_{z} \qquad & \bar{D}_- &= -\partial_{\bar{\theta}^-} - i \theta^- \partial_{z}
\end{aligned}
\ee
Since they commute with $Q_\pm, \bar{Q}_\pm$, we can use them to constrain the components of a superfield. For example, a \textit{chiral superfield} $\Phi(z, \bar{z}; \theta^\pm, \bar{\theta}^\pm)$ is required to satisfy $\bar{D}_{\pm}\Phi=0$ and its complex conjugate \textit{anti-chiral superfield} $\bar{\Phi}$ satisfies $D_{\pm}\bar{\Phi}=0$. Similarly, a \textit{twisted chiral superfield} $\tilde{\Phi}$ satisfies $\bar{D}_+ \tilde{\Phi} = D_- \tilde{\Phi}=0$ (and similarly for the complex conjugate field). 

It is convenient to introduce the shifted coordinates $y= z + i \theta^{-}\bar{\theta}^{-}$, $\bar{y}= \bar{z} - i \theta^{+}\bar{\theta}^{+}$. The chirality constraint $\bar{D}_{\pm}\Phi=0$ implies that the superfield depends on the fermionic coordinates in via $y, \bar{y}$ and $\theta^\pm$. Thus, we can express a chiral superfields as%
\footnote{The expression for chiral super fields in terms of the shifted showcases the simplicity of a chiral multiplet that can be somewhat hidden in its full component expansion. The full expansion is
	\[
	\begin{aligned}
		\Phi &= \phi + \theta^+ \psi_+ + \theta^- \psi_- + \theta^+ \theta^- F + \theta^+ \bar{\theta}^+(-i \partial_{\bar{z}}\phi) + \theta^- \bar{\theta}^-(i \partial_z\phi)\\
		& \qquad + \theta^+ \theta^- \bar{\theta}^+(i \partial_{\bar{z}} \psi_-) + \theta^+ \theta^- \bar{\theta}^-(i \partial_z \psi_+) + \theta^+\theta^-\bar{\theta}^+\bar{\theta^-} (-\partial_{\bar{z}} \partial_z\phi).
	\end{aligned}
	\]
}: %
\be
\Phi(y, \bar{y}; \theta^{\pm}, \bar{\theta}^{\pm}) = \phi(y, \bar{y}) + \theta^+ \psi_+(y, \bar{y}) + \theta^- \psi_-(y, \bar{y}) + \theta^+ \theta^- F(y, \bar{y}).
\ee
We also note that a twisted chiral superfield has a similar expansion in terms of the shifted coordinates $\tilde{y} = z - i \theta^- \bar{\theta}^-, \bar{y} = \bar{z} - i \theta^+ \bar{\theta}^+$:
\be
\tilde{\Phi}(\tilde{y}, \bar{y}; \theta^{\pm}, \bar{\theta}^{\pm}) = v(\tilde{y}, \bar{y}) + \theta^{+}\bar{\chi}_{+}(\tilde{y}, \bar{y}) + \bar{\theta}^- \chi_-(\tilde{y}, \bar{y})+ \theta^+ \bar{\theta}^- E(\tilde{y}, \bar{y})
\ee
The conjugate superfields have a similar component expansions. We think of the chiral superfields $\Phi^n$ and its conjugate $\bar{\Phi}^{\bar{n}}$ as a map from superspace to some complex target space $M$ (composed with a choice of local complex coordinates). The twisted versions have a similar interpretation.

Using these above component expression of chiral superfields, it is straight-forward to compute the action of the supercharges $Q_\pm, \bar{Q}_\pm$; twisted chiral superfields, and their conjugates, are treated similarly. First, when acting on a chiral superfield, i.e. in the coordinates $(y, \bar{y}; \theta^\pm, \bar{\theta}^\pm)$, the vector fields $Q_\pm, \bar{Q}_\pm$ are given as follows:
\be
\begin{aligned}
	Q_+ & = \partial_{\theta^+} \qquad & \bar{Q}_{+} &= -\partial_{\bar{\theta}^+} - 2 i \theta^+ \partial_{\bar{z}}\\
	Q_- & = \partial_{\theta^-} \qquad & \bar{Q}_{-} &= -\partial_{\bar{\theta}^-} + 2 i \theta^- \partial_{z}\\
\end{aligned}
\ee
Since the chiral superfield only depends on $\bar{\theta}^\pm$ through $y, \bar{y}$, we are safe in ignoring the first term of $\bar{Q}_{\pm}$. We then define the action of $Q_\pm$ on the components $\phi, \psi_\pm, F$ via the formula
\be
\begin{aligned}
	Q_\pm \Phi & = \psi_\pm \pm \theta^{\mp} F\\
	& := (Q_\pm \phi) - \theta^+ (Q_\pm \psi_+) - \theta^- (Q_\pm \psi_-) + \theta^+\theta^- (Q_\pm F)\\
\end{aligned}
\ee
from which it follows that the action of $Q_\pm$ on the component field is as follows:
\be
\label{eq:Qchiral}
\begin{aligned}
	& & & \hspace*{-2cm} Q_\pm \phi = \psi_{\pm}\\
	Q_\pm \psi_\pm & = 0 \qquad & Q_\pm \psi_\mp & = \mp F\\
	& & & \hspace*{-2cm} Q_\pm F = 0\\
\end{aligned}
\ee
Similarly, the action of $\bar{Q}_\pm$ on the component fields is:
\be
\label{eq:Qbarchiral}
\begin{aligned}
	\bar{Q}_+ \phi & = 0 \qquad & \bar{Q}_- \phi & = 0\\
	\bar{Q}_+ \psi_+ & = 2i \partial_{\bar{z}} \phi \qquad & \bar{Q}_- \psi_+ & = 0\\
	\bar{Q}_+ \psi_- & = 0 \qquad & \bar{Q}_- \psi_- & = -2i \partial_{z} \phi\\
	\bar{Q}_+ F & = -2 i \partial_{\bar{z}} \psi_- \qquad & \bar{Q}_- F & = -2 i \partial_{z} \psi_+\\
\end{aligned}
\ee

The action of the supercharges $Q_\pm, \bar{Q}_\pm$ on a twisted chiral superfield $\tilde{\Phi}$ can be found in a similar fashion. A slick way to determine it is to use the mirror automorphism described in the previous section: the mirror automorphism naturally acts on the odd coordinates of superspace by exchanging $\theta^- \leftrightarrow \bar{\theta}^-$. In particular, we see that a chiral multiplet is transformed into a twisted chiral multiplet with $\nu = \phi$, $\bar{\chi}_- = \psi_+$, $\chi_- = \psi_-$, and $E = F$. Thus, the action of $Q_+$ and $\bar{Q}_-$ (resp. $\bar{Q}_+$ and $Q_-$) on the components of a twisted chiral multiplet are given by Eq. \eqref{eq:Qchiral} (resp. Eq. \eqref{eq:Qbarchiral}) with this identification.

From this rudimentary analysis, we find an instance of mirror symmetry at the level of supermultiplets: a chiral multiplet is mirror to a twisted chiral multiplet. Admittedly, this mirror symmetry is not much deeper than the mirror symmetry described in Section \ref{sec:MSalgs} above -- twisted chiral multiplets are essentially defined to be mirror to chiral multiplets. More generally, given a theory of chiral multiplets and vector multiplets there is a trivially mirror theory of twisted chiral multiplets and twisted vector multiplets defined by the same data. Most statements of mirror symmetry, however, are much more interesting: they exchange, e.g., a theory of chiral multiplets (and vector multiplets) with another theory of chiral multiplets (possibly without vector multiplets)!

\subsubsection{Landau-Ginzburg models}
\label{sec:MSsigmaLG}
In addition to concisely expressing supermultiplets, superspace is a useful tool for writing manifestly supersymmetric action functions as integrals over superspace. For example, if $K(S(z,\bar{z}; \theta^\pm, \bar{\theta}^\pm))$ is an arbitrary differentiable function of the superfields $S$, it follows that
\be
\int d^2z d^4 \theta\, K\big(S(z,\bar{z}; \theta^\pm, \bar{\theta}^\pm)\big)\,,
\ee
where $\int d^4\theta\, K$ extracts the $\theta^+\theta^-\bar{\theta}^+\bar{\theta}^-$ term of $K$, is automatically supersymmetry invariant: the first term in the $Q_\pm$ (resp. $\bar{Q}_\pm$) variation of $K$ removes $\theta^\pm$ (resp. $\bar{\theta}^\pm$), and hence vanishes upon integration over the fermionic coordinates $\int d^4\theta$; the second term survives the fermionic integration but the result is a total derivative, and hence vanishes upon integration over the bosonic coordinates.  This type of expression is called a \textit{D-term}.

It is worth noting that if the D-term is a function of only chiral multiplets (or only anti-chiral multiplets) the resulting D-term is a total derivative. More generally, in a theory of (bosonic) chiral multiplets $\Phi^n$, $n = 1, \ldots, N$ (and their conjugate anti-chiral multiplets $\bar{\Phi}^{\bar{n}}$) parameterizing some complex manifold $M$, $K$ has an interpretation as a \textit{K{\"a}hler potential}. The shift of the K{\"a}hler potential by a holomorphic function of the chirals $\Phi^n$ and its conjugate 
\be
K(\Phi^n, \bar{\Phi}^{\bar{n}}) \to K(\Phi^n, \bar{\Phi}^{\bar{n}}) + \delta(\Phi^n) + \bar{\delta}(\bar{\Phi}^{\bar{n}})
\ee is a \textit{K{\"a}hler transformation}: the complex target space of 2d $\mathcal{N}=(2,2)$ chiral multiplets is naturally a K{\"a}hler manifold! Explicitly performing the fermionic integration, such a D-term gives
\be
\begin{aligned}
	\int d^2z d^4 \theta K(\Phi^n, \bar{\Phi}^{\bar{n}}) & = \int d^2z \bigg[g_{n \bar{n}}\big(\partial_{\bar{z}}\phi^n \partial_z \bar{\phi}^{\bar{n}} + \partial_z\phi^n \partial_{\bar{z}}\bar{\phi}^{\bar{n}} + i\bar{\psi}^{\bar{n}}_- \mathcal{D}_{\bar{z}} \psi^n_- - i\bar{\psi}^{\bar{n}}_+ \mathcal{D}_z \psi^n_+ \big)\\
	& \hspace{-1cm} + g_{n \bar{n}}\big(F^n - \Gamma^n{}_{ml} \psi^m_+ \psi^l_-\big)\big(\bar{F}^{\bar{n}} - \Gamma^{\bar{n}}{}_{\bar{m}\bar{l}}\bar{\psi}^{\bar{m}}_- \bar{\psi}^{\bar{l}}_+\big) + R_{n\bar{n}m\bar{m}} \psi^n_+ \psi^m_- \bar{\psi}^{\bar{n}}_- \bar{\psi}^{\bar{m}}_+\bigg]
\end{aligned}	
\ee
where $g_{n\bar{n}}:= \partial_{\phi^n}\partial_{\bar{\phi}^{\bar{n}}}K$ is the K{\"a}hler metric; $\Gamma^n{}_{ml}$ and $\Gamma^{\bar{n}}{}_{\bar{m}\bar{l}}$ are the Christoffel symbols for $g_{n\bar{n}}$; $\mathcal{D}_z, \mathcal{D}_{\bar{z}}$ is the pullback of the corresponding Levi-Civita connection, such that, e.g., 
\be
\mathcal{D}_z \psi^n_{+} = \partial_z \psi^n_+ + \Gamma^n{}_{ml} (\partial_z \phi^m) \psi^l_+\,, \qquad \mathcal{D}_{\bar{z}} \psi^n_{-} = \partial_{\bar{z}} \psi^n_- + \Gamma^n{}_{ml} (\partial_{\bar{z}} \phi^m) \psi^l_-\,;
\ee
and $R_{n\bar{n}m\bar{m}}$ the Riemann tensor for $g_{n\bar{n}}$.

If we interpret $\phi, \bar{\phi}$ as a map $\phi:\Sigma \to M$ from the worldsheet $\Sigma$ to the K{\"a}hler target $M$, the remaining fields also admit a clean geometric description.%
\footnote{The following descriptions arise from positing that the chiral superfields $\Phi^n$ transform as holomorphic coordinates on the complex target space. In particular, under a coordinate transformation $\phi^n \to \phi'^{n}(\phi)$, the chiral superfield (in the shifted coordinates) transforms as follows:
	\[
	\begin{aligned}
		\Phi'^n(\Phi) & = \phi'^n(\phi) + \theta^+\bigg(\frac{\partial\phi'^n}{\partial\phi^m}\psi^m_+\bigg) + \theta^-\bigg(\frac{\partial\phi'^n}{\partial\phi^m}\psi^m_-\bigg) + \theta^+\theta^-\bigg(\frac{\partial\phi'^n}{\partial\phi^m} F^m - \frac{\partial^2\phi'^n}{\partial\phi^m\partial\phi^l}\psi^m_+\psi^l_-\bigg)\\
		&:= \phi'^n + \theta^+ \psi'^n_+ + \theta^- \psi'^n_- + \theta^+ \theta^- F'^n\\
	\end{aligned}
	\]
	In particular, the fermions $\psi^n_\pm$ transform as holomorphic tangent vectors. The bosons $F^n$ do not transform as a tensor on the target space, but the shifted field $F^n - \Gamma^n{}_{ml}\psi^m_+\psi^l_-$ transforms as a holomorphic tangent vector.} %
We focus on the chiral multiplet fields, as the anti-chiral multiplet fields are obtained by conjugation. First, the fermions $\psi^n_\pm$ are naturally identified as left/right-handed spinors on $\Sigma$ valued in the pullback of the holomorphic tangent bundle to $M$: $\psi_+ \in \phi^* T^{(1,0)} M \otimes K_\Sigma{}^{1/2}$ and $\psi_- \in \phi^* T^{(1,0)} M \otimes K_\Sigma{}^{-1/2}$. The complex boson $F^n$ doesn't transform as a tensor under coordinate transformations of on the target, but $F^n - \Gamma^n{}_{ml} \psi^m_+ \psi^l_-$ is naturally a section of pullback of the holomorphic tangent bundle $\phi^*T^{(1,0)}M$.

The second type of supersymmetric terms we consider only integrates over half of the fermionic coordinates, but supersymmetry restricts the allowed integrand. In particular, if $W(\Phi^n)$ is a holomorphic function of chiral superfields $\Phi^n$, it follows that
\be
\int d^2z d^2 \theta\, W(\Phi^n)\big|_{\bar{\theta}^\pm = 0}\,,
\ee
where $\int d^2 \theta\, W|_{\bar{\theta}^\pm = 0}$ extracts the $\theta^+ \theta^-$ term of $W$, is supersymmetry invariant via a similar argument to the $D$-term: the variation with respect to $Q_\pm$ removes a $\theta^\pm$, hence the result vanishes upon fermionic integration; the variation with respect to $\bar{Q}_\pm$ doesn't vanish under fermionic integration, but the result is a total derivative. This type of expression is called an \textit{F-term}, and $W$ is called the \textit{superpotential}. It is important to note that the invariance of the F-term with respect to $\bar{Q}_\pm$ requires the fact that the superpotential $W$ is itself a chiral superfield, e.g. a holomorphic function of chiral superfields. Explicitly performing the fermionic integral, the superpotential/F-term contributes
\be
\int d^2z d^2\theta W(\Phi^n)\big|_{\bar{\theta}^\pm = 0} = \int d^2z \big[F^n W_n - W_{nm}\psi^n_+ \psi^m_- + \bar{F}^{\bar{n}} \bar{W}_{\bar{n}} - \bar{W}_{\bar{n}\bar{m}}\psi^n_- \psi^m_+\big],
\ee
where $W_n = \partial_{\phi^n} W$, $W_{nm} = \partial_{\phi^n} \partial_{\phi^m} W$, and so on. We can similarly construct a supersymmetric \textit{twisted F-term} as the integral of a holomorphic function $\tilde{W}(\tilde{\Phi}^{\tilde{n}})$, the \textit{twisted superpotential}, of twisted chiral superfields:
\be
\int d^2z d^2 \tilde{\theta}\, \tilde{W}(\tilde{\Phi}^{\tilde{n}})\big|_{\bar{\theta}^+ = \theta^- = 0}\,,
\ee
where $\int d^2 \tilde{\theta} \tilde{W}|_{\bar{\theta}^+ = \theta^- = 0}$ extracts the $\theta^+ \bar{\theta}^-$ term of $\tilde{W}(\tilde{\Phi}^{\tilde{n}})$.

When we combine the D-term and F-term in theories of chiral (and anti-chiral) multiplets, we find a family of 2d $\mathcal{N}=(2,2)$ theories labeled by a K{\"a}hler manifold $M$ and holomorphic superpotential $W: M \to \CC$:
\be
S = \int d^2z d^4\theta\, K + \tfrac{1}{2}\bigg(\int d^2z d^2 \theta\, W\big|_{\bar{\theta}^\pm=0} + \textrm{c.c.}\bigg)\,.
\ee
These theories often go by the name \textit{Landau-Ginzburg models}, but are merely a dimensional reduction of the 4d Wess-Zumino model \cite{WZ74, Z79}. It is worth noting that the complex bosons $\bar{F}^{\bar{n}}$ are auxiliary fields and their equations of motion specialize them, after dualizing with the K{\"a}hler metric, to
\be
\bar{F}_n - \Gamma_{n \bar{m} \bar{l}} \bar{\psi}^{\bar{m}}_- \bar{\psi}^{\bar{l}}_+ = W_n.
\ee
Written as a (pulled-back) holomorphic 1-form, this reads $\bar{F} = \partial_M W$, where $\partial_M = d\phi^n \partial_{\phi^n} $ the holomorphic exterior derivative on $M$. The conjugate fields $F^n$ are similarly specialized: their equations of motion read (after shifting and dualizing) $F = \bar{\partial}_M \bar{W}$, where $\bar{\partial}_M$ is the anti-holomorphic exterior derivative, or \textit{Dolbeault differential}, on $M$.

We can ask when this Landau-Ginzburg model preserves the two $U(1)$ $R$-symmetries of the 2d $\mathcal{N}=(2,2)$ algebra. The D and F terms must preserve the symmetries independently, since they do not mix under the vector or axial rotations. In the case of the D-term, $d^4\theta$ is invariant under both axial and vector $R$-symmetry rotations, so as long as one can make a charge assignment for the chiral superfields such that $K(\Phi^n, \bar{\Phi}^{\bar{n}})$ has total charge 0 under each symmetry, the D-term will preserve both. For example, if $K$ is only a function of the combinations $\Phi^n \bar{\Phi}^{\bar{n}}$, it will be invariant under any charge assignment.

The superpotential term is more interesting. Here, $d^2\theta$ transforms with charge $-2$ under the vector $R$-symmetry $U(1)_V$ and charge $0$ under the axial $U(1)$. Therefore, in order to preserve the vector $R$-symmetry, the superpotential has to have overall charge $2$ under the vector $R$-symmetry $U(1)_V$ and charge $0$ under the axial $R$-symmetry $U(1)_A$. For the axial symmetry, it is common to just assign all chiral fields $U(1)_A$ charge 0. For the vector symmetry, an overall charge 2 is possible if we take the form of the superpotential to be \textit{quasihomogeneous} (of degree 2): for some choice of $U(1)_V$ charges $q_n$ for $\Phi^n$ we have $W(\lambda^{q_n}\Phi^n) = \lambda^2 W(\Phi^n)$.

So far, this has been a classical analysis, but there are interesting anomalies that may prevent the symmetries from being preserved at the quantum level, i.e. the path integral measure may not be invariant. We will simply state the conclusion of this analysis and refer to (e.g.) \cite{HKKPTVVZ03} for details. Since the fermions are charged under the $R$-symmetries, one must undertake a study of fermionic zero modes in the path integral measure. It turns out that the superpotential is not perturbatively renormalized, so symmetry-breaking corrections cannot be generated in perturbation theory; thus, the vector $R$-symmetry $U(1)_V$ is a quantum symmetry so long as we choose a quasihomogenous superpotential (of degree 2).

The axial $R$-symmetry is not as lucky, and often suffers from an anomaly. An application of the beautiful Atiyah-Singer index theorem enables us to state the result geometrically: the K{\"a}hler manifold $M$ must have a vanishing first Chern class $c_1(M) = 0$. In particular, if $\mathcal{X}$ is a Calabi-Yau manifold and $W:\mathcal{X} \to \CC$ a quasihomogeneous superpotential, then both the vector and axial symmetries are non-anomalous symmetries. Conversely, the vector/axial $R$-symmetries will be broken/anomalous for Landau-Ginzburg models with a general K{\"a}hler target and generic superpotential.

We end this section by noting that that mirror symmetry of Landau-Ginzburg theories does not preserve the presence of superpotentials, nor does it preserve non-trivial curvature. For example, Section 2.2.2 of the classic work \cite{HV00} provides three simple examples of the various possibilities, e.g. a non-linear sigma model with target $\PP^{N-1}$ (without superpotential) is mirror to a theory of $N-1$ chirals $\Phi^n$ (valued in $\CC^*$) with superpotential $W = \Phi^1 + \ldots + \Phi^{N-1} + q (\Phi^1\ldots\Phi^{N-1})^{-1}$, also called $A_{N-1}$ Toda theory.%
\footnote{We note that this instance of mirror symmetry was known prior to \cite{HV00} from topologically twisted theory, see e.g. \cite{CV93, EHY95, EHX97, G95, G96, Gi97}. The work \cite{HV00} extended this to the full, physical theory by summing over instanton contributions in (a gauged linear $\sigma$-model realization of) the $\PP^{N-1}$ theory and thereby identifying the mirror Landau-Ginzburg model.} %
This latter example is particularly interesting because $c_1(\PP^{N-1}) = N \neq 0$, hence the sigma model has an anomalous axial $R$-symmetry $U(1)_A$, which is mirror to the superpotential of $A_{N-1}$ Toda theory being inhomogenous, hence the Landau-Ginzburg model breaks $U(1)_V$.%
\footnote{It turns out that there is a non-trivial discrete subgroup of these $R$-symmetry groups that remains in the quantum theory. In the Landau-Ginzburg model, there is a $\ZZ_{2N}$ subgroup of $U(1)_V$ that is unbroken: the superpotential transforms homogeneously (with weight 2) under $\Phi^n \to \zeta \Phi^n$ for $\zeta$ an $N$-th root of unity, i.e. $\Phi^n$ has $\ZZ_{2N}$ charge $2$. Although we cannot use the discrete vector $R$-symmetry to perform an $A$ twist of the Landau-Ginzburg model, it does the refine the $\ZZ_2$ grading of the $B$ twist to a $\ZZ_{2N}$ grading. This is mirror to a non-anomalous $\ZZ_{2N}$ axial $R$-symmetry of the $\PP^{N-1}$ sigma model and the corresponding $\ZZ_{2N}$ grading present on its $A$ twists.} 

\subsection{Twisting the worldsheet}
\label{sec:worldsheettwist}
Let us work now with a special case of a 2d $\mathcal{N}=(2, 2)$ quantum field theory with obvious relevance to string theory: a superconformal field theory with central charge $c=3d$. For a superconformal sigma model, $d$ denotes the dimension of the target space, so that $c=9$ could describe a Calabi-Yau threefold. There are no central terms allowed in the algebra since both vector and axial $R$-symmetries are preserved. Although mirror symmetry is a duality of the physical string theory, topologically twisting the theories will enable us to extract computable observables on both sides of a mirror duality. 

\subsubsection{Chiral and twisted chiral rings}
As we saw above, the topological supercharges $Q_A$ and $Q_B$ are exchanged under the mirror map. For superconformal theories, both $U(1)$ $R$-symmetries are preserved, and so both $Q_A$ and $Q_B$ lead to $\ZZ$-graded topological twists. We can consider the $Q$-cohomology ($Q = Q_A$ or $Q_B$) of states or operators for either of these supercharges -- in the case of most interest, where the theory is superconformal, there will of course be an isomorphism between states and operators. Our focus in this section, and throughout these notes, will be on the operatorial point of view.  

A local operator $O$ is called chiral\footnote{We follow the terminology and notation of \cite{HKKPTVVZ03} throughout this section.} if it satisfies $[Q_B, O] = 0$ and twisted chiral if $[Q_A, O]=0$; in the language of Section \ref{sec:twistinto}, chiral operators are $Q_B$-closed while twisted chiral operators are $Q_A$-closed. For example, the analysis at the end of Section \ref{sec:MSchirals} implies that the lowest component of a free chiral multiplet (resp. free twisted chiral multiplet) is a chiral (resp. twisted chiral) local operator. Importantly, since $[Q_A, Q_A] = [Q_B, Q_B] = 0$ it follows that $[Q_B, O']$ is trivially chiral and $[Q_A, O']$ is trivially twisted chiral for any local operator $O'$; we define the \textit{chiral ring} to be the collection of chiral local operators $O$ modulo those local operators that are trivially chiral, i.e. the chiral ring is simply the $Q_B$-cohomology of local operators. Similarly, we define the \textit{twisted chiral ring} to be the $Q_A$-cohomology of local operators. 

As we will see shortly, the chiral ring and twisted chiral ring naturally have the structures of graded-commutative algebras; in fact, in Section \ref{sec:sec-prod} we will see that they also have natural (shifted) Poisson brackets. We immediately see that if $\mathcal{T}$ and $\tilde{\mathcal{T}}$ are mirror theories, then there must be a suitable algebra isomorphism between, e.g., the chiral ring of $\mathcal{T}$ and the twisted chiral ring of $\tilde{\mathcal{T}}$. Thus, if we wish to check the putative mirror symmetry of $\mathcal{T}$ and $\tilde{\mathcal{T}}$, the (potentially very hard) task of identifying all local operators in $\mathcal{T}$ with local operators in $\tilde{\mathcal{T}}$ can be first checked by the (easier and necessary, but not sufficient) task of matching their chiral ring and twisted chiral ring.

Let us now show that the chiral ring naturally has the structure of a graded-commutative algebra; the analogous result for the twisted chiral ring follows simply by replacing $Q_B$ with $Q_A$. First, we note that the $Q_B$-cohomology of local operators doesn't depend on the insertion points of the local operators. Given a chiral operator $O$, a short computation shows that its translations $\partial_z, \partial_{\bar{z}}$ on the worldsheet are $Q_B$-exact (similarly for twisted chiral operators and $Q_A$-exactness), e.g.:
\[
\tfrac{i}{2}\partial_{\bar{z}} O = [P_{\bar{z}}, O] = [[\bar{Q}_+, Q_+], O] = [Q_B, [Q_+, O]].
\]

Given two chiral operators, their product is also a chiral operator (similarly for twisted chirals). One just defines the product by colliding the operators, or taking them to coincident points. The lack of dependence on position arising from topological invariance means that the product must be nonsingular (up to $Q_B$-exact terms) as the points become coincident. Thus, taking $Q_B$-cohomology reduces the full operator product of these chiral operators to an \textit{ordinary} product, and the operator algebra to an algebra in the usual sense! Moreover, the $Q_B$-exactness of translations of chiral operators implies that it doesn't matter what order we collide a collection of chiral local operators: the algebraic product of chiral operators is associative. Since there is always the trivial local operator $1$, which we assume is not $Q_B$-exact, we find that this is a unital, associative algebra. Finally, since there is no preferred direction to perform the product, the associative product must moreover be graded-commutative -- the only signs that appear in commuting to chiral operators are from the fermionic parity of the operators.

\subsubsection{The chiral ring and Dolbeault cohomology}
\label{sec:CRdolbeault}
Let's start by analyzing local operators in the $B$-twist of a 2d $\mathcal{N}=(2,2)$ sigma model with Calabi-Yau target $\mathcal{X}$, i.e. the theory's chiral ring. We assume that there are $N$ chiral multiplets $\Phi^n$ (and their conjugate anti-chiral multiplets $\bar{\Phi}^{\bar{n}}$) with vanishing axial $R$-charge and vector $R$-charge. Using the vanishing axial $R$-charge of our chirals, it is a straightforward procedure to reorganize the fields based off of their $B$-twisted spin $J_B = J + \tfrac{1}{2}R_A$ and cohomological grading $C_B = -R_V$; we organize the spins $J$, $R$-charges, twisted spin $J_B$ and cohomological grading in Table \ref{table:Bcharges}.

\begin{table}[h!]
	\centering
	\begin{tabular}{c|c|c|c|c|c|c}
		& $\phi$   & $\bar{\phi}$  & $\psi_\pm$  & $\bar{\psi}_\pm$ & $F$ & $\bar{F}$ \\ \hline
		$J$ & $0$ & $0$ & $\pm \tfrac{1}{2}$ & $\pm \tfrac{1}{2}$ & $0$   & $0$\\
		$(R_A, R_V)$ & $(0,0)$ & $(0,0)$ & $(\pm 1, 1)$ & $(\mp 1, -1)$ & $(0,2)$   & $(0,-2)$\\
		$J_B$ & $0$ & $0$ & $\pm 1$ & $0$ & $0$   & $0$\\
		$C_B$ & $0$ & $0$ & $-1$ & $1$ & $-2$   & $2$\\
	\end{tabular}
	\caption{Spin $J$, axial and vector $R$-charges $(R_A, R_V)$, twisted spin $J_B = J + \tfrac{1}{2}R_A$, and cohomological grading $C_B = -R_V$ for a $B$-twisted chiral and anti-chiral multiplet with vanishing $R$-charges.}
	\label{table:Bcharges}
\end{table}

We now organize our $B$-twisted data into some natural geometric objects. First, we find that the bosons $\phi, \bar{\phi}, \bar{F}$ remain scalars on the worldsheet; we continue to think of $\phi, \bar{\phi}$ as the holomorphic/anti-holomorphic parts of a map $\phi:\Sigma \to \mathcal{X}$ from the worldsheet $\Sigma$ into the Calabi-Yau target $\mathcal{X}$, and dualize the auxiliary boson $\bar{F}$ to a scalar valued in the pullback of the holomorphic cotangent bundle $\bar{F} \in \phi^* T^*_{(0,1)}\mathcal{X}$.%
\footnote{Really, it's $\bar{F}_n - \Gamma_{n \bar{m}\bar{l}} \bar{\psi}^{\bar{m}}_- \bar{\psi}^{\bar{l}}_+$ that yields this $\phi^* T^*_{(1,0)}\mathcal{X}$-valued scalar. By abuse of notation, we call this shifted field dualized with the K{\"a}hler metric by the same name. The same holds to for the worldsheet 2-form $F$ described below.} %
The fermions in the chiral multiplet $\psi^n_+, \psi^n_-$ are naturally reorganized as a 1-form $\rho^n = \tfrac{1}{2i}(-\psi^n_- dz + \psi^n_+ d\bar{z})$%
\footnote{We take the worldsheet differential forms $dz, d\bar{z}$ to be fermionic. We also find it convenient to give them cohomological degree 1
	\[[C_B, dz] = dz \qquad [C_B, d\bar{z}] = d\bar{z}\]
	so that the worldsheet exterior derivative $d_\Sigma = dz \partial_z + d\bar{z} \partial_{\bar{z}}$ is a (fermionic) derivation of cohomological degree 1. Consequently, $\rho$ is naturally a bosonic worldsheet 1-form of cohomological degree 0.} %
on the worldsheet, still valued in the pullback of the holomorphic tangent bundle, $\rho \in \Omega^1(\Sigma, \phi^*T^{(1,0)}\mathcal{X})$, whereas the fermions in the anti-chiral multiplet $\bar{\psi}^{\bar{n}}_\pm$ become worldsheet scalars. We find it convenient to organize them into a (fermionic, cohomological degree 1) $\phi^*T^{(0,1)}\mathcal{X}$-valued scalar $\bar{\psi}^{\bar{n}} = \bar{\psi}^{\bar{n}}_+ + \bar{\psi}^{\bar{n}}_- \in \phi^*T^{(0,1)}\mathcal{X}$ and a (fermionic, cohomological degree 1) $\phi^*T^*_{(1,0)}\mathcal{X}$-valued scalar $\bar{\zeta}_n = g_{n \bar{n}}(\bar{\psi}^{\bar{n}}_+ - \bar{\psi}^{\bar{n}}_-)$. Finally, the boson $F$ naturally becomes a (bosonic, cohomological degree 0) worldsheet 2-form valued in the pullback of the holomorphic tangent bundle $F = -\tfrac{1}{4} F dz d\bar{z} \in \Omega^2(\Sigma, T^{(1,0)}\mathcal{X})$. With these field redefinitions, we find that the action of $Q_B$ is given by
\be
\label{eq:QBreorg}
\begin{aligned}
	Q_B \phi = 0 \qquad Q_B \rho = d_\Sigma \phi \qquad Q_B F = d_\Sigma \rho \hspace{1cm}\\
	Q_B \bar{\phi} = \bar{\psi} \qquad Q_B \bar{\psi} = 0 \qquad Q_B \bar{\zeta} = \bar{F} \qquad Q_B \bar{F} = 0\\
\end{aligned}
\ee
where $d_\Sigma \phi \in \Omega^1(\Sigma, \phi^* T^{(1,0)}\mathcal{X})$ is the (worldsheet) differential of (the holomorphic part of) the map $\phi: \Sigma \to \mathcal{X}$, and $d_\Sigma \rho \in \Omega^2(\Sigma, \phi^* T^{(1,0)}\mathcal{X})$ is the (worldsheet) exterior derivative $d_\Sigma = dz \partial_z + d\bar{z} \partial_{\bar{z}}$ of the 1-form $\rho$.

We can build local operators in the $B$-twist as $Q_B$-closed functions of the worldsheet 0-form fields. Namely, functions of the bosonic fields $\phi, \bar{\phi}$ parameterizing a map $\phi: \Sigma \to \mathcal{X}$, and the section $\bar{F}$ of the pulled-back holomorphic cotangent bundle $\phi^*T^*_{(1,0)}\mathcal{X}$ and the fermionic fields $\bar{\psi}$, valued in $\phi^*T^{(0,1)}\mathcal{X}$, and $\bar{\zeta}$, valued in $\phi^*T^*_{(0,1)}\mathcal{X}$. We can identify functions of these bosonic and fermionic fields as sections of the complex of $(0,\bullet)$-forms with values in exterior powers of the holomorphic tangent bundle $\Omega^{(0,\bullet)}(\mathcal{X}, \bigwedge^\bullet T^{(0,1)}\mathcal{X})$ (pulled back to the worldsheet along $\phi:\Sigma \to \mathcal{X}$). In particular, we have the following identifications:
\be
\bar{\psi}^{\bar{n}} \leftrightarrow d \bar{\phi}^{\bar{n}} \qquad \bar{\zeta}_n \leftrightarrow \partial_{\phi^n}
\ee
We need not consider functions dependent on $\bar{F}$ due to the equation of motion $\bar{F} = 0$: this implies the Ward identity $\langle \bar{F}_n \ldots \rangle = 0$. (If there were a superpotential, we would similarly replace $\bar{F}_n$ by $W_n$.) Using this identification, the supercharge $Q_B$ is reinterpreted as the \textit{Dolbeault differential} on the above complex: $Q_B \leftrightarrow \bar{\partial}_{\mathcal{X}} = d\bar{\phi}^{\bar{n}} \partial_{\bar{\phi}^{\bar{n}}}$. Thus, we conclude that local operators in the $B$-twist are labeled by elements of the \textit{Dolbeault cohomology} of the vector bundle $\bigwedge^\bullet T^{(1,0)}\mathcal{X}$:
\be
\textrm{vector space of local operators: } H^{(0,\bullet)}(\mathcal{X}, \mbox{$\bigwedge$}^\bullet T^{(1,0)}\mathcal{X}) := H\big(\Omega^{(0,\bullet)}(\mathcal{X}, \mbox{$\bigwedge$}^\bullet T^{(1,0)}\mathcal{X}), \bar{\partial}_\mathcal{X}\big).
\ee
As predicted, these local operators are $\ZZ^2$-graded: a $(0,p)$ form valued in $\bigwedge^q T^{(1,0)}\mathcal{X}$ has cohomological degree (the negative of its vector $R$-charge) $q+p$ and internal degree (essentially%
\footnote{The operators with homogeneous $R_A$ charge, as defined above, are built from the original fermionic fields $\bar{\psi}^{\bar{n}}_\pm$. The internal degree presented here is nonetheless a symmetry of the $B$-twisted theory; it conjugates $-R_A$ with the rotation $\bar{\psi}^{\bar{m}}_\pm \to \tfrac{1}{\sqrt{2}}(\bar{\psi}^{\bar{m}}_+ \pm \bar{\psi}^{\bar{m}}_-)$. This rotation of the fermionic field space does not commute with usual rotations since it mixes fields of different spin, but it \textit{does} commute with twisted rotations and hence is a valid internal symmetry of the twisted theory.} %
the negative of its axial $R$-charge) $p-q$.

Let's consider these cohomology groups for affine space: $\mathcal{X} = \CC^N$. Every $\bar{\partial}_{\CC^N}$-closed $(0,p)$ form is $\bar{\partial}_{\CC^N}$-exact by a holomorphic version of the Poincar{\'e} lemma. In particular, the cohomology group $H^{(0,p)}(\CC^N, \bigwedge^q T^{(1,0)}\CC^N)$ vanishes unless $p = 0$. The $\bar{\partial}_{\CC^N}$-closed $(0,0)$ forms are simply holomorphic functions of $\phi$, therefore we conclude that $H^{(0,0)}(\CC^N, \bigwedge^q T^{(1,0)}\CC^N)$ consists of holomorphic \textit{polyvector fields}:
\be
H^{(0,\bullet)}(\CC^N, \mbox{$\bigwedge$}^\bullet T^{(1,0)}\CC^N) = H^{(0,0)}(\CC^N, \mbox{$\bigwedge$}^\bullet T^{(1,0)}\CC^N) \cong \{f(\phi) \partial_{\phi^{n_1}} \wedge \ldots \wedge \partial_{\phi^{n_k}} \} \cong \CC[\phi^n, \bar{\zeta}_n].
\ee

The fixed locus of $Q_B$ consists of \textit{locally constant} maps from the worldsheet to the target, i.e. the $Q_B$-invariant field configurations have $d_\Sigma \phi = 0$. This implies that insertions of the above local operators are independent of their position on the worldsheet. Moreover, the space of constant maps is $\mathcal{X}$ itself, and the pointlike maps admit no nonperturbative stringy corrections. See Appendix \ref{sec:localization} for a brief review of localization, and its role in computing supersymmetric observables such as correlation functions of chiral ring elements. Roughly speaking, computing correlation functions in the $B$-model involves integrating Dolbeault cohomology classes $[\omega] \in H^{(0,\bullet)}(\mathcal{X}, \bigwedge^\bullet T^{(1,0)}\mathcal{X})$ over $\mathcal{X}$, which is a classical geometry problem, unlike the much more subtle $A$-model side we will soon address. Thus, the vector space of local operators in the $B$-twist agrees with the (classical) wedge product of differential forms -- it doesn't receive any quantum corrections!

Indeed, a more detailed analysis shows that the entire $B$-model action is $Q_B$-exact. It is a little bit less easy to see that the $B$-model only depends on the \textit{complex structure} of the target manifold, but this turns out to be true: the B-model is simultaneously a topological theory on the worldsheet and ``half-topological'' theory in the target space, depending on the geometric deformations that are complementary to those of the $A$-model. One way to see this is to note that we can contract a (representative of a) cohomology class $\omega \in \Omega^{(0,q)}(\mathcal{X}, \bigwedge^p T^{(1,0)}\mathcal{X})$ over $\mathcal{X}$ with the holomorphic volume form $\Omega$ of the Calabi-Yau to get an ordinary differential form of Dolbeault degree $(n-p,q)$; we can integrate such a differential form over $\mathcal{X}$ to get a vanishing answer only when $p = q = n$, and the holomorphic volume form depends on the choice of complex structure.

\subsubsection{The twisted chiral ring and quantum cohomology}
\label{sec:TCRqcoho}

%In the A-twisted theory, $Q_A$ now has spin 0. The physical operators in the A-twist are those commuting with $Q_A$: the operators in the twisted chiral ring. In the B-twisted theory, $Q_B$ has spin 0, and the physical operators are the chiral ring operators. 
%
%The twist modifies the stress-energy tensor of the theory, shifting it by suitable combinations of background R-symmetry currents:
%\[
%T^{tw}_{\mu\nu} = T_{\mu \nu} + {1 \over 4}(\epsilon^{\lambda}_{\mu}\partial_{\lambda}J^R_{\nu} + \epsilon^{\lambda}_{\nu}\partial_{\lambda}J^R_{\mu}).
%\]
%This twisted stress-energy tensor is $Q$-exact, $T^{tw}_{\mu \nu} = \left\lbrace Q, G_{\mu \nu} \right\rbrace$, meaning that the metric dependence of the twisted theory is cohomologically trivial. Both the A and B-twisted theories are topological field theories.

Let's now move to the A-twist of our 2d $\mathcal{N}=(2,2)$ Calabi-Yau sigma model. We continue to take the superpotential to be zero, and view it as a theory of multiple chiral multiplets with a suitable K{\"a}hler potential; we also take the chiral multiplets to have vanishing vector and axial $R$-charges. The corresponding twisted spin and cohomological grading are given in Table \ref{table:Acharges}.

\begin{table}[h!]
	\centering
	\begin{tabular}{c|c|c|c|c|c|c}
		& $\phi$   & $\bar{\phi}$  & $\psi_-, \bar{\psi}_+$  & $\psi_+, \bar{\psi}_-$ & $F$ & $\bar{F}$ \\ \hline
		$J_A$ & $0$ & $0$ & $0$ & $1,-1$ & $1$   & $-1$\\
		$C_A$ & $0$ & $0$ & $1$ & $-1$ & $0$   & $0$\\
	\end{tabular}
	\caption{Twisted spin $J_A = J + \tfrac{1}{2}R_V$, and cohomological grading $C_A = -R_A$ for an $A$-twisted chiral and anti-chiral multiplet with vanishing $R$-charges.}
	\label{table:Acharges}
\end{table}

In terms of component fields, we have the bosons $\phi, \bar{\phi}$, which continue to furnish a map $\phi: \Sigma \rightarrow \mathcal{X}$. Just as in the $B$-twist, two of the fermionic worldsheet spinors, this time $\psi_-$ and $\bar{\psi}_+$, become (fermionic, degree 1) worldsheet scalars that geometrically realize sections of (the pullback of) the holomorphic tangent bundle $\chi:= \psi_- \in \phi^*T^{(1,0)}\mathcal{X}$ and the anti-holomorphic tangent bundle $\bar{\chi}:= \bar{\psi}_+ \in \phi^*T^{(0,1)}\mathcal{X}$. The other two fermions $\psi_+$ and $\bar{\psi}_-$, as well as the auxiliary bosons $F, \bar{F}$, are naturally identified as components of holomorphic/anti-holomorphic world sheet 1-forms: the fermions become $\bar{\xi}:= -\frac{1}{2i} \bar{\psi}_- dz$ and $\xi:= \frac{1}{2i} \psi_+ d\bar{z}$ (both bosonic, degree 0) and the bosons become $\bar{F} = -\frac{1}{2i} \bar{F} dz$ and $F = \frac{1}{2i} F d\bar{z}$ (fermionic, degree 1). It follows that the action of $Q_A$ on these fields is given by
\be
\label{eq:QAreorg}
\begin{aligned}
	Q_A \phi & = \chi \qquad & Q_A \chi & = 0 \qquad & Q_A \xi &= \bar{\partial}_\Sigma \phi + F \qquad & Q_A F & = \bar{\partial}_\Sigma \chi\\
	Q_A \bar{\phi} & = \bar{\chi} \qquad & Q_A \bar{\chi} & = 0 \qquad & Q_A \bar{\xi} & = \partial_\Sigma 
	\bar{\phi} - \bar{F} \qquad & Q_A \bar{F} & = -\partial_\Sigma \bar{\chi}
\end{aligned}
\ee
where $\partial_\Sigma, \bar{\partial}_\Sigma$ are the worldsheet Dolbeault operators.

We can obtain the vector space of local operators as in the $B$-twist. The only Lorentz-invariant options must be built from the twisted worldsheet scalars $\phi, \bar{\phi}, \chi, \bar{\chi}$. Any other putative scalars that might be constructed by contraction of indices with the worldsheet metric turn out to be $Q$-exact by $Q$-exactness of the worldsheet metric. Since the fermionic fields $\chi, \bar{\chi}$ take values in the holomorphic and anti-holomorphic tangent bundles of $\mathcal{X}$, we can identify them with forms:
\be
\chi^n \leftrightarrow d \phi^n \qquad \bar{\chi}^{\bar{n}} \leftrightarrow d\bar{\phi}^{\bar{n}}
\ee
Then $Q_A$ can be naturally identified with the de Rham differential $d_\mathcal{X} = d\phi^n \partial_{\phi^n} + d\bar{\phi}^{\bar{n}} \partial_{\bar{\phi}^{\bar{n}}}$ acting on general differential forms on $\mathcal{X}$! Thus, local operators of the $A$-model can be identified with the de Rham cohomology of the target manifold:
\be
\textrm{vector space of local operators: } H^{\bullet}(\mathcal{X}) := H\big(\Omega^{(\bullet,\bullet)}(\mathcal{X}), d_\mathcal{X}\big).
\ee
Just as in the $B$-twist, the $A$-twist has a $\ZZ^2$-grading: a $(p,q)$-form has cohomological degree $C_A$ (the negative of axial $R$-charge) $p+q$ and internal degree (the negative of the vector $R$-charge) $p-q$.

Maps $\phi: \Sigma \to \mathcal{X}$ in the $Q_A$ localization locus (c.f. Appendix \ref{sec:localization}) satisfy $\partial_{\bar{z}}\phi = 0$, so the topological $A$-model localizes to \textit{holomorphic} maps to the target. More explicitly, one can write the twisted sigma model Lagrangian as a $Q_A$-exact term plus a term like $\int_{\phi(\Sigma)} \omega$, an integral of the target space (complexified) K{\"a}hler class over the image of the worldsheet, which only depends on the homology class of the map. This term only depends on the K{\"a}hler class of the target manifold, but not its complex structure, and the homology class $\beta = [\phi(\Sigma)]$ of the worldsheet. Thus, the $A$-model is not only topological on the worldsheet, but it is half-topological in the target space. All of the complex structure dependence is $Q_A$-exact.

When $\Sigma = S^2$, these maps compute, via correlation functions between chiral ring elements, the intersection numbers of cycles in the target manifold $\mathcal{X}$. However, it turns out that nontrivial \textit{worldsheet instantons} can contribute to these observables. These are nonperturbative effects in the worldsheet sigma model so the $A$-model is the ``hard'' side of the mirror symmetry duality. From correlation functions in the $A$-model, these quantum-corrected intersection numbers are related to enumerative invariants known as \textit{Gromov-Witten invariants} (roughly, counts of holomorphic curves in the target manifold), a rich industry in algebraic geometry in its own right. These quantum corrections also deform the classical wedge product of differential forms: the twisted chiral ring is identified with the \textit{quantum cohomology} of $\mathcal{X}$. See, e.g., \cite[Chapters 21-30]{HKKPTVVZ03} for a comprehensive introduction to this subject. Mirror symmetry is useful in part because computations in the simpler $B$-model side can be used to extract these intricate invariants on the $A$-side.

\subsubsection{Deformations}
We will discuss a bit more in the sequel the deformations on which the topological (i.e. chiral and twisted chiral) rings depend in the topological $A$- and $B$-models. As we have seen, in the former case, they are a function of the K{\"a}hler moduli of the target space and in the latter case, the complex structure moduli, as one might expect for quantities that are exchanged under mirror symmetry. In either case, the deformations span the tangent bundle of the respective moduli space; the chiral rings furnish interesting bundles over geometric moduli spaces including the tangent bundles to the moduli space as subbundles.

\subsection{A brief introduction to homological mirror symmetry}
\label{sec:homMSintro}
In this section we attempt to provide a brief introduction to the basic ingredients of homological mirror symmetry: the categories of $A$- and $B$-branes in the twisted Calabi-Yau sigma models described above. We start by reviewing why topological boundary conditions form a category and then go on to describe the categories of branes in twisted sigma models. Our aim is to provide an intuition for why various structures arise rather than provide a thorough introduction. For more details, we encourage the reader to consult the comprehensive textbooks \cite{HKKPTVVZ03, ABCDKMGSSW09} and references therein.

\subsubsection{Branes and categories in 2d TQFT}
\label{sec:branecat}
We start by introducing an object central to any 2d TQFT: its \textit{category of boundary conditions}. A category is a mathematical object that is at the heart of many mathematical disciplines. In brief, a category $\mathcal{C}$ is a collection of \textit{objects} $\textrm{Obj}(\mathcal{C})$ and, for each order pair of objects $(B_1, B_2) \in \textrm{Obj}(\mathcal{C})$, \textit{morphisms} $\textrm{Hom}_{\mathcal{C}}(B_1, B_2)$ together with a rule to \textit{compose} morphisms
\be
\begin{aligned}
	\textrm{Hom}_\mathcal{C}(B_1, B_2) && \times && \textrm{Hom}_\mathcal{C}(B_2, B_3) && \to &&  \textrm{Hom}_\mathcal{C}(B_1, B_3)\\
	\rotatebox{90}{$\in$} \hspace{0.75cm} &&  && \rotatebox{90}{$\in$} \hspace{0.75cm} &&  &&  \rotatebox{90}{$\in$} \hspace{1cm}\\
	O_{12}\hspace{0.5cm} && && O_{23}\hspace{0.5cm} && \mapsto && O_{23} O_{12}\hspace{0.5cm}
\end{aligned}
\ee
Moreover, the composition of morphisms is required to be associative%
\footnote{As we mention in Section \ref{sec:hom-alg}, associativity of morphisms can be relaxed to associativity \textit{up to homotopy}. In this section, we restrict our attention to strictly associative composition of morphisms.}: %
if we are given morphisms $O_{ij}~\in~\textrm{Hom}_{\mathcal{C}}(B_i, B_j)$, then $O_{34} (O_{23} O_{12}) = (O_{34} O_{23}) O_{12}$. Some examples of categories the reader may be familiar with are:
\begin{enumerate}
	\item \textbf{Set}: the category of sets; objects of \textbf{Set} are sets, and morphisms are functions
	\item \textbf{Top}: the category of topological spaces; objects of \textbf{Top} are topological spaces, and morphisms are homeomorphisms
	\item \textbf{Grp}: the category of groups; objects of \textbf{Grp} are groups, and morphisms are group homomorphisms
	\item $\textrm{Rep}(G)$: the category of representations of a group $G$; objects are representations of $G$, and morphisms are intertwining operators
\end{enumerate}

We claim that boundary conditions in 2d TQFTs have the structure of a category. More generally, extended operators with support on 1-dimensional submanifolds of spacetime, called \textit{line operators} or \textit{line defects}, in any $d$-dimensional TQFT have such a structure. Yet more generally, operators with $k$-dimensional support in such a theory have the structure of a \textit{$k$-category}, which requires the data of \textit{$2$-morphisms} between any pair of morphisms (also called $1$-morphisms) between the same two objects ($0$-morphisms), and so on up to \textit{$k$-morphisms} between appropriate pairs of $(k-1)$-morphisms.

\begin{figure}[h!]
	\centering
	\raisebox{-2cm}{\includegraphics{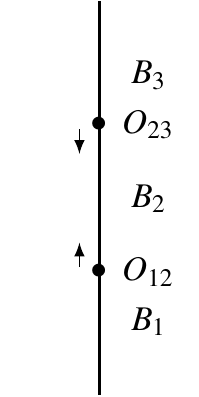}} \qquad $\rightsquigarrow$ \qquad \raisebox{-2cm}{\includegraphics{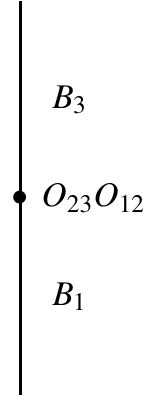}}
	\caption{An illustration of the composition of morphisms in the category of boundary conditions in a 2d TQFT induced by collision of boundary local operators.}
	\label{fig:bdycat}
\end{figure}

The category \textbf{Bdy} of boundary conditions in a 2d TQFT $\mathcal{T}$ arises as follows: objects in \textbf{Bdy} are the boundary conditions $B$ allowed by $\mathcal{T}$. Given two boundary conditions $B_1, B_2$, a morphism $O_{12} \in \textrm{Hom}_{\textbf{Bdy}}(B_1, B_2)$ are local operators on the boundary that interpolate between $B_1$ and $B_2$. The composition of morphisms is induced from colliding boundary local operators; see Figure \ref{fig:bdycat}. There are analogous pictures for the category of line operators in higher dimensional TQFTs. The higher categorical structure of higher dimensional defects arises from considering junctions of various dimension; for example, surface operators (2-dimensional support) form a 2-category whose 1-morphisms are line operators joining two surfaces operators, and whose 2-morphisms are local operators joining such line operators. See, e.g., \cite{K10} and references therein for a more thorough presentation.

\begin{figure}[h!]
	\centering
	\raisebox{-2cm}{\includegraphics{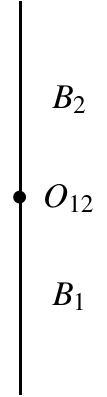}} $\overset{\textrm{state/operator}}{\longleftrightarrow}$ \raisebox{-2cm}{\includegraphics{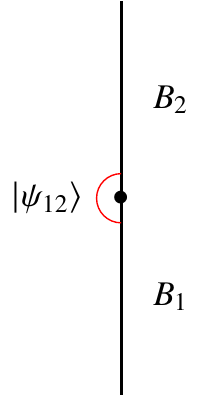}} \qquad $\overset{\textrm{exp/log}}{\longleftrightarrow}$ \qquad \raisebox{-2cm}{\includegraphics{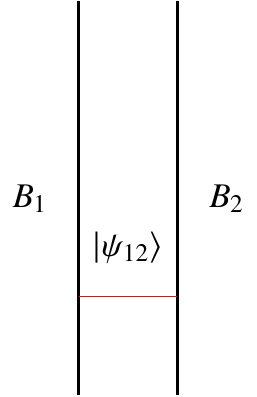}}
	\caption{Three interpretations for the morphism space $\textrm{Hom}_{\textbf{Bdy}}(B_1, B_2)$. Left: local operators interpolating between $B_1$ and $B_2$. Middle: states on a semicircle anchored on the two boundary conditions $B_1, B_2$. Right: states on a strip with boundary conditions $B_1, B_2$ at the two ends.}
	\label{fig:bdystateoperator}
\end{figure}

We can obtain two more interpretations for this space of morphisms by some standard TQFT manipulations. Our second description uses the state-operator correspondence: a boundary local operator $O_{12} \in \textrm{Hom}_{\textbf{Bdy}}(B_1, B_2)$ produces a state on the semicircular arc (or any other homologous cycle) surrounding the insertion, and, conversely, a state on such a semicircle produces a local operator in the limit of an infinitesimal arc; see the middle of Figure \ref{fig:bdystateoperator}. Our third description comes from using the topological nature of the theory to deform this half-space to a strip via the complex logarithm. Under this mapping, the semicircular arc gets mapped to a horizontal line in the strip, thus we map states on the semicircle to states on this strip; see the right of Figure \ref{fig:bdystateoperator}.

\begin{figure}[h!]
	\centering
	\includegraphics{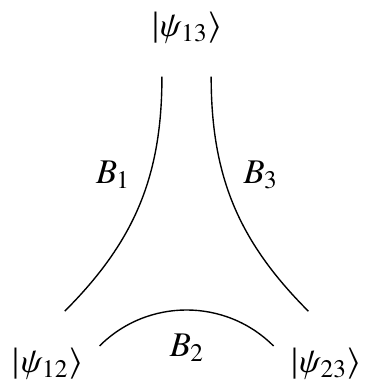}
	\caption{An illustration of the composition of morphisms in the category of boundary conditions in terms of the joining of open strings.}
	\label{fig:stripcomp}
\end{figure}

This final description naturally lends itself to a string-theoretic interpretation: the space of morphisms between two branes $B_1, B_2$ is identified with the space of open string states, with ends satisfying the corresponding boundary conditions. The composition of morphisms in this interpretation comes from joining of opening strings; see Figure \ref{fig:stripcomp}. 

\begin{figure}[h!]
	\centering
	\includegraphics{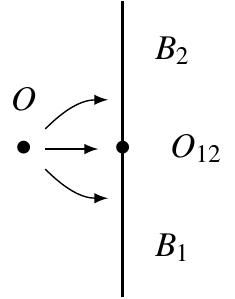}
	\caption{An illustration showing that local operators at the junction of two boundary conditions must commute with the action of bulk local operators. Moreover, they themselves furnish the structure of a module for the algebra of bulk local operators.}
	\label{fig:opcommute}
\end{figure}

It turns out that the category of boundary conditions is rich enough to be able to reconstruct the closed string sector described in Section \ref{sec:worldsheettwist} by way of its \textit{Hochschild cohomology}, c.f. \cite[Section 2.2.3]{ABCDKMGSSW09} or \cite{KR04} and references therein.%
\footnote{Strictly speaking, it is \textit{cyclic cohomology} which is relevant to the topological string. Cyclic cohomology corresponds to the cohomology of cyclically symmetric Hochschild cochains; this requirement of cyclic symmetry is related to the fact that string theory couples the TQFT to topological gravity, whence we must integrate over the moduli of boundary insertions. In the absence of a coupling to gravity, it is the full cohomology that is relevant. \label{fn:cyclic}} %
The argument goes as follows.%
\footnote{The perspective we take is in the spirit of \cite[Section 2.2.3]{ABCDKMGSSW09}. Another useful perspective on the identification of bulk local operators and Hochschild cohomology goes by way of the \textit{folding trick}, c.f. \cite{KR04}. This corresponds to viewing bulk local operators in a 2d theory $\mathcal{T}$ as local operators bound to the ``trivial interface'' between $\mathcal{T}$ and itself. Equivalently, bulk local operators can be viewed as boundary local operators on the \textit{diagonal brane} in $\mathcal{T} \times \mathcal{T}^{-}$, where $\mathcal{T}^{-}$ is obtained from $\mathcal{T}$ by reflecting across the boundary.} %
If $O$ is a bulk local operator and $B$ is a topological brane, we can bring $O$ to the boundary to get a (possibly vanishing) boundary local operator $O|_{B}$, i.e. a morphism $O|_{B} \in \textrm{Hom}_{\textbf{Bdy}}(B,B)$. Moreover, the space of morphisms $\textrm{Hom}_{\textbf{Bdy}}(B_1,B_2)$ between branes $B_1$ and $B_2$ is naturally a module for bulk local operators via collision, i.e. we get a map $O: \textrm{Hom}_{\textbf{Bdy}}(B_1,B_2) \to \textrm{Hom}_{\textbf{Bdy}}(B_1,B_2)$, where $O_{12} \mapsto O \cdot O_{12}$. We can think of the boundary local operator $O|_{B}$ as $O \cdot \textrm{id}_B$, where $\textrm{id}_B$ is the identity operator on $B$. Since we were free to collide the bulk local operator anywhere on the boundary, it follows that these two operations are intertwined:
\be
\label{eq:bulkaction}
O|_{B_2} O_{12} = (O \cdot O_{12}) = O_{12} O|_{B_1}.
\ee
See Figure \ref{fig:opcommute}.

Categorically speaking, the collection of maps $\{O|_{B}\}$ satisfying these relations is a \textit{natural transformation} of the \textit{identity endofunctor} $\textrm{id}_{\textbf{Bdy}}: \textbf{Bdy} \to \textbf{Bdy}$. Let's unpack this statement. First, a (covariant) \textit{functor} is a type of map between categories $\mathcal{F}: \mathcal{C}_1 \to \mathcal{C}_2$ (a.k.a. a 1-morphism in the 2-category of categories). The data of $\mathcal{F}$ includes a map on objects $B$
\be
B \in \textrm{Obj}(\mathcal{C}_1) \mapsto \mathcal{F}(B) \in \textrm{Obj}(\mathcal{C}_2)
\ee
and on morphisms between any two objects $B_1, B_2$
\be
O_{12} \in \textrm{Hom}_{\mathcal{C}_1}(B_1, B_2) \mapsto \mathcal{F}(O_{12}) \in \textrm{Hom}_{\mathcal{C}_2}(\mathcal{F}(B_1),\mathcal{F}(B_2))\,.
\ee
Moreover, these maps are compatible with composition of morphisms; given $O_{12} \in \textrm{Hom}_{\mathcal{C}_1}(B_1, B_2)$ and $O_{23} \in \textrm{Hom}_{\mathcal{C}_1}(B_2, B_3)$, we have the following equality in $\textrm{Hom}_{\mathcal{C}_2}(\mathcal{F}(B_1), \mathcal{F}(B_3))$:
\be
\mathcal{F}(O_{23} O_{12}) = \mathcal{F}(O_{23}) \mathcal{F}(O_{12})\,.
\ee
An \textit{endofunctor} is simply a functor from a category to itself; the \textit{identity endofunctor} $\textrm{id}_{\textbf{Bdy}}$ is the endofunctor of \textbf{Bdy} defined by doing nothing
\be
\textrm{id}_{\textbf{Bdy}}(B) = B \qquad \textrm{id}_{\textbf{Bdy}}(O_{12}) = O_{12}.
\ee
Finally, a \textit{natural transformation} is a map between functors $\eta: \mathcal{F} \to \mathcal{G}$ (a.k.a. a 2-morphism in the 2-category of categories). This is given by the data of maps $\eta_B: \mathcal{F}(B) \to \mathcal{G}(B)$ for every object $B \in \textrm{Obj}(\mathcal{C}_1)$ that intertwine maps of morphisms
\be
\eta_{B_2} \mathcal{F}(O_{12}) = \mathcal{G}(O_{12}) \eta_{B_1} \in \textrm{Hom}_{\mathcal{C}_2}(\mathcal{F}(B_1),\mathcal{G}(B_2)).
\ee
Taking the functors $\mathcal{F}$ and $\mathcal{G}$ to be the identify endofunctor, we see that Eq. \eqref{eq:bulkaction} is exactly the statement that the bulk local operator $O$ defines a natural transformation from $\textrm{id}_{\textbf{Bdy}}$ to itself.

Consider the case where the category has (or, more generally, is generated by) a single boundary condition $B$. It thus suffices to consider the algebra $A$ of local operators on $B$, i.e. $A = \textrm{Hom}_{\textbf{Bdy}}(B,B)$. Then, for every bulk local operator $O$, we have an element $O|_B \in A$; Eq. \eqref{eq:bulkaction} implies that $O|_B$ commutes with every other element of $A$, i.e. $O|_B$ belongs to the \textit{center} of $A$, $O|_{B} \in Z(A)$. This cannot be everything since, e.g., a non-trivial bulk local operator may vanish when brought to the boundary. In a fully derived setting, e.g. when we consider a TQFT obtained via twisting or in the BV formalism, the center of $A$ is merely the zeroth cohomology group of the full Hochschild cohomology of $A$ (with coefficients in $A$), or simply the \textit{derived center} of $A$: $Z(A) \cong HH^0(A) \subset HH^\bullet(A)$. There is a natural generalization to theories with more elaborate categories of topological branes, resulting in the Hochschild cohomology of the category of branes $HH^\bullet(\textbf{Bdy})$. The identification of bulk local operators with the Hochschild cohomology of the category of topological branes is expected to hold quite generally. For example, it is known to recover the \textit{entire} $\sigma$-model chiral ring described in Section \ref{sec:CRdolbeault} from the category of $B$-branes, see e.g. \cite[Section 2.5.3]{ABCDKMGSSW09}, and Kontsevich conjectured the same should be true for the (much more difficult) category of $A$-branes \cite{K94}, see also \cite{GW09, Gaiotto:2021kma}. We describe the construction of Hochschild cohomology of the category of branes, focusing on its relation to topological descent, in Section \ref{sec:hochschild}.

Before moving on to any explicit considerations, we note that if two theories $\mathcal{T}, \tilde{\mathcal{T}}$ are mirror to one another, it immediately follows that (so long as the corresponding twist is well posed) the corresponding categories of boundary conditions are exchanged. For example, the category of topological $A$-branes \textbf{Bdy}${}_A$ in the theory $\mathcal{T}$ should be equivalent to the category of topological $B$-branes $\tilde{\textbf{Bdy}}_B$ in the theory $\tilde{\mathcal{T}}$, and similarly for \textbf{Bdy}${}_B$ and $\tilde{\textbf{Bdy}}_A$. The \textit{homological mirror symmetry conjecture} of Kontsevich \cite{K94} realizes a mathematically precise incarnation of this idea, together with all of the necessary homotopy-theoretic considerations we discuss in Section \ref{sec:homotopy}.

\subsubsection{$B$-branes and coherent sheaves}
\label{sec:Bbranes}
With the knowledge that boundary conditions in 2d TQFTs have the structure of a category, we now wish to describe the category of boundary conditions in $B$-twisted $\mathcal{N}=(2,2)$ theories. We focus on the case of sigma-models, and leave many of the details to Appendix \ref{sec:MFintro}. In that Appendix, we provide a detailed description of ($\tfrac{1}{2}$-BPS) boundary conditions in Landau-Ginzburg models (focusing on the case of a flat target $\mathcal{X} = \CC^N$) in terms of coupling to an auxiliary $\mathcal{N}=2$ quantum mechanical system living on the boundary, also known as \textit{Chan-Paton factors}.

Our first goal is to describe the supersymmetric boundary conditions of the untwisted theory that are compatible with the $Q_B$-twist, i.e. describe the objects in category of $B$-branes \textbf{Bdy}${}_B$. Suppose we prescribe that the boundary values of the bosons $\phi|, \bar{\phi}|$ lie in a submanifold $\mathcal{S} \hookrightarrow \mathcal{X}$, locally cut out by equations $f_i(\phi|, \bar{\phi}|) = 0$. To be invariant under the action of $Q_B \sim \bar{\partial}_{\mathcal{X}}$, we find that these must be \textit{holomorphic} constraints: $\bar{\partial}_{\mathcal{X}}f_i(\phi|, \bar{\phi}|) = 0,$ so that $\mathcal{S}$ is a \textit{holomorphic} submanifold of $\mathcal{X}$. 

The boundary conditions of the remaining fields are determined uniquely by requiring the boundary conditions preserves the full 1d $\mathcal{N}=2$ supersymmetry algebra generated by $Q_B$ and $Q_B^\dagger = Q_+ + Q_-$. For example, if $\phi^1| = \bar{\phi}^{\bar{1}}| = 0$ we must have $\bar{\psi}^{\bar{1}}_+| + \bar{\psi}^{\bar{1}}_-| = 0$ (since $Q_B \bar{\phi}^{\bar{1}} = \bar{\psi}^{\bar{1}} \sim \bar{\psi}^{\bar{1}}_+ + \bar{\psi}^{\bar{1}}_-$), and $\psi^1_+| + \psi^1_-| = 0$ (since $Q_B^\dagger \phi^1 = 2i \rho_t^1 \sim \psi^1_+ + \psi^1_-$). In particular, the fermionic scalar $\bar{\psi}|$ and the bosonic 1-form (pulled back to the boundary) $\rho|$ must be normal to $\mathcal{S}$, and $\bar{\zeta}|$ must be tangent to $\mathcal{S}$.

As described in Appendix \ref{sec:MFintro}, these holomorphic constraints are naturally imposed by coupling to boundary Fermi multiplets. In addition to naturally imposing the above holomorphic constraints, these boundary Fermi multiplets can enrich the holomorphic submanifold with finite-dimensional complexes of holomorphic vector bundles, i.e. with finite-dimensional $\ZZ$-graded (given by $R$-symmetry) holomorphic vector bundles with differential $\delta$. Very roughly speaking, this data encodes a \textit{coherent sheaf}%
\footnote{Somewhat less roughly, note that to each open set $D$ of $\mathcal{X}$ there is an ring $\CC[D]$ of holomorphic functions on $D$. A coherent sheaf $\mathcal{E}$ is an assignment of a finitely generated $\CC[D]$-module $\mathcal{E}(D)$ compatible with gluing on overlaps of open sets. Given a vector bundle $E \to \mathcal{X}$, we can get a coherent sheaf whose corresponding modules are the spaces of holomorphic sections over $D$ $\mathcal{E}(D) = \Gamma_D(E)$ with module structure given by multiplying holomorphic sections by holomorphic functions. Similarly, holomorphic functions on a complex submanifold $\mathcal{S}$ yield such a module via multiplication by functions pulled back along the inclusion $\mathcal{S} \hookrightarrow \mathcal{X}$.} %
over the target space $\mathcal{X}$
\be
B \in \textrm{Obj}(\textbf{Bdy}_B) \leftrightarrow \textrm{ coherent sheaf } \mathcal{E} \in \textrm{Obj}(\textrm{Coh}(\mathcal{X}))\,.
\ee
Moreover, the states of the BPS Hilbert space (on a half-line of a flat half-worldsheet) are identified with (derived) global sections of this coherent sheaf, i.e. sheaf cohomology $H^{(0,\bullet)}(\mathcal{X}, \mathcal{E})$: $Q_B$ localizes to constant maps, so BPS configurations are determined by their value on the boundary, and acts on the boundary fluctuations as the Dolbeault differential. See Appendix \ref{sec:MFintro} for a more algebraic treatment of the origin of coherent sheaves (and more generally matrix factorizations) in this context.

Now that we know that $B$-branes are labeled by coherent sheaves, we move to the problem of determining the space of boundary local operator interpolating between two boundary conditions $B_1, B_2$ labeled by coherent sheaves $\mathcal{E}, \mathcal{F}$. Such a boundary local operator should produce a map from BPS states before the junction to BPS states after, i.e. a map from one space of sections (even better: a map of the underlying complexes) to the other. Moreover, it should commute with multiplying sections by holomorphic functions (i.e. $H^{(0,0)}(\mathcal{X})$), viewed as collision with bulk local operators; c.f. Figure \ref{fig:opcommute}. Thus, we conclude that local operators at the interface between the two boundary conditions $B_1, B_2$ labeled by coherent sheaves $\mathcal{E}$ and $\mathcal{F}$ are can be identified with \textit{morphisms of coherent sheaves}. In fact, all morphisms of coherent sheaves, especially those that have non-trivial $R$-charge/cohomological degree, can be expressed in this fashion. (See e.g. \cite[Chapter 3]{ABCDKMGSSW09}). There is a natural differential on these morphisms, corresponding physically to the action of $Q_B$ on these local operators, and we identify the physical operators with $Q_B$-cohomology classes of morphisms (also called \textit{extensions} or \textit{derived morphisms})
\be
\textrm{Hom}_{\textbf{Bdy}}(B_1, B_2) \cong \textrm{Ext}_{\textrm{Coh}(\mathcal{X})}(\mathcal{E}, \mathcal{F}).
\ee
Finally, we note that collision of local operators on the boundary agrees with the usual composition of morphisms of coherent sheaves.

As an example, consider the $B$-twist of a single free chiral multiplet $\Phi$, i.e. our K{\"a}hler sigma model with target $\CC$. Coherent sheaves on $\CC$ can be identified with modules for the polynomial algebra $\CC[\phi]$. A Dirichlet boundary condition can be engineered by coupling to a boundary Fermi multiplet with $J$-term $J = \Phi$, resulting in the coherent sheaf/complex of $\CC[\phi]$-modules $\CC[\phi, \gamma]$, where $\gamma$ has $R$-charge $-1$ and $\phi$ acts by multiplication, and differential $Q_B \phi = 0, Q_B \gamma = \phi$. Maps from $\CC[\phi, \gamma]$ to itself that commute with multiplication by $\phi$ are generated by $\phi$ (multiply by $\phi$), $\gamma$ (multiply by $\gamma$), and $\partial_\gamma$ (differentiate with respect to $\gamma$): Hom${}_{\CC[\phi]-\textrm{mod}}(\CC[\phi, \gamma], \CC[\phi, \gamma]) = \CC[\phi, \gamma, \partial_\gamma]$. The action of $Q_B$ on these local operators is
\be
Q_B \phi = 0 \qquad Q_B \gamma = \phi \qquad Q_B \partial_\gamma = 0\,,
\ee
from which we conclude that Hom${}_{\textbf{Bdy}}($Dir, Dir$) \cong \CC[\partial_\gamma]$: the algebra of local operators on this Dirichlet boundary condition is simply an exterior algebra in the boundary fermion $\partial_\gamma \sim \bar{\gamma}$.

Two important aspects to take into account, and about which we will say very little, are the related notions of \textit{universality} and \textit{stability}. Universality dictates that we should deem equivalent any two UV boundary conditions, described, e.g., in term of a complex of vector bundles, that flow to the same IR boundary condition. For example, two such boundary conditions will necessarily have the same BPS state spaces (their sheaf cohomology). However, we should not deem equivalent any two coherent sheaves $\mathcal{E}, \mathcal{F}$ with the same cohomology, nor should we do this with boundary conditions $B_1, B_2$. Instead, we require something slightly stronger: we require a morphism/local operator $O: \mathcal{E} \to \mathcal{F}$ such that the induced map on cohomology is an isomorphism. Such a morphism is called a \textit{quasi-isomorphism}. Thus, considerations of universality imply that $B$-branes are labeled by coherent sheaves \textit{up to quasi-isomorphism}. For example, the above coherent sheaf $\CC[\phi, \gamma]$ is quasi-isomorphic to the module $\CC$ where $\phi$ acts as zero, thereby establishing an equivalence (at least as topological $B$-branes) between a Dirichlet boundary condition (the module $\CC$) and an enriched Neumann boundary condition (the module $\CC[\phi, \gamma]$).%
\footnote{We could have similarly considered the $\CC[\phi]$-module $\CC$, where $\phi$ acts as 0. Since this isn't a projective module, we need to resolve it $\CC$ in order to compute the extension group Ext$(\CC,\CC)$. The answer given here can be computed as the $Q_B$ cohomology of any of the three Hom${}_{\CC[\phi]-\textrm{mod}}(\CC[\phi, \gamma], \CC)$, Hom${}_{\CC[\phi]-\textrm{mod}}(\CC, \CC[\phi, \gamma])$, or Hom${}_{\CC[\phi]-\textrm{mod}}(\CC[\phi, \gamma], \CC[\phi, \gamma])$. We chose the third option to emphasize the natural algebra structure.} %
This identification is sometimes called a \textit{flip}, c.f. \cite{DGG14, DGP18}.

It is possible to build a theory of categories based upon complexes up to quasi-isomorphism, more generally called \textit{triangulated categories}, leading the notion of the (bounded) \textit{derived category} $D^b\mathcal{C}$ of a(n abelian) category $\mathcal{C}$. In the present context, we conclude that the category of $B$-branes in a $B$-twisted sigma model with target $\mathcal{X}$ is the (bounded) derived category of coherent sheaves on $\mathcal{X}$, or simply the \textit{derived category} of $\mathcal{X}$:
\be
\textbf{Bdy}_B \cong D^b \textrm{Coh}(\mathcal{X}) = D^b(\mathcal{X})\,.
\ee

Stability, on the other hand, says that many of the $B$-branes described above, although being perfectly good boundary conditions for the topologically twisted theory, are physically unstable. Indeed, the data required to define a boundary condition prior to the twist is much larger than simply a complex of vector bundles. For example, the physical brane requires a connection on the vector bundle satisfying the Hermitian Yang-Mills equations; an arbitrary complex of vector bundles may not admit such a connection. It is possible to translate this physical notion of stability to the derived category, and more general triangulated categories, although we do not do so here. See \cite[Chapter 5]{ABCDKMGSSW09}, and references therein, for more details.

\subsubsection{$A$-branes and the Fukaya category}
\label{sec:Abranes}

We now move on to the category of boundary conditions compatible with the topological $A$ twist of our 2d $\mathcal{N}=(2,2)$ sigma model. As compared to the $B$ twist, which is nearly classical, the $A$ twist is highly quantum, much more difficult, and we will correspondingly say much less. Our hope here it to provide intuition for how the Fukaya category arises in the context of topological $A$-branes in sigma models. For a more thorough introduction to Fukaya categories, we recommend the physical approach of the books \cite{HKKPTVVZ03,ABCDKMGSSW09} as well as the notes of Auroux \cite{A13}.

We again assume that the bosonic fields $\phi, \bar{\phi}$ are constrained to a submanifold $\mathcal{S}$ of our Calabi-Yau target $\mathcal{X}$ (equipped with the trivial vector bundle and connection). Similar considerations to those in Section \ref{sec:Bbranes} show that such a submanifold must be \textit{Lagrangian}, i.e. middle dimensional such that the K{\"a}hler/symplectic form $\omega$ vanishes on $\mathcal{S}$,
for it to be compatible with the 1d $\mathcal{N}=2$ supersymmetry algebra generated by $Q_A$ and $Q_A^\dagger = Q_+ + \bar{Q}_-$.%
\footnote{Depending on the curvature of the Hermitian connection on the vector bundle $E \to \mathcal{S}$, we can enlarge the allowed type of boundary manifold $\mathcal{L}$ to be \textit{coisotropic}, i.e. $\mathcal{S}$ is locally given by equations $f_a(\phi, \bar{\phi})$, $a=1,...,p < N$, whose Poisson brackets vanish on $\mathcal{S}$.}
However, there can be an anomaly in the axial $R$-symmetry for a general Lagrangian: the axial $R$-symmetry is non-anomalous if and only if the Lagrangian has vanishing \textit{Maslov class} $\mu(\mathcal{S}):= \mu(\mathcal{S}, \mathcal{S}) \in H^1(\mathcal{S},\ZZ)$; see, e.g., \cite[Chapter 39.3]{HKKPTVVZ03} or references therein. As described in Section \ref{sec:gradings}, the lack of the axial $R$-symmetry simply implies that the twisted theory will not have the full $\ZZ$ (or $\ZZ^2$) grading expected classically. We also note that there is a notion of universality amongst topological $A$-branes: many seemingly different Lagrangians can give the same $A$-brane. In particular, the $A$-brane should be invariant under small deformations of the underlying Lagrangian -- infinitesimal deformations of the submanifold $\mathcal{S}$ that preserve the Lagrangian property are (at least locally) given by \textit{Hamiltonian flow}, i.e. by a vector field locally of the form $V^I \sim \omega^{IJ}\partial_J H$ with $\omega^{IJ}$ the inverse of the symplectic form. Therefore, we should deem equivalent two Lagrangians that are related by a global Hamiltonian flow, also called \textit{Hamiltonian isotopy}:
\be
B \in \textrm{Obj}(\textbf{Bdy}_A) \leftrightarrow \begin{aligned}
	\textrm{ Lagrangian } \mathcal{L} \textrm{ with } \mu(\mathcal{L})= 0\\
	\textrm{up to Hamiltonian isotopy} \hspace{0.25cm}\\
\end{aligned}
\ee

Now let's consider the space of morphisms between two boundary conditions $B_1, B_2$ labeled by Lagrangians $\mathcal{L}_1, \mathcal{L}_2$. We will use the description of this morphism space in terms of the supersymmetric ground states on a strip as in the right of Figure \ref{fig:bdystateoperator}. As described in Appendix \ref{sec:morseintro}, we can use Witten's interpretation of Morse theory \cite{WittenMorse} (see also \cite[Section 10]{GMW115} for a very readable review) in 1d $\mathcal{N}=2$ quantum mechanics to deduce the answer. In particular, perturbative ground states are given simply by critical points of the Morse potential $h'(X) = 0$. In the present context, specializing to flat $\CC^N$ with $\omega = \delta_{\bar{n} n} d\bar{\phi}^{\bar{n}} \wedge d\phi^n$ for simplicity, we find a Morse potential of the form
\be
h = 2i \int ds\, \bar{\phi} \partial_s \phi\,,
\ee
whose critical points are simply constant maps $\partial_s \phi = \partial_s \bar{\phi} = 0$. If we assume that $\mathcal{L}_1, \mathcal{L}_2$ intersect transversely, or at least that we can choose isotopic Lagrangians $\mathcal{L}_1', \mathcal{L}_2'$ that do intersect transversely, then the perturbative ground states are simply labeled by the points in the intersection $L_1 \cap L_2$.

\begin{figure}[h!]
	\centering
	\raisebox{-2cm}{\includegraphics{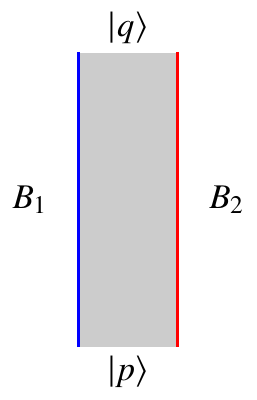}} \qquad $\cong$ \qquad \raisebox{-1.75cm}{\includegraphics{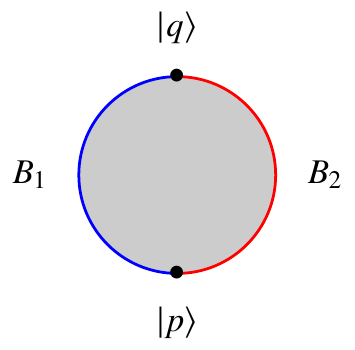}} \qquad $\overset{\phi}{\to}$ \qquad \raisebox{-2cm}{\includegraphics{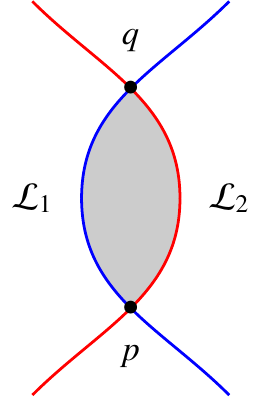}}
	\caption{An illustration of an instanton correction corresponding to tunneling from the perturbative ground state $|p\rangle$ to the perturbative ground state $|q\rangle$. Under the map $\phi: \Sigma \to \mathcal{X}$ coming from the chiral multiplets, this instanton is interpreted as a holomorphic disk being bounded by the Lagrangians $\mathcal{L}_1, \mathcal{L}_2$ describing the boundary conditions $B_1, B_2$ on the boundaries of the strip.}
	\label{fig:stripdisk}
\end{figure}

In addition to perturbative considerations, there are non-perturbative corrections to this coming from tunneling. In particular, there is a differential on the space of critical points counting gradient flow lines, also called \textit{Morse flows}, that asymptote to critical points. In general, such flows are solutions to the gradient flow equation $\partial_t \phi^n + g^{n \bar{n}} \partial_{\bar{n}} h = 0$; in the present context, this translates to counting holomorphic maps! See Figure \ref{fig:stripdisk} for an illustration.

It is not immediately clear that such a count even makes sense -- a priori, the moduli space of holomorphic disks as above may have an infinite number of points! Moreover, it is not even clear that $Q_A$ is a differential, i.e. it squares to zero. As described in the Appendix \ref{sec:morseintro}, the $\ZZ$-grading afforded by the $R$-symmetry comes to the rescue. In the finite dimensional context, the $\ZZ$-grading given to a critical point $p$ of the superpotential $d_M h(p) = 0$ is given by its \textit{Morse index} $\mu(p)$. Explicitly, the Morse index $\mu(p)$ equal to the number of negative eigenvalues of the Hessian. (We assume that the Hessian of $h$ is nondegenerate at each critical point $p$. We can always perturb $h$ so that this is the case.) As explained in Appendix \ref{sec:morseintro}, the Morse index corresponds to the $R$-charge of the perturbative ground state $|p\rangle$. In terms of gradient flow, this counts the number of ``downwards flow'' directions. Since the supercharge $Q$ has $R$-charge $+1$, it follows that we need only consider instantons/Morse flows from the critical point $p$ to critical points $q$ with $\mu(q) = \mu(p) + 1$. By carefully analyzing the Morse flow equation, it is possible to show that the moduli space of Morse flows asymptoting to $q$ (resp. $p$) at $+\infty$ (resp. $-\infty$) has (expected%
\footnote{We note that the analysis of the ordinary differential equations gives us the expected dimension of the moduli space of solutions as the index of a suitable Dirac operator. Nonetheless, if the Morse function and/or metric are degenerate it is possible that the dimension of the moduli space is different from the expected dimension. After a suitable deformation, the moduli space can made to have actual dimension equal to the expected dimension.}) %
dimension $\mu(q) - \mu(p)$. Thus, the moduli space of flows for $\mu(q) = \mu(p) + 1$ has an infinite number of points, but this is to be expected -- we can always shift the ``time'' parameter of the flow to get another solution. After quotienting by these translations, we see that the \textit{reduced moduli space} of flows has (expected) dimension 0, i.e. it is simply a number of points. In particular the differential $Q_A$ simply counts the number of these points \textit{with signs}%
\footnote{We refer to \cite[Appendix F]{GMW115} for the precise and careful treatment of these subtle sign issues.}: %
\be
Q_A |p \rangle = \sum_{q,\, \mu(q) = \mu(p)+1} \begin{pmatrix}\# \textrm{ Morse flows }\\ p \to q \end{pmatrix} |q\rangle
\ee

In order to check that this defines a differential, i.e. $Q_A^2 = 0$, we have to count (with signs) the flows from a given critical point $p$ to another critical point $r$ with $\mu(r) = \mu(p) +2$ that pass through a third critical point $q$ with $\mu(q) = \mu(p) + 1$. Note that the (reduced) moduli space of gradient flows from $p$ to $r$ has (expected) dimension $2-1 = 1$. As described in the Appendix, a connected component of this (reduced) moduli space necessarily has the topology of $\RR$, with the $\pm\infty$ limits exactly corresponding to these composed flows $p \to q \to r$. In particular, a given composed flow $p \to q \to r$ is always paired with another composed flow $p \to q' \to r$ (with $\mu(q') = \mu(p)+1$) with the opposite sign, hence the full count of these composed flows vanishes, i.e. $Q^2_A|p\rangle = 0$.

We can formally apply these arguments to our $A$ twisted sigma model: we have perturbative ground states $|p\rangle$ labeled by the intersection points $p \in \mathcal{L}_1 \cap \mathcal{L}_2$ of the (transversally intersecting) Lagrangians $\mathcal{L}_1, \mathcal{L}_2$; the Morse index of the the intersection point is given by the \textit{Maslov index} of the intersection point; and the Morse flow equations correspond to maps of holomorphic strips/disks with boundaries on the Lagrangians $\mathcal{L}_1, \mathcal{L}_2$. We conclude that the space of morphisms Hom$_{\textbf{Bdy}_A}(B_1, B_2)$ between boundary conditions labeled by (transversely intersecting) Lagrangians $\mathcal{L}_1, \mathcal{L}_2$ is 
\be
\textrm{Hom}_{\textbf{Bdy}_A}(B_1, B_2) = H(\bigoplus_{p \in \mathcal{L}_1 \cap \mathcal{L}_2} | p \rangle, Q_A) \qquad Q_A |p \rangle = \sum_{q, \mu(q) = \mu(p)+1} \begin{pmatrix}\# \textrm{holo. strips/disks}\\ p \to q \end{pmatrix} |q\rangle\,.
\ee
The complex on the right hand side is called the \textit{Floer complex}, with cohomology called \textit{Floer cohomology}. Although the Floer complex depends on many explicit choices, such as the explicit representatives of a given Hamiltonian isotopy class, Floer cohomology does not.

\begin{figure}[h!]
	\centering
	\raisebox{-2cm}{\includegraphics{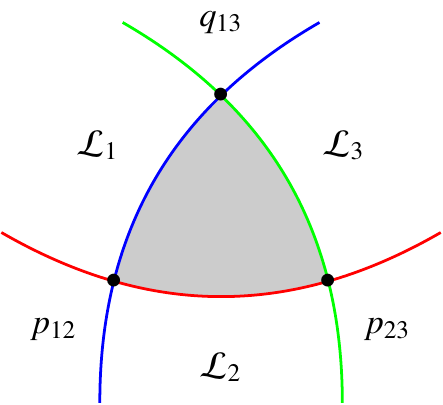}} \hspace{2cm} \raisebox{-1.75cm}{\includegraphics{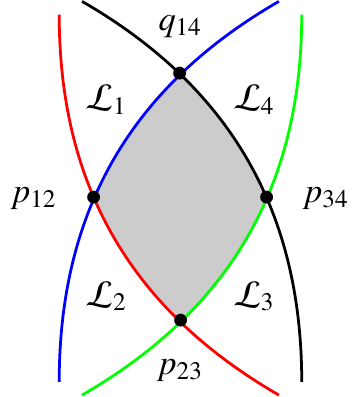}}
	\caption{Illustrations of the holomorphic disks that contribute to the Floer products. Left: A holomorphic disk that contributes to the coefficient of $|q_{13}\rangle$ in the binary Floer product $|p_{12} \rangle \star |p_{23} \rangle$. Right: An illustration of a holomorphic disk that contributes to the ternary Floer product $\mu_3(p_{12}, p_{23}, p_{34})$.}
	\label{fig:Fukayacomp}
\end{figure}

Given three Lagrangians $\mathcal{L}_1$, $\mathcal{L}_2$, and $\mathcal{L}_3$, with transverse pairwise intersections, there is a natural product structure, called the \textit{Floer product}, on the Floer complex that will lead to the composition of morphisms in the category of boundary conditions in the $A$ twist. Roughly speaking, given intersection points $p_{12} \in \mathcal{L}_1 \cap \mathcal{L}_2$, $p_{23} \in \mathcal{L}_2 \cap \mathcal{L}_3$, and $q_{13} \in \mathcal{L}_1 \cap \mathcal{L}_3$, the coefficient of $|q_{23}\rangle$ in $|p_{23}\rangle \star |p_{12}\rangle$ is given by a (signed) count of holomorphic disks with three marked points:
\be
|p_{23}\rangle \star |p_{12}\rangle = \sum_{q_{13}} \begin{pmatrix}\# \textrm{holo. disks}\\ p_{12}, p_{23} \to q_{13} \end{pmatrix} |q_{13}\rangle,
\ee
where the sum is over intersection points $q_{13} \in \mathcal{L}_1 \cap \mathcal{L}_3$ with $\mu(q_{13}) = \mu(p_{12}) +  \mu(p_{23})$. See the left of Figure \ref{fig:Fukayacomp} for an illustration. The Floer product induces an associative composition on the level of Floer cohomology, thereby defining our category of $A$-branes: the objects are Lagrangians $\mathcal{L}$ with vanishing Maslov class $\mu(\mathcal{L})= 0$ (up to Hamiltonian isotopy), with morphisms between $\mathcal{L}_1$ and $\mathcal{L}_2$ given by the Floer cohomology of the intersection, and composition of morphism given by the Floer product; the resulting category is known as the \textit{Fukaya category of $\mathcal{X}$}.
\be
\textbf{Bdy}_A \cong \textrm{Fuk}(\mathcal{X}).
\ee

It is important to note that the Floer product fails to be associative in a highly controlled fashion. In particular, it is \textit{associative up to homotopy} or \textit{homotopy associative}. For example, given four Lagrangians $\mathcal{L}_1, \ldots, \mathcal{L}_4$ and $Q_A$-closed intersection points $p_{12}, p_{23}, p_{34}$, the compositions $p_{12} \star (p_{23} \star p_{34})$ and $(p_{12} \star p_{23}) \star p_{34}$ differ by the $Q_A$ variation of a (formal sum of) intersection point(s) $\mu_3(p_{12}, p_{23}, p_{34}) \in \mathcal{L}_1 \cap \mathcal{L}_4$. Schematically, we count (with signs) holomorphic disks that connect these fixed points and are bounded by the Lagrangians $\mathcal{L}_1, \ldots, \mathcal{L}_4$:
\be
\mu_3(p_{12}, p_{23}, p_{34}) = \sum_{q_{14}} \begin{pmatrix}\# \textrm{holo. disks}\\ p_{12}, p_{23}, p_{34} \to q_{14} \end{pmatrix} |q_{14}\rangle
\ee
where the sum is over intersection points $q_{14} \in \mathcal{L}_1 \cap \mathcal{L}_4$ with $\mu(q_{14}) = \mu(p_{12}) + \mu(p_{23}) +  \mu(p_{34}) - 1$. See the right of Figure \ref{fig:Fukayacomp} for an illustration. There is an entire tower of higher order products $\mu_n(p_{12}, ..., p_{n n+1})$ obtained by counting holomorphic disks asymptoting to the critical points; together with the Floer product $\mu_2 = \star$ and the ternary product $\mu_3$, this tower of operations contain a great deal of information and give the Fukaya category the structure of an \textit{$A_\infty$ category}. The data of these homotopies have a natural physical interpretation in terms of \textit{descent} that we turn to in the next section.

\section{Homotopy \& further developments}
\label{sec:homotopy}
In this section we will discuss additional features and enrichments of twisted theories. These considerations will lead to the appearance of homotopy algebras, which loosely speaking govern the structure of operator products modulo homotopies. To access the homotopic structures we are most interested in, we will go beyond the algebra of local operators at the level of cohomology, and instead use chain-level considerations to extract more refined information via a procedure called \textit{descent}. One consequence of such chain-level considerations is the introduction of a natural class of $Q$-closed extended operators, which are crucial ingredients for matching the physics between two dual (twisted) theories. These extended operators first appeared in Witten's study of Donaldson invariants of 4-manifolds, in which he computed 2-form descendants from nontrivial 2-cycles in the 4-manifold \cite{W88}. Descent, and the related notion of \textit{secondary products} which we also review, also have origins in foundational studies of the string worldsheet, e.g. \cite{LZ, Getzler, WZ92}. We will explain the descent procedure and discuss the algebraic structure underlying the existence of these extended operators, as well as the consequences they have for local operators themselves. These have played an important role in the mathematics of the topological string and string field theory, and also feature in interesting dualities.

Some appearances of homotopy algebras are well-known, especially in mirror symmetry, which arises from topological twists on the worldsheet. For example, mathematical statements about how mirror symmetry exchanges $A$- and $B$-branes are often phrased in terms of homotopy algebras and categories, c.f. the \emph{homological mirror symmetry conjecture} of Kontsevich \cite{K94}. Even outside of twisted theories, homotopy Lie algebras, i.e. $L_\infty$ algebras, arise as a fundamental ingredient in string field theories \cite{SZ89, WZ92, Z93} and field theory \cite{HZ17}. Other homotopy algebras, in particular in holomorphically twisted theories, are currently active areas of research in mathematics and physics; see e.g. \cite[Chapter 5]{CG16} for a general discussion of holomorphic field theories or \cite[Section 2]{CDG20} for a detailed discussion of mixed holomorphic-topological theories in 3d.

\subsection{Topological descent}
\label{sec:topdescent}
Let us continue to consider the case of a \textit{topologically} twisted theory (including the choice of twisting homomorphism $\iota$). As before, the $A$- and $B$-models will serve as our primary examples. In this subsection, we will continue to largely follow the notation and presentation of \cite{HKKPTVVZ03}. In Section \ref{sec:worldsheettwist}, we briefly derived the algebra of local, i.e. zero-form, operators in $Q$-cohomology, or chiral and twisted chiral rings, with respect to our now-scalar supercharges $Q_B$ and $Q_A$. The other supercharges also have their spin modified by the twist: they become one-forms. This feature enables us to consider a broader class of form-valued observables in topologically twisted theories. Roughly speaking, they are obtained by descending local operators to differential form-valued operators, and then integrating those forms over suitable spacetime submanifolds. This procedure is called \textit{topological descent}.%
\footnote{You may have seen a version of the descent equations before, when studying the $2n+2$-form-valued anomaly polynomial of a $2n$-dimensional field theory. A version of the descent equations on the anomaly polynomial can be used to produce a $2n$-form local counterterm on spacetime. Ambiguity of the counterterm is manifested as cohomological exactness from the descent point of view. See, e.g, \cite{Harvey:2005it} for details.}

Let us see this explicitly. A basic twisted\footnote{Since in this analysis we are not demanding the stronger condition of $Q$-exactness of the stress tensor, these theories may be called weakly topologically-twisted.} super-Poincar{\'e} algebra in arbitrary dimension will have the following form: there is the nilpotent twisting supercharge $[Q,Q]=0$; there are the momenta $P_\mu \sim -i \partial_\mu$ generating translations along $x^\mu$ that commute amongst themselves $[P_\mu, P_\nu] = 0$ and are $Q$-closed $[Q, P_\mu]=0$; and there are supercharges $Q_\mu$, necessarily commuting with the momenta $[Q_\mu, P_\nu] = 0$ and amongst themselves $[Q_\mu, Q_\nu] = 0$, that ensure the momenta are $Q$-exact $[Q, Q_{\mu}] = i P_{\mu}$. Given a general operator $O(x)$, we can define its 1-form \textit{topological descendant}:
\be
O^{(1)}(x):= -dx^{\mu} O_{\mu}(x), \ \ O_{\mu}(x) = [Q_{\mu}, O(x)].
\ee
Using the twisted Poincare algebra, the 1-form valued operator $O^{(1)}$ satisfies the following 
\be
\begin{aligned}
	QO^{(1)} & = -[Q, dx^\mu [Q_{\mu}, O]] = dx^\mu \big(i [P_{\mu},O] - [Q_\mu, [Q, O(x)]]\big)\\
	& = d O + (Q O)^{(1)},
\end{aligned}
\ee
where $d = dx^\mu \partial_\mu$ is the spacetime exterior derivative. We see that if $O$ is $Q$-closed, $Q O = 0$, its first descendant $O^{(1)}$ isn't $Q$-closed but instead contains explicit information about the position dependence of $O$. In particular, if we choose a path $\gamma_{x\to y}$ from $x$ to $y$ we can consider the extended operator
\be
O(\gamma_{x\to y}) = \int_{\gamma_{x\to y}} O^{(1)}.
\ee
The descent equation, together with Stokes' theorem, implies that the $Q$-variation of $O(\gamma_{x \to y})$ realizes the position independence of $O$ in $Q$-cohomology, i.e. $Q O(\gamma_{x\to y}) = O(y) - O(x)$.

We can continue this process for higher-form descendants:
\be
\begin{aligned}
	O^{(2)}(x) &= {1 \over 2}dx^{\mu} \wedge dx^{\nu} [Q_{\mu}, [Q_{\nu}, O(x)]]\\
	&\vdots \\
	O^{(k)}(x) &= {(-1)^k \over k!}dx^{\mu_1} \wedge \ldots \wedge dx^{\mu_k} [Q_{\mu_1}, [\ldots, [Q_{\mu_k}, O]]].
\end{aligned}
\ee
Again, these descendants are not $Q$-closed even if $O$ is $Q$-closed; instead, they satisfy the \textit{descent equations}:
\be
\label{eq:descent}
Q O^{(k)} = d O^{(k-1)} + (Q O)^{(k)}.
\ee
We can integrate the $k$-form descendant $O^{(k)}(x)$ over a $k$-dimensional submanifold $\gamma$ of spacetime to get an extended operator $O(\gamma)$ supported on $\gamma$; the descent equations ensure that, when $O$ is $Q$-closed, the resulting operator only depends on the \textit{homology} class of the integration cycle up to $Q$-exact terms: Stokes' theorem implies $Q(O(\gamma)) = O(\partial \gamma)$. In particular, if we integrate these descendants over nontrivial $k$-\textit{cycles}, i.e. $\gamma$ with $\partial \gamma = 0$, the resulting extended operator $O(\gamma)$ is necessarily $Q$-closed.

\subsubsection{2d $A$- and $B$-models}
\label{sec:2dAB}
In a two-dimensional theory like the $B$-model, there are 1- and 2-form descendants that one could study on worldsheets of various topology, given by
\be
\begin{aligned}
	O^{(1)} &= -\tfrac{1}{2i}dz [Q_-, O] + \tfrac{1}{2i} d\bar{z} [Q_+, O] \\
	O^{(2)} &= \tfrac{1}{4}dz \wedge d\bar{z} [Q_-, [Q_+, O]]
\end{aligned}
\ee
and in the $A$-model
\be
\begin{aligned}
	O^{(1)} &= -\tfrac{1}{2i} dz [\bar{Q}_-, O] + \tfrac{1}{2i} d\bar{z} [Q_+, O]\\
	O^{(2)} &= \tfrac{1}{4}dz \wedge d\bar{z} [\bar{Q}_-, [Q_+, O]]\\
\end{aligned}
\ee
The $d$-form descendants in a $d$-dimensional TQFT turn out to govern interesting deformations of the original TQFT. The reason is simple: one can can integrate the $d$-form descendants over (Euclidean) spacetime and add the result to the original action (multiplied by arbitrary coupling constants). When $O$ is $Q$-closed, so too is the resulting integrated descendant (on a closed spacetime, or given a suitable boundary condition) and therefore it represents a consistent deformation of the original action, at least classically.

For example, in the B-model, which depends on chiral parameters, a term corresponding to a deformation of a superpotential can be obtained by looking at the 2-form descendant of a chiral operator $\delta W(\phi)$. Explicitly, one can write a deformation of the superpotential as $\int d^2 z d^2 \theta \delta W(\Phi)$, which can be integrated over the superspace coordinates to give a term proportional to $\int d^2z [ Q_-, [Q_+, \delta W(\phi) ]] \sim \int (\delta W(\phi))^{(2)}$, which is nothing but the 2-form descendant of the chiral operator corresponding to the superpotential deformation. One can perform similar manipulations in the B-model to show that D-term and twisted F-term deformations are $Q_B$-exact, and do not correspond to descendants. As one might expect, an analogous statement holds in A-model: 2-form descendants of twisted chiral operators govern deformations of the model, while D and F-term deformations are $Q_A$-exact. 

One of the beautiful facts about these twisted theories, as emphasized for instance in \cite{HKKPTVVZ03} and references therein, is that twisting enables us to focus on a tractable, finite subset of possible deformations. General correlation functions in the full physical (i.e. untwisted) theory will depend on an infinite number of parameters. A consequence of the cohomological properties of twisting is that twisted correlation functions will depend on the finite number of parameters (F-term or twisted F-term, respectively) we consider that perturb our worldsheet theory. 

The coupling constants which multiply such deformation terms can be interpreted as local coordinates on the moduli space of the physical theory. In the case of $\mathcal{N}=(2,2)$ superconformal field theories, the conformal manifold, which is spanned by exactly marginal deformations, locally factorizes as a product of two (K{\"a}hler) manifolds associated to the chiral and twisted chiral deformations. These are precisely the deformations accessed by the B- and A-twists, respectively; when the superconformal field theories are sigma models with Calabi-Yau target, the two local factors coincide with the complex structure and K{\"a}hler moduli spaces of the Calabi-Yau geometry. Some references exploring the geometry of these moduli spaces from properties of the chiral rings include \cite{Cecotti:1991me, deBoer:2008ss, Gomis:2015yaa, Gomis:2016sab}.%
\footnote{Note that not all twists localize on operators which have the interpretation as sections of vector bundles over moduli space. For example, $1/4$-BPS operators in a 2d $\mathcal{N}=(4,4)$ superconformal field theory, which may be naturally studied in the holomorphic, or half-twist, furnish \textit{sheaves} over moduli space: the rank, or number of such operators, may jump discontinuously on subloci of the moduli space where the theory acquires an enhanced symmetry. 
	
	For an even simpler illustration of this phenomenon closely connected with the principal subject of these notes, consider a Landau-Ginzburg model with a single chiral multiplet $\Phi$ (the K{\"a}hler target is the flat complex plane $\CC$) and a superpotential of order $n$ $W = \Phi^n + \ldots$. For generic superpotential, this theory has $n-1$ vacua labeled by the distinct critical points of $W$, leading to $n-1$ elementary $A$-branes. However, the conformal point $W = \Phi^n$ has $\binom{n}{2}$ elementary $A$-branes, corresponding to the $\ZZ/n\ZZ$ enhanced symmetry. See \cite[Section 5.2]{HIV00} for more details about this example.}%

\subsubsection{3d $A$- and $B$-models}
\label{sec:3dAB}
The 3d $\mathcal{N}=4$ super Poincar\'{e} algebra is in many ways similar to the 2d $\mathcal{N}=(2,2)$ algebra described in great detail in these notes. The algebra has 8 supercharges which we denote $Q^{a \dot{a}}_\alpha$, where $\alpha \in \{\pm\}$ is an Spin$(3) \cong SU(2)$ spinor index and $a \in \{\pm\}$, $\dot{a} \in \{\dot{\pm}\}$ are $SO(4) \cong (SU(2)_A \times SU(2)_B)/\ZZ_2$ $R$-symmetry indices. In the absence of central charges, the anti-commutators can be expressed as
\be
[Q^{a \dot{a}}_\alpha, Q^{b \dot{b}}_\beta] = \epsilon^{ab} \epsilon^{\dot{a} \dot{b}} (\sigma^\mu)_{\alpha \beta} P_\mu
\ee
for $(\sigma^\mu)^\alpha{}_\beta$ the Pauli matrices. (3d spinor indices are raised/lowered with the Levi-Civita tensor $\epsilon^{\alpha \beta}$.) In both supersymmetry algebras, the $R$-symmetry group is (locally) isomorphic to two copies of the Euclidean spin group $\mathfrak{g}_R \cong \textrm{spin}(d) \oplus \textrm{spin}(d)$ with a \textit{mirror automorphism} exchanging the two $R$-symmetries. As a result, each of these supersymmetry algebras admits two distinct fully topological twists that are exchanged under this automorphism of the algebra. In analogy with the 2d setting, the two topological twists of 3d $\mathcal{N}=4$ theories are called%
\footnote{Historically, what we call the $B$-twist is known as the \textit{Rozansky-Witten twist}, after the introduction of the 3d TQFT known as Rozansky-Witten theory \cite{RW97}, but this twist was also studied by Blau and Thompson in pure gauge theory \cite{BT97}. What we call the $A$-twist is historically called the \textit{mirror Rozansky-Witten twist} and is a dimensional reduction of the 4d Donaldson-Witten twist of \cite{W88}. The names we use in the present paper reflect the fact that the 3d $A$ and $B$ twists become the 2d $A$ and $B$ twists, respectively, on 2d $\mathcal{N}=(2,2)$ boundaries of the 3d bulk. Similarly, the 3d $A$ and $B$ twists can lie on the boundary of $A$ and $B$ twists of 4d $\mathcal{N}=4$, themselves being the $\psi \to 0,\infty$ limits of the \textit{Kapustin-Witten} or \textit{Langlands twists} of \cite{KW07}.}
the $A$ and $B$ twists with twisting supercharges given by, e.g., $Q_A = Q^{+\dot{+}}_+ + Q^{-\dot{+}}_-$ and $Q_B = Q^{+\dot{+}}_+ + Q^{+\dot{-}}_-$.%
\footnote{In fact, the supercharge $Q_H = Q^{+\dot{+}}_+$ is itself nilpotent and leads to a mixed holomorphic-topological, in analogy with the half-twist supercharge $Q_H = \bar{Q}_+$ of 2d $\mathcal{N}=(2,2)$ described in Section \ref{sec:nilpot}.}

There are two rings of protected operators arising from the $Q_B$ and $Q_A$ cohomologies of local operators; in the 3d $\mathcal{N}=4$ setting, these local operators realize the algebras of holomorphic functions on the (hyperk{\"a}hler) \textit{Higgs branch} and \textit{Coulomb branch} of the vacuum moduli space, respectively. In a fashion completely analogous to the chiral ring of standard 2d $\mathcal{N}=(2,2)$ theories, the Higgs branch chiral ring of standard%
\footnote{By standard we are referring to 3d $\mathcal{N}=4$ theories of hypermultiplets minimally coupled to $\mathcal{N}=4$ vector multiplets. Just as in 2d, the above mirror automorphism implies the existence of twisted versions of these multiplets that can be used to construct $\mathcal{N}=4$ theories. For simplicity, we also restrict to the case where the hypermultiplets transform in a linear representation of \textit{cotangent type} $T^*R:= R \oplus \bar{R}$ for $R$ a unitary representation of the gauge group and $\bar{R}$ its conjugate.} %
3d $\mathcal{N}=4$ gauge theories receives no quantum corrections. On the other hand, the Coulomb branch and the Coulomb branch chiral ring, analogously to the K{\"a}hler moduli space of the A-model parameterized by the twisted chiral ring, receives myriad nonperturbative corrections. Nonetheless, it has recently enjoyed a complete mathematical definition \cite{braverman1604coulomb, Braverman:2016wma}. Descendants of these local operators span the tangent bundles of the corresponding moduli spaces.%
\footnote{We remark that even if one restricts attention, as we have, to twists which are sensitive to the \textit{topological rings} which capture the geometry of the vacuum manifold, one can encounter extended, rather than local, BPS operators in more general contexts (e.g. \cite{Cecotti:2013mba}).} %

There is a(n infrared) 3d duality, called \textit{3d mirror symmetry} to distinguish it from the usual notion of mirror symmetry, which
relates two $\mathcal{N}=4$ theories $\mathcal{T}_{\textrm{3d}} \leftrightarrow \tilde{\mathcal{T}}_{\textrm{3d}}$ in a way that intertwines the action of supersymmetry by the aforementioned mirror automorphism of the 3d $\mathcal{N}=4$ algebra. In particular, it equates the Coulomb branch of $\mathcal{T}_{\textrm{3d}}$ with the Higgs branch of its mirror $\tilde{\mathcal{T}}_{\textrm{3d}}$ (and vice versa) \cite{Intriligator:1996ex}. Just as with usual mirror symmetry, this duality often exchanges something nearly classical (the Higgs branch) with something highly quantum (the Coulomb branch). This 3d field theoretic duality enjoys beautiful string \cite{deBoer:1996mp, deBoer:1996ck} and M-theory \cite{Porrati:1996xi} uplifts.

\subsubsection{The secondary product}
\label{sec:sec-prod}
\begin{figure}[h!]
	\centering
	\includegraphics{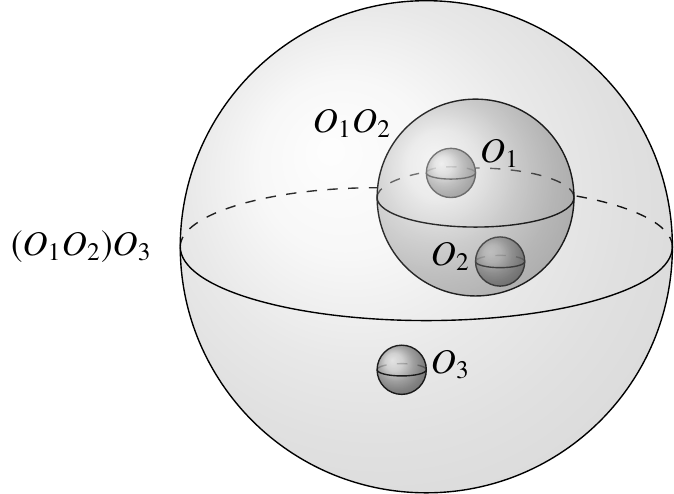}
	\caption{An illustration of the iterated product $(O_1 O_2) O_3$ in a $d$-dimensional TQFT induced by the inclusion of three small balls into two larger balls.}
	\label{fig:topOPE}
\end{figure}

We saw above that topological descendants contain information about the position-dependence of correlation functions of local operators in a $d$-dimensional topologically twisted theory --- we can use them to construct explicit homotopies between different configurations of operators. The (cohomological) position-independence of correlation functions of $Q$-closed local operators implies that the collision of local operators induces an algebra structure on the $Q$-cohomology of local operators in such a twisted theory: given two $Q$-closed local operators $O_1, O_2$ their product is another $Q$-closed local operator obtained by placing $O_1$ at $x_1$ and $O_2$ at $x_2 \neq x_1$ and then taking the limit $x_1 \to x_2$ (up to $Q$-exact terms). Phrased differently, this product is induced by including two disjoint open $d$-balls into a larger open $d$-ball; see Figure \ref{fig:topOPE} for an illustration corresponding to the product $(O_1 O_2) O_3$.

It is important to note that we are also free to collide in any order we would like, up to $Q$-exact terms. In particular, as briefly described in Section \ref{sec:worldsheettwist} for $d = 2$, given three local operators as in Figure \ref{fig:topOPE} we could have equivalently collided $O_2$ with $O_3$ first and then collide the result with $O_1$ to get the local operator $O_1 (O_2 O_3)$. Up to $Q$-exact terms, these two operators must be the same $(O_1 O_2) O_3 - O_1 (O_2 O_3) = Q(\ldots)$, implying that collision gives the $Q$-cohomology of local operators the structure of an \textit{associative algebra}. More generally, we see that collision of $Q$-closed local operators is associative \textit{up to homotopy}; we will return to these homotopies in Section \ref{sec:hom-alg}. 

When $d\geq2$, any configuration of two distinct points (better: open balls) in $\RR^d$ can be continuously deformed into any other, and so the $Q$-cohomology of local operators in $d \geq 2$ is necessarily a commutative (associative) algebra. On the other hand, in $d = 1$ we cannot continuously deform a configuration with $x_2 < x_1$ to a configuration with $x_1 < x_2$, thus there are two possible collisions that need not be equal. Consequently, the $Q$-cohomology of local operators in $d = 1$ has the structure of a not-necessarily-commutative associative algebra.

Even though the $Q$-cohomology of local operators in $d \geq 2$ is commutative, it turns out that topological descent induces an additional so-called \textit{secondary product} on the $Q$-cohomology of local operators. %(If one does not pass to cohomology and instead works at \textit{chain level}, viewing the Hilbert space as a complex with differential $Q$, the secondary product can interact with nonvanishing higher, $n$-ary products to form an $L_{\infty}$ algebra.)
In this subsection, we will review this idea, following the presentations of \cite{BBBDN18, OY19}. For $Q$-closed local operators $O_1$ and $O_2$, the secondary product $\{O_1, O_2\}$ is defined by placing $O_2$ at $x_2$ and then integrating the $(d-1)$-form operator $O_1^{(d-1)}$ over a $(d-1)$-sphere $S^{d-1}$ surrounding $x_2$:
\be
\{O_1, O_2\} = \int_{S^{d-1}_{x_2}} O_1^{(d-1)} O_2(x_2).
\ee
This is another local operator, since it is supported inside a single sufficiently large ball, and it is again $Q$-closed so we may again view it as representing a $Q$-cohomology class.

More generally, we consider the configuration space of two points (or, better, small open disks) on $\RR^d$, denoted $\mathcal{C}_{\RR^d}(2)$; topological descent allows us to define a ``product'' for every homology class of this configuration space $H_\bullet(\mathcal{C}_{\RR^d}(2))$, c.f. \cite[Section 1.3]{CG16} and \cite[Section 3.2.1]{BBBDN18}. Given an $n$-dimensional chain $\Gamma$, we define the (degree $-n$) product $\star_\Gamma(O_1, O_2)$ as
\be
\star_\Gamma(O_1, O_2) = \int_{\Gamma} (O_1(x_1) O_2(x_2))^{(n)}\,,
\ee
where we integrate over the $n$-th descendant over the space of configurations determined by $\Gamma$. A quick application of Stokes' theorem and the descent equation implies that%
\footnote{If $\Gamma$ is an $n$-dimensional chain, then the operation of integrating a form over $\Gamma$, i.e. $\omega \mapsto \int_{\Gamma} \omega$, is naturally a degree $-n$ map when we correlate the parity of total degree (including form degree) and Fermionic parity. Since $Q$ is degree $1$, it follows that we acquire a factor of $(-1)^n$ from passing $Q$ through the integral $\int_\Gamma$.}%
\be
\begin{aligned}
	Q(\star_\Gamma(O_1, O_2)) & = (-1)^n \int_{\Gamma} \bigg(d (O_1 O_2)^{(n-1)} + \big((Q O_1) O_2 + (-1)^{F(O_1)} O_1 (QO_2)\big)\bigg)\\
	& = (-1)^n \bigg(\star_{\partial \Gamma}(O_1, O_2) + \star_{\Gamma}(Q O_1, O_2) + (-1)^{F(O_1)} \star_{\Gamma}(O_1, Q O_2)\bigg)\,.\\
\end{aligned}
\ee
In particular, we see that if $\Gamma$ is a cycle and $O_1, O_2$ are $Q$-closed, then so too is the product $\star_\Gamma(O_1, O_2)$. Moreover, if two chains $\Gamma, \Gamma'$ are homologous $\Gamma' = \Gamma + \partial \Gamma''$, for $\Gamma''$ some $n+1$ chain, then the two products (on $Q$-closed local operators) differ by $Q (\star_{\Gamma''}(O_1, O_2))$, hence agree at the level of $Q$-cohomology.

The configuration space $\mathcal{C}_{\RR^d}(2)$ is homotopic to a $(d-1)$-sphere $S^{d-1}$, whose homology has a degree $0$ generator, corresponding to the usual collision product $(O_1, O_2) \mapsto O_1 O_2$, and a degree $(d-1)$ generator, corresponding to the secondary product $(O_1, O_2) \mapsto \{O_1, O_2\}$.%
\footnote{When $d = 1$, the sphere $S^{d-1}$ is a simply pair of points $[1], [-1]$ and the corresponding products are the maps $(O_1, O_2) \to O_1 O_2$ and $(O_1, O_2) \to (-1)^{F(O_1) F(O_2)}O_2 O_1$. The analog of the top dimensional cycle is the formal difference of these two points $[S^0] = [+1] - [-1]$ and so the analog of the secondary product is the usual graded commutator 
	\[ \star_{[S^0]}(O_1, O_2) = O_1 O_2 - (-1)^{F(O_1) F(O_2)} O_2 O_1\,.\]
	
	This phenomenon is related to quantization of the secondary product in 3d TQFTs via an Omega-background, c.f. \cite[Section 6]{BBBDN18}. More precisely, turning on an Omega-background requires that we consider the $U(1)$-equivariant homology of the configuration space $\mathcal{C}_{\RR^3}(2) \sim S^2$ (with respect to rotation around a fixed axis in $\RR^3$). This is generated (as a ring over polynomials in the equivariant parameter $\varepsilon$ and its inverse $\varepsilon^{-1}$) by the fixed points/poles $[N] \sim [1], [S] \sim [-1]$; the relation between the secondary product and the products induces by $[N]$ and $[S]$ is exactly as that between the Poisson bracket and the graded commutator
	\[[N]-[S] = \varepsilon [S^2] \leftrightarrow \star_{[N]}(O_1, O_2) - \star_{[S]}(O_1, O_2) = \varepsilon \star_{S^2}(O_1, O_2)\,.\]} %
Since there are no other homology classes in this configuration space, these are the only two ``products'' which survive at the level of cohomology. (In the next section we will also find use of more general chains, rather than simply cycles, on these configuration spaces.) Together with the collision product, this secondary product endows the $Q$-cohomology of local operators with the structure of a $\ZZ/2\ZZ$-graded \textit{Poisson algebra}, whose Poisson bracket has degree $1-d \mod 2$. (In many examples with a $U(1)$ R-symmetry that gives a fermion number, this can be enhanced to a $\mathbb{Z}$-graded algebra, just as in our previous discussion of twisting.)

Let's look at a quick example for illustration, described in more detail in \cite[Section 4]{BBBDN18}. We will consider the 2d $B$-model with target some Calabi-Yau manifold $\mathcal{X}$, on a flat worldsheet. We already deduced that the space of local operators is isomorphic to the Dolbeault cohomology of polyvector fields on $\mathcal{X}$. In other words, the $Q$-cohomology is isomorphic to the $\bar{\partial}_{\mathcal{X}}$-cohomology of $(0,q)$-forms valued in arbitrary exterior powers of the holomorphic tangent bundle $\bigwedge^p(T^{(1,0)}\mathcal{X})$. The chiral ring elements are holomorphic functions on $\mathcal{X}$, which is the degree-0 part of the cohomology. The $B$-model does not get any instanton corrections, unlike the $A$-model, so the primary product on cohomology is just the usual geometric wedge product of polyvector fields, c.f. Section \ref{sec:CRdolbeault}:
\be
[O_1]\cdot[O_2] = [O_1 \wedge O_2].
\ee
The secondary product in this case is rather famous in string theory and other contexts, and is called a Gerstenhaber bracket. In geometric language, it is called the Schouten-Nijenhuis bracket of polyvector fields, which generalizes the Lie bracket on ordinary vector fields. Let's consider the target to be $\mathcal{X} = \mathbb{C}^N$ for simplicity. The twisting supercharge is $Q = Q_B := \bar{Q}_+ + \bar{Q}_-$ and the ``vector'' supercharge in the previous notation can be given by a vector with components $-\tfrac{1}{2i} Q_-, \tfrac{1}{2i} Q_+$. On flat $\mathbb{C}^N$, polyvector fields are generated by holomorphic functions $f(\phi)$ and holomorphic vector fields $g^n(\phi) \partial_{\phi^n}$. We will again denote the (fermionic) coordinate vector field by $\bar{\zeta}_n := \partial_{\phi^n}$. To obtain the secondary bracket $\left\lbrace \bar{\zeta}_n, \phi^m \right\rbrace$ we need the one-form descendant for $\bar{\zeta}_n$: 
\be
\bar{\zeta}^{(1)}_n = -dx^\mu [Q_{\mu}, \bar{\zeta}_n] = -(\star d \bar{\phi}_n)\,.
\ee
Here, $d$ is the de Rham differential as before, and $\star$ is the Hodge star operator. Observe also that $d \bar{\zeta}_n^{(1)} = d \star d\bar{\phi}_n = {\delta S \over \delta \phi^n}$, and so is proportional to the equation of motion for $\phi^n$. 
If we treat this as an operator, standard arguments in Euclidean QFT show that ${\delta S \over \delta \phi}(z, \bar{z})$ is zero up to contact terms, such that in any correlation function the operator product ${\delta S \over \delta \phi}(z, \bar{z}) \phi(w, \bar{w})$ is equivalent to inserting a delta function two-form: $\delta^2(z - w, \bar{z}- \bar{w})$ (provided they are each separate from any other operators in the correlation function). One can just show this from integration by parts:
\be
\begin{aligned}
	\int \mathcal{D}\phi \mathcal{D}\bar{\phi}e^{-S}&{\delta S \over \delta \phi}(z, \bar{z}) \phi(w, \bar{w}) \\
	&= \int \mathcal{D}\phi \mathcal{D}\bar{\phi}\left( -{\delta \over \delta \phi(z, \bar{z})}(\phi(w, \bar{w})e^{-S}) + \delta^2(z-w, \bar{z}-\bar{w}) e^{-S}\right) \\
	&= \int \mathcal{D}\phi \mathcal{D}\bar{\phi}\delta^2(z-w, \bar{z}-\bar{w}) e^{-S}.
\end{aligned}
\ee

By the definition of the secondary bracket, we have
\be
\begin{aligned}
	\left\lbrace \bar{\zeta}_n, \phi^m \right\rbrace &=\left[ \oint_{S^{1}_{w, \bar{w}}}\bar{\zeta}_n^{(1)} \phi^m(w, \bar{w}) \right] \\
	&=\left[\int_{D^2_{w, \bar{w}}} d \bar{\zeta}_n^{(1)}\phi^m(w, \bar{w}) \right],
\end{aligned}
\ee
where we use square brackets to denote cohomology classes, and in the second line we used Stokes's theorem. Using the result we just derived and inserting the delta-function two-form into the final line gives
\be
\left\lbrace \bar{\zeta}_n, \phi^m \right\rbrace = \delta_n^m. 
\ee
We could also have obtained this result by performing descent on $\phi^m$, and this is an instructive way to check the answer. Similar manipulations show that the secondary bracket between two $\phi$'s or two $\bar{\zeta}$'s vanishes, due to the lack of contact terms between the operators and the resulting equations of motion. One can check that this secondary bracket is the geometric Schouten-Nijenhuis bracket on polyvector fields up to a fermion parity factor $\left\lbrace a, b \right\rbrace_{SN} = (-1)^{F(a)-1} \left\lbrace a, b \right\rbrace$, and that in particular it satisfies the properties that it is (graded) antisymmetric, is a (graded) derivation on each argument, and satisfies the (graded) Jacobi identity.

\subsection{Homotopy-algebraic considerations}
\label{sec:hom-alg}
Already we see that algebraic structures in twisted supersymmetric quantum field theories can be extremely rich. In fact, chain-level algebraic structures can be even richer, as one might expect since the $Q$-exact terms would not have yet been discarded. The relevant class of algebraic structures that we need in this context are sometimes called \textit{higher algebras} or \textit{homotopy algebras}. These are deep, and often forbidding, structures in both mathematics and physics, so we will spend most of our time with perhaps the most prominent such example: the $A_{\infty}$ algebra. Roughly speaking, an $A_{\infty}$ algebra is an associative algebra \textit{up to homotopies}; similar algebras like $L_{\infty}$ algebras satisfy the axioms of a Lie algebra only \textit{up to homotopies}, as we will see. These extra structures can transfer in intricate ways across dualities.

\subsubsection{Topological quantum mechanics and $A_\infty$ algebras}
\label{sec:Ainfty}
Let us start in the simplest possible setting: with a twisted quantum mechanical ($1$-dimensional) theory. We saw in Section \ref{sec:topdescent} that the $Q$-cohomology of local operators is a not-necessarily-commutative algebra by considering the configuration space of two points in $\RR$. We also saw that considering the configuration space of three points to implies that the collision product of ($Q$-cohomology classes of) local operators is moreover associative. In particular, given three $Q$-closed local operators $O_1$, $O_2$, and $O_3$ it follows that the \textit{associator} $(O_1 O_2) O_3 - O_1 (O_2 O_3)$ is trivial in $Q$-cohomology; we now study how this associator is trivialized.

First, let's rephrase the binary collision in terms of the homology of the configuration space of two (distinct) points (or better, open intervals) on $\RR$. There are two components of this configuration space, corresponding to $x_1 < x_2$ and $x_1 > x_2$, we will focusing on the component with $x_1 > x_2$, which can be identified with a half-space. We can always use the overall translations to set $x_1 = - x_2$, and so once we've chosen an overall scale (the value of $x_1 - x_2$), the \textit{reduced} configuration space is simply a single point $\mathcal{K}_2$. The (degree 0) product $O_1 O_2$ arises from the $x_1 - x_2 \to 0$ limit of an integral over this reduced configuration space:
\be
O_1 O_2 = \lim\limits_{x_1 - x_2 \to 0} \int_{\mathcal{K}_2} (O_1(x_1) O_2(x_2))^{(0)} = \lim\limits_{x_1 - x_2 \to 0} \int_{\mathcal{K}_2} O_1(x_1) O_2(x_2)
\ee
More precisely, we should choose a small, positive parameter $\epsilon$ used for point-splitting and instead take the $x_1-x_2 \to \epsilon$; the (regulated) product $O_1 O_2$ is obtained by removing the terms singular in $\epsilon$, and then taking the $\epsilon \to 0$ limit. This is what we mean by the above $x_1 - x_2 \to 0$ limit. See, e.g., \cite[Chapter 2]{P198} for more details about point-splitting of local operators. In particular, it is best to think of local operators as small, open intervals rather than simply points. Either way, the supercharge $Q$ is a fermionic derivation of this product:
\be
Q(O_1 O_2) = (Q O_1) O_2 + (-1)^{F(O_1)} O_1 (Q O_2)\,.
\ee
The same analysis applies to the component with $x_1 < x_2$, where we get the (degree 0) product in the opposite order $-(-1)^{F(O_1) F(O_2)} O_2 O_1$ (with an appropriate sign for the parity of the operators).

The configuration space of three points (or open intervals) on $\RR$ has six components, corresponding to the $6 = 3!$ possible orders. After translating a configuration to $x_1 = - x_3$ and choosing the overall scale $x_1 - x_3$ we find a reduced configuration space with $x_1 > x_2 > x_3$ that is simply an interval
\be
\mathcal{K}_3(\epsilon) = \{x_2 \in \RR| x_3 + \epsilon \leq x_2 \leq x_1 - \epsilon\}.
\ee
We include explicitly here the point-splitting parameter $\epsilon$, and require $x_1-x_3 \geq 3 \epsilon$.%
\footnote{The factor of 3 here ensures that the following configuration space has non-trivial extent; any separation (strictly) larger than $2\epsilon$ works equally well and ensures that all operator insertions are separated by more than $\epsilon$.} %
The boundaries of this interval correspond to the two ways to collide the operators $O_1$, $O_2,$ and $O_3$: we could either first collide $O_2$ and $O_3$ then collide the result with $O_1$ to get $O_1 (O_2 O_3)$ (the boundary at $x_2 = x_3+\epsilon$), or we could first collide $O_1$ and $O_2$ then collide the result with $O_3$ to get $(O_1 O_2) O_3$ (the boundary at $x_2 = x_1 - \epsilon$). 

The points of $\mathcal{K}_3(\epsilon)$ should be thought of as constituting a path (better: a homotopy) between the configurations corresponding to these two possible collisions. With this in mind, we define a ternary operation $\mu_3(O_1, O_2, O_3)$ as an integral over $\mathcal{K}_3(\epsilon)$, taking the ($\epsilon$-regulated) $x_1 - x_3 \to 0$ limit:
\be
\mu_3(O_1, O_2, O_3)  = \lim\limits_{x_1-x_3 \to 0} \int_{\mathcal{K}_3(\epsilon)} (O_1(x_1) O_2(x_2) O_3(x_3))^{(1)} = \lim\limits_{x_1-x_3 \to 0} \int_{\mathcal{K}_3(\epsilon)} O_1(x_1) O_2^{(1)}(x_2) O_3(x_3)\,.
\ee
Here, only the terms where we descend $O_2$ survive the integration over $\mathcal{K}_3$ because $x_1, x_3$ are held fixed; there are other homologous chains in $\mathcal{C}_\RR(3)$ that would yield an equally good product $\mu_3'$. Stokes' theorem and the descent equations imply the $Q$-variation of this local operator is exactly the associator $(O_1 O_2) O_3 - O_1 (O_2 O_3)$. More generally, if the $O_i$ are not $Q$-closed, the descent equation Eq. \eqref{eq:descent} implies
\be
\label{eq:Ainfrel3}
\begin{aligned}
	Q\mu_3(O_1, O_2, O_3) & = (O_1 O_2) O_3 - O_1 (O_2 O_3)\\ 
	& \hspace{-2cm} - \mu_3(Q O_1, O_2, O_3) - (-1)^{F(O_1)}\mu_3(O_1, Q O_2, O_3) - (-1)^{F(O_1) + F(O_2)}\mu_3(O_1, O_2, Q O_3)\,,
\end{aligned}
\ee
where the first line contains the above boundary contributions and the second line comes from the action of $Q$ on the $O_i$. Equation \eqref{eq:Ainfrel3} is known as the third $A_\infty$ relation.

We can collect the above data (the differential $\mu_1(O_1) = QO_1$, the product $\mu_2(O_1,O_2)$, and the ternary operation $\mu_3(O_1, O_2, O_3)$) diagrammatically in terms of \textit{rooted trees}:
\[
\mu_1 \leftrightarrow \raisebox{-0.5cm}{\includegraphics{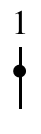}} \hspace{2cm} \mu_2 \leftrightarrow \raisebox{-0.5cm}{\includegraphics{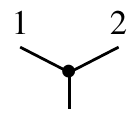}} \hspace{2cm}
\mu_3 \leftrightarrow \raisebox{-0.5cm}{\includegraphics{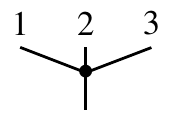}}
\]
The upper \textit{leaves} are interpreted as inputs to the corresponding operation, with the lower \textit{root} being the output. The above relations can then be expressed in terms these diagrams via \textit{grafting} the root of one tree onto a leaf of a second. The relations saying $Q$ is both a differential and a derivation of the product (also known as the first and second $A_\infty$ relations) can be written as follows:
\[
\raisebox{-0.75cm}{\includegraphics{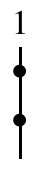}} = 0 \hspace{2cm} \raisebox{-0.75cm}{\includegraphics{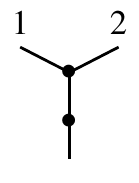}} = \raisebox{-0.75cm}{\includegraphics{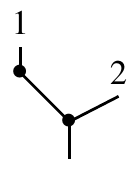}} + \raisebox{-0.75cm}{\includegraphics{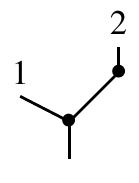}}
\]
The third $A_\infty$ relation can similarly be expressed in terms of trees:
\[
\raisebox{-0.75cm}{\includegraphics{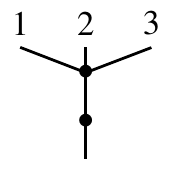}} = \raisebox{-0.75cm}{\includegraphics{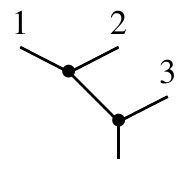}} - \raisebox{-0.75cm}{\includegraphics{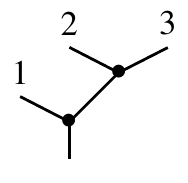}} - \raisebox{-0.75cm}{\includegraphics{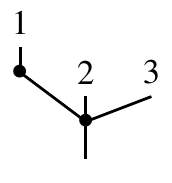}} - \raisebox{-0.75cm}{\includegraphics{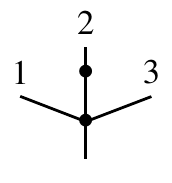}} - \raisebox{-0.75cm}{\includegraphics{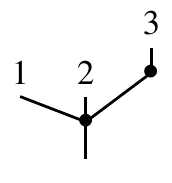}}
\]

\begin{figure}[h!]
	\centering
	\includegraphics{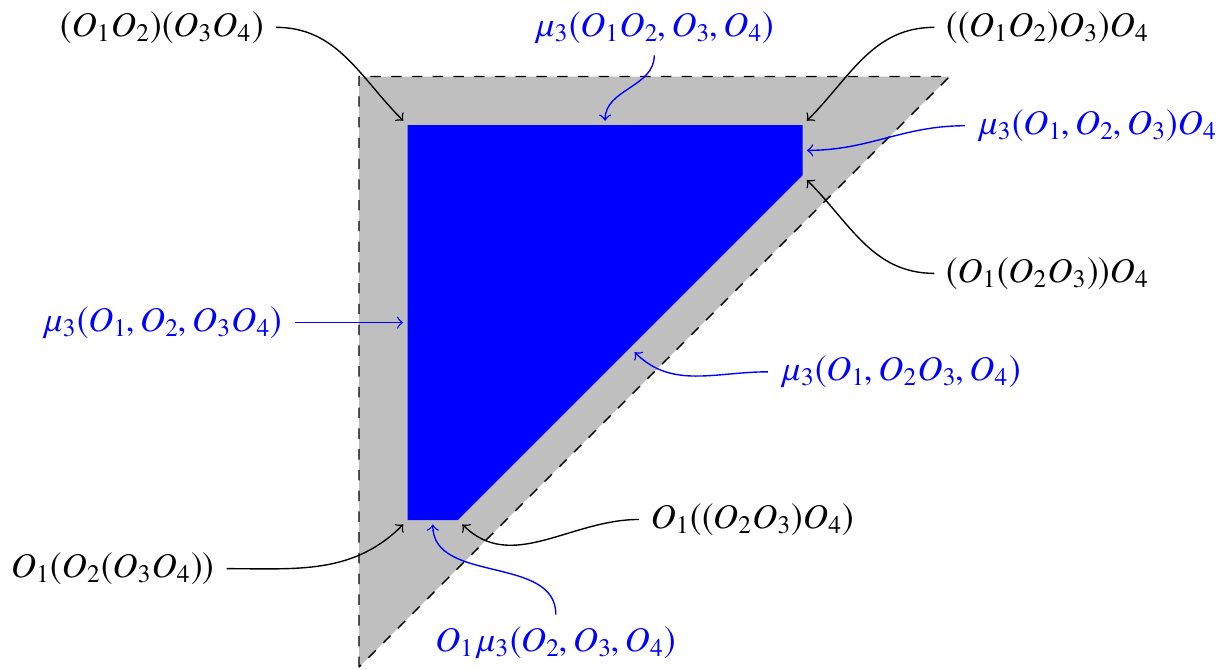}
	\caption{The configuration space of $4$ points $x_1, x_2, x_3$ and $x_4$ on $\RR$ for fixed $x_1, x_4$. The vertical direction corresponds to the value of $x_2$ and the horizontal direction to $x_3$. The 4-th Sashtev polytope $K_4$ is identified with the subspace $\mathcal{K}_4(\epsilon)$ and shaded in blue. We decorate the corners of $\mathcal{K}_4(\epsilon)$ with the corresponding placement of parentheses in the product $O_1 O_2 O_3 O_4$ and each edge with the corresponding algebraic homotopy.}
	\label{fig:mu4-pent}
\end{figure}

We can continue this process to higher $n$-ary operations, i.e. to a larger number of input local operators: for a collection of $n$ local operators $O_1, ..., O_n$ there are many ways to do pairwise collisions. We can construct an ($n-2$)-dimensional polytope, called the $n$-th ``Sashtev polytope'' or ``associahedron'' $K_n$, whose vertices are these choices, whose edges connect vertices related by a single one of the homotopies above, and whose higher faces correspond to homotopies of compositions of such homotopies and so on. In terms of configuration space of open intervals on $\RR$, we set to $x_1 = - x_n$ (with $x_1 - x_n > (2n-3) \epsilon$) and consider
\be
\mathcal{K}_n(\epsilon) = \big\{(x_2, ..., x_{n-1}) \in \RR^{n-2} | x_i - x_j \geq [2(j-i)-1]\epsilon, 1 \leq i < j \leq n  \big\}\,,
\ee
which is homeomorphic to $K_n$. We saw above that $K_2$ is a single point and $K_3$ is an interval. Similarly, there are five ways to collide four local operators, and we can arrange them as vertices in a pentagon $K_4$; we illustrate $\mathcal{K}_4(\epsilon)$ in Figure \ref{fig:mu4-pent}.

For each $n$ we construct a new local operator $\mu_n(O_1, ..., O_n)$ via topological descent (suppressing the insertion points $x_1, \ldots, x_n$)
\be
\mu_n(O_1, \ldots, O_n) = \lim\limits_{x_1 - x_n \to 0} \int_{\mathcal{K}_n(\epsilon)} (O_1 \ldots O_n)^{(n-2)} = \lim\limits_{x_1 - x_n \to 0} \int_{\mathcal{K}_n(\epsilon)} O_1 O_2^{(1)} \ldots O_{n-1}^{(1)} O_n.
\ee
This operation involves taking $n-2$ descendants, and therefore has degree $2-n$, i.e. if $O_i$ has degree $r_i$ then $\mu_n(O_1, ..., O_n)$ has degree $2 - n + r_1 + \ldots + r_n$. Just as $\mu_3(O_1, O_2, O_3)$ witnessed the $Q$-exactness of the associator $(O_1 O_2) O_3 - O_1 (O_2 O_3)$, so too does the operator $\mu_n(O_1, ..., O_n)$ witness the $Q$-exactness a higher analog of the associator:
\be
\label{eq:Ainfrelations}
\begin{aligned}
	Q \mu_n(O_1, ..., O_n) & = \sum (-1)^{i+jk+1} (\pm 1) \mu_{i+k+1}\big(O_1, ..., O_i, \mu_j(O_{i+1},..., O_{i+j}), O_{i+j+1}, ..., O_n)\\
	& + (-1)^n \mu_n(Q O_1, ..., O_n) + ... + (-1)^{n}(\pm1)\mu_n(O_1, ..., Q O_n) \,.
\end{aligned}\,,
\ee
where $(\pm1)$ is the Koszul sign%
\footnote{The Koszul sign assignment is such that for maps $f:A \to X$ and $g: B \to X$ with $g$ homogeneous of degree $F(g)$ we have $$(f\otimes g)(a\otimes b) = (-1)^{F(a)F(g)} f(a) \otimes g(b)$$ where $a$ is homogeneous of degree $F(a)$. The signs follow from noting the differential $Q$ has degree $1$ and $\mu_j$ has degree $2-j$. Explicitly, the $(\pm1)$ on the first line is $(-1)^{F(O_1) + ... + F(O_{n-1})}$ and the $(\pm1)$ on the second line is $(-1)^{(2-j)(F(O_1) + ... + F(O_{i}))}$.} %
and the sum is over $i, j, k$ with $i+j+k = n$, $i+k \geq 1$, and $j \geq 2$. Equation \eqref{eq:Ainfrelations} is known as the \textit{$n$-th $A_\infty$ relation}, and the operations $\mu_n$ equip local operators with the structure of an $A_\infty$ algebra. If we represent the $n$-ary operation $\mu_n$ by a rooted tree with $n$ leaves, this relation can be expressed as diagrammatically as:
\[
\raisebox{-0.75cm}{\includegraphics{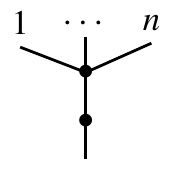}} = \sum (-1)^{i+jk+1} \raisebox{-1.5cm}{\includegraphics{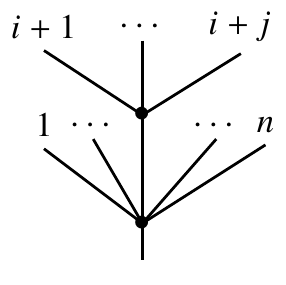}},
\]
This relation is also often stated simply as a sum of trees like those on the right-hand side without the restriction $i+k \geq 1$; we have isolated the $i = k = 0$ term on the left-hand side to emphasize its interpretation as a homotopy for some higher associator.

We see that an $A_{\infty}$ algebra, sometimes called a \textit{homotopy associative algebra}, is a (co)chain complex $(A, d_A)$ endowed with a family of operations $\mu_n$ satisfying the aforementioned relations. If all $\mu_n = 0, n\geq 3$, then $A$ a \textit{differential-graded associative algebra} or DGA: $A$ is an associative algebra, which also has a differential $d_A$ that acts as a derivation of the associative product. It turns out that \textit{even the cohomology} of a DGA inherits the structure of an $A_{\infty}$ algebra that is $A_\infty$ equivalent to that of $(A, d_A)$ (see \emph{e.g.} \cite[Sect. 3.2]{Keller} and references therein).

Roughly speaking, the higher operations on $H^\bullet(A)$ measure the failure of algebraic relations satisfied by cohomology classes to be satisfied by their chain-level lifts. The homology $H^\bullet(A)$ is an example of a \textit{minimal model} for the $A_\infty$ algebra $(A, d_A)$, i.e. it has trivial differential but higher operations aplenty. The DGA $(A, d_A)$ on the other hand is said to be \textit{anti-minimal}, having non-trivial differential but trivial higher operations. These higher operations on $H^\bullet(A)$ are called $A_{\infty}-$Massey products. These have the virtue of being very explicit, but they depend on many choices. Mathematically, we say that this process is unique \textit{up to quasi-isomorphism}. A quasi-isomorphism is simply a chain map that induces an isomorphism on cohomology.  

A simple example of a DGA comes from ordinary $\mathcal{N}=2$ supersymmetric quantum mechanics. $\mathcal{N}=2$ supersymmetry ensures the existence of two fermionic operators $Q, Q^\dagger$ satisfying $\left\lbrace Q^{\dagger}, Q \right\rbrace~=~2H$. We will take the operators of the supersymmetric quantum mechanics to form the operator algebra $A$, which is necessarily associative. We will assume $A$ is $\mathbb{Z}$-graded by the fermion number operator $F$, so that we can write $A = \oplus_{j \in \mathbb{Z}}A^j$ with $A^j$ being the elements of $A$ satisfying $[F, A^j] = jA^j$. We will be interested in only one of the two supercharges, $Q$, which commutes with the Hamiltonian and sends $A^j \xrightarrow{[Q, -]} A^{j+1}$, so we say $Q$ is a derivation of degree 1. Indeed, it acts as a derivation on the operator product: 
\be
[Q, ab] = [Q, a]b + (-1)^{F(a)}a[Q, b].
\ee
Here and throughout this discussion, we will use $[- , -]$ to indicate a graded commutator (i.e. it denotes either an anticommutator or commutator depending on the parity of the operators). With respect to the privileged supercharge $Q$, we see that $A$ is a DGA with differential given by $d_A = [Q, -]$. Similarly, the Hilbert space of states, being a representation of the algebra with compatible grading and differential, is a \textit{DG-module}. The $a \in A$ satisfying $[Q, a] = 0$ are often called ($\tfrac{1}{2}$-)\textit{BPS operators} and we will denote the corresponding cohomology class by $[a] \in H^\bullet(A)$.

The higher operations on the cohomology of a DGA can be built from homotopies between products (and higher associators) of lifts of cohomology classes and the lifts of their products (and higher associators). There are many choices involved, but they lead to equivalent, i.e. quasi-isomorphic, $A_\infty$ algebras. %
%Dyckerhoff studied examples of these higher products in the context of matrix factorizations and branes in the 2d $B$-model in \cite[Section 5.6]{DyckerhoffGen}, but the computations quickly become impractical.
%
First, choose a map $f_1: H^\bullet(A) \to A$ that lifts every cohomology class to a cocycle representing it. It follows that for any $[a], [b]$ the difference $f_1([a] [b]) - f_1([a]) f_1([b])$ is trivial in cohomology by construction (i.e. the difference is $Q$-exact) and we \textit{choose} a map $f_2: H^\bullet(A)^{\otimes 2} \to A$ that witnesses this $Q$-exactness:
\be
Q f_2([a],[b]) = f_1([a] [b]) - f_1([a]) f_1([b])\,.
\ee
The triple product $\mu_3([a],[b],[c])$ of elements $[a],[b],$ and $[c]$ then satisfies
\be
\label{eq:mu3lift}
\begin{aligned}
	f_1(\mu_3([a],[b],[c])) & = f_2([a],[b])f_1([c]) -(-1)^{F([a])} f_1([a]) f_2([b],[c])\\
	& \qquad + f_2([a] [b], [c]) - f_2([a],[b][c]) + Q(...)\,,
\end{aligned}
\ee
i.e. $\mu_3([a],[b],[c])$ is the cohomology class of the right-hand side.

This process continues. Given the lifts $f_i$ up to $i = n-2$ and products $\mu_i$ up to $i = n-1$, one chooses an $n-1$-ary lift $f_{n-1}: H^{\bullet}(A)^{\otimes n-1} \to A$ that witnesses the $Q$-exactness of an expression involving a difference between $f_1(\mu_{n-1}([a], [b], ...))$ and an (explicit) expression involving and the lower $i$ lifts and products. The $n$-ary product $\mu_{n}([a], [b], ...)$ is then given as the cohomology class of an (explicit) expression involving $f_1, f_2, ..., f_{n-1}$ and the lower $i$ products. See \emph{e.g.} \cite{Kadeishvili} for more details. The collection of maps $f_i$ define an \textit{$A_\infty$ morphism} from $\{H^\bullet(A), \mu_n\}$ to $(A, Q)$ that is the identity map on $H^\bullet(A)$ at the level of cohomology. Again, this is a \textit{quasi-isomorphism}.

To summarize what we have seen so far: local operators in $\mathcal{N}=2$ quantum mechanics, with a choice of twisting supercharge $Q$, are naturally endowed with the structure of a DGA. Passing to $Q$-cohomology classes, or passing to the twisted theory, involves ignoring a great deal of the physical theory. However, much of this lost information can still be encoded in higher $n$-ary operations on cohomology classes: these higher operations are powerful enough to recover the physical local operators up to $A_\infty$ equivalence.

\subsubsection{Example: Landau-Ginzburg $B$-model}
\label{sec:Ainf-xyz}
Let's compute an explicit example of the above higher products in a somewhat more interesting context: the algebra of local operators on the 1d boundary of a 2d theory. This will also illustrate how A$_{\infty}$ algebras can transfer across dualities via quasi-isomorphisms. We consider the 2d Landau-Ginzburg model of three chiral superfields $\Phi^n = \phi^n + ...\,$, $n=1,2,3$, and superpotential $W=\Phi^1 \Phi^2 \Phi^3$, and then take the $B$-twist. The category of boundary conditions compatible with the $B$-twist, a.k.a. the category of \textit{$B$-branes}, was proposed by Kontsevich to be the (derived) category of \textit{matrix factorizations} of the superpotential $W$ \cite[Section 7]{KapustinLi}. We provide a lightning review of how matrix factorizations arise as a description of the category of $B$-branes in Landau-Ginzburg models in Appendix \ref{sec:MFintro}. In the present example, this (derived) category is generated by three objects: the matrix factorization  $E = \phi^1, J = 2 \partial_{\phi^1}W$ and its permutations. In particular, any matrix factorization $(E^a, J_a)$ can be realized as a direct sum of these three matrix factorizations, or perhaps a deformation and/or summand thereof.

We consider the matrix factorization with $E = (\phi^1, \phi^2, \phi^3)$ and $J = \tfrac{2}{3}(\phi^2\phi^3,\phi^1\phi^3, \phi^1\phi^2);$ this choice is related to a Dirichlet boundary condition%
\footnote{Dirichlet matrix factorizations of this sort were studied in detail by Dyckerhoff (who called them \textit{stabilized residue fields}) and are known to generate the derived category of matrix factorizations by themselves when $W$ has isolated singularities \cite[Theorem 4.1]{DyckerhoffGen}. Moreover, its endomorphism DGA $A$ is Koszul dual to the \textit{curved associative algebra} $R_W = (\CC[\phi^a|], W)$ \cite[Proposition 3.2]{TuMFKoszul}. Unfortunately, the present $W$ is far too singular for these results to apply -- the hypersurface $W = 0$ is singular along each of the divisors $\phi^1 = \phi^2 = 0$, $\phi^1 = \phi^3 = 0$, and $\phi^2 = \phi^3 = 0$.} %
on the bulk chiral multiplets -- this $E$ is designed to force $\phi^n = 0$. Concretely, we find that the DGA of local operators on this Dirichlet boundary condition is given by
\be
A = \CC[\phi^n, \bar{\gamma}_n, \partial_{\bar{\gamma}_n}] \qquad Q = \phi^n \bar{\gamma}_n + \tfrac{2}{3} W_n \partial_{\bar{\gamma}_n},
\ee
where $W_n := \partial_{\phi^n} W$. It is worth noting that this matrix factorization is preserved by the $S_3$ symmetry permuting the $n$ index. A straightforward computation shows that the three fermionic operators 
\be
\psi_1 = \bar{\gamma}_1 - \tfrac{1}{3}(\phi^2 \partial_{\bar{\gamma}_3} + \phi^3 \partial_{\bar{\gamma}_2}) \qquad \psi_2 = \bar{\gamma}_2 - \tfrac{1}{3} (\phi^1 \partial_{\bar{\gamma}_3} + \phi^3 \partial_{\bar{\gamma}_1}) \qquad \psi_3 = \bar{\gamma}_3 - \tfrac{1}{3} (\phi^2 \partial_{\bar{\gamma}_3} + \phi^3 \partial_{\bar{\gamma}_2})
\ee
are $Q$-closed but not $Q$-exact. Moreover, they generate $Q$-cohomology. The non-vanishing anti-commutators are
\be
\{\psi_1, \psi_2\} = -\tfrac{2}{3} \phi^3 \qquad \{\psi_1, \psi_3\} = -\tfrac{2}{3} \phi^2 \qquad \{\psi_2, \psi_3\} = -\tfrac{2}{3} \phi^1,
\ee
which are all $Q$-exact: $[\psi_n] [\psi_m] = - [\psi_m] [\psi_n]$. In particular the $[\psi_n]$ generate an exterior algebra: $H^\bullet(A) \cong \bigwedge^\bullet \CC^3$. Nonetheless, we now show that there is a non-trivial ternary product $\mu_3([\psi_1], [\psi_2], [\psi_3])$.

In order to compute the ternary product $\mu_3([\psi_1], [\psi_2], [\psi_3])$, we will construct the lifts $f_1, f_2$. First, we choose the following unary lifts of cohomology classes
\be
\begin{aligned}
	f_1(1)= 1 \qquad f_1([\psi_n]) = \psi_n \qquad f_1([\psi_1] [\psi_2]) = \psi_1 \psi_2 + \tfrac{1}{3} \phi^3\\
	f_1([\psi_1][\psi_2][\psi_3]) = \psi_1 \psi_2 \psi_3 + \tfrac{1}{3} \big(\phi^1 \psi_1 - \phi^2 \psi_2 + \phi^3 \psi_3\big) \quad
\end{aligned}
\ee
where the remaining lifts are obtained by acting with the $S_3$.%
\footnote{It is important that these lifts are compatible with the algebraic relations on $H^\bullet(A)$. In particular, we can not freely lift $[\psi_1][\psi_2]$ and $[\psi_2] [\psi_1]$: they are constrained to satisfy $f_1([\psi_1][\psi_2]) = - f_1([\psi_2][\psi_1])$. In essence, this choice of $f_1$ corresponds lifting to the ``normal-ordered'' cochain, e.g. $f_1([\psi_1][\psi_2]) =\tfrac{1}{2}(\psi_1 \psi_2 - \psi_2 \psi_1) = \psi_1 \psi_2 + \tfrac{1}{3} \phi^3.$}
To construct the binary lift $f_2$, we need to choose a homotopy between $f_1([a][b])$ and $f_1([a])f_1([b])$. For example, we find that $f_1([\psi_1][\psi_2]) - f_1([\psi_1])f_1([\psi_2]) = \tfrac{1}{3}\phi^3 = \tfrac{1}{3} Q \partial_{\gamma_3}$ implying that we can choose $f_2([\psi_1], [\psi_2]) = \tfrac{1}{3} \partial_{\gamma_3}$. The remaining binary lifts relevant for the triple product $\mu_3([\psi_1], [\psi_2], [\psi_3])$ are chosen to be $f_2([\psi_2], [\psi_3]) = \tfrac{1}{3} \partial_{\gamma_1}$ and
\be
\begin{aligned}
	f_2([\psi_1], [\psi_2][\psi_3]) = f_2([\psi_1], [\psi_2])\psi_3 - f_2([\psi_1], [\psi_3]) \psi_2,\\
	f_2([\psi_1][\psi_2], [\psi_3]) = f_2([\psi_2], [\psi_3])\psi_1 - f_2([\psi_1], [\psi_3]) \psi_2\,.
\end{aligned}
\ee
Using the above lifting data, Equation \eqref{eq:mu3lift} yields
\be
\mu_3([\psi_1], [\psi_2], [\psi_3]) = \tfrac{1}{3}
\ee
after a short computation.

\begin{figure}[h!]
	\centering
	\includegraphics[scale=2]{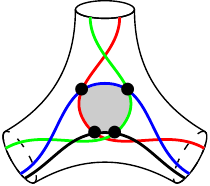}
	\caption{The $3$-punctured sphere and the three Lagrangians (red, blue $\sim$ black, green) that generate its wrapped Fukaya category. The shaded disk contributes to an $A_\infty$ triple product that is mirror to the non-zero triple product $\mu_3(\psi_1, \psi_2, \psi_3)$.}
	\label{fig:xyz-mirror}
\end{figure}

This $B$-twisted Landau-Ginzburg model is mirror to an $A$-twisted sigma model whose target is the 3-punctured sphere \cite{HV00, AAK12}. The (wrapped) Fukaya category of the $3$-punctured sphere, i.e. the category of boundary conditions in the mirror $A$-twisted theory, is generated by the three Lagrangians illustrated in Figure \ref{fig:xyz-mirror}, each being mirror to one of the above generators. See \cite{Se11, Sh11, AAEKO13} for complementary analyses of homological mirror symmetry for this example and its generalizations. The work \cite{AAEKO13} provides a concrete description of (a minimal model for the $A_\infty$ structure on) this category and, in particular, shows that there is a non-vanishing ternary operation $\mu_3$ (in this minimal model) -- this non-vanishing ternary operation receives contributions from the holomorphic disk drawn in Figure \ref{fig:xyz-mirror}. Physically, holomorphic discs in the $A$-model coincide with the contributions of worldsheet instantons with boundaries on the corresponding Lagrangians, and hence are more difficult to compute directly than their $B$-model counterparts; for more details, see e.g. \cite{ABCDKMGSSW09}. The triple product computation we performed above is mirror to this non-vanishing ternary operation.

\subsubsection{Bulk local operators and Hochschild cohomology}
\label{sec:hochschild}
Before turning to the mixed holomorphic-topological setting, let's return to the question of recovering the algebra of bulk local operators from the knowledge of the category of topological boundary conditions. As mentioned in Section \ref{sec:branecat}, the algebra of local operators is expected to be able to be recovered from the \textit{Hochschild cohomology} of the category of branes. We now show how this relation arises from topological descent by constructing a map from bulk local operators to the Hochschild cochains that intertwines the action of the twisting supercharge $Q$ and the differential on the Hochschild complex. Hochschild cohomology naturally has the structure of a Gerstenhaber algebra \cite{G63}, and the Gerstenhaber bracket is expected to recover the secondary product described in Section \ref{sec:sec-prod}, and the (now proven) \textit{Deligne conjecture} states that this can be enriched to a full $E_2$ algebra; see e.g. \cite[Section 1.2]{BBBDN18} and references therein.

We start by reviewing the notion of Hochschild cohomology of a category with (or generated by) a single object $B$, and assume that the endomorphisms $\textrm{Hom}_{\textbf{Bdy}}(B,B)$ of this object are given by the DGA $(A, Q_\partial)$ (rather than, say, a full $A_\infty$ algebra). We are then interested in the Hochschild cohomology of $A$ (with coefficients in $A$), denoted $HH^\bullet(A) = HH^\bullet(A,A)$. A Hochschild $n$-cochain is a multi-linear map from $n$ copies of $A$ to itself:
\be
HC^n(A) = \textrm{Hom}_{\CC}(A^{\otimes n}, A).
\ee
$HC^\bullet(A)$ has two natural gradings: we give an $n$-cochain $\phi \in HC^n(A)$ \textit{Hochschild degree} $n$ and we say it has \textit{internal degree} $|\phi|$ if
\be
F(\phi(O_1, ..., O_n)) = |\phi| + F(O_1) + \ldots F(O_n)
\ee
where, as usual, $F$ measures the degree in $A$. Similarly, there are two natural differentials. First is the differential $Q_\partial^*$ induced from $A$ that increases internal degree by $1$ but leaves the Hochschild degree unchanged:
\be
\begin{aligned}
	(Q_\partial^* \phi)(O_1, \ldots, O_n) &= Q_\partial \phi(O_1, \ldots, O_n) -(-1)^{|\phi|} \phi(Q_\partial Q_1, \ldots O_n)\\
	& \qquad + \ldots + (-1)^{|\phi| + F(O_1) + \ldots} \phi(O_1, \ldots, Q_\partial O_n)\big)\,.\\
\end{aligned}
\ee
The second differential $\delta$ leaves the internal degree unchanged and increases the Hochschild degree:
\be
\begin{aligned}
	(\delta \phi)(O_1, \ldots, O_{n+1}) & = (-1)^{|\phi| F(O_1)} O_1 \phi(O_2, \ldots, O_{n+1}) - \phi(O_1 O_2, \ldots, O_{n+1})\\
	& + \ldots + (-1)^{n} \phi(O_1, \ldots, O_{n} O_{n+1}) + (-1)^{n+1} \phi(O_1, \ldots, O_n) O_{n+1}\,.\\
\end{aligned}
\ee
Moreover, it is straightforward to check that $Q_\partial^*$ and $\delta$ anticommute with one another, so the Hochschild complex of a DGA is naturally a bicomplex. Hochschild cohomology is the cohomology of this bicomplex with respect to the full differential $Q_\partial^* + \delta$; Hochschild cohomology is naturally $\ZZ$ graded by the sum of Hochschild degree and internal degree, which we call degree or total degree and denote by $F$: $F(\phi) = |\phi| + n$. Note that $0$-cochains are identified with elements of $A$ in a way consistent with out conventions for $F$. It will be this grading that is correlated with the $\ZZ$ grading of our TQFT.

A general Hochschild cohomology class of a given degree will be a sum of elements with different internal and Hochschild degrees, and we should expect this to be the case when identifying bulk local operators. In particular, if the bulk local operator $O$ has degree $F(O)$, we should expect to realize it as a formal sum of cochains with total degree $F(O)$:
\be
O \leftrightarrow hc_O = \sum_n hc^n_O, \qquad hc^n_O \in HC^n(A), \quad |hc^n_O| = F(O) - n\,.
\ee
We will identify the action of the twisting supercharge $Q$ on the local operator $O$ with the action of the total differential $Q_\partial^* + \delta$ on the Hochschild cochain $hc_O$
\be
hc^n_{QO} = Q_\partial^* hc^n_O + \delta hc^{n-1}_O\,,
\ee
with the convention $hc^{-1}_O = 0$. Note that both terms on the right hand side have Hochschild degree $n$ and internal degree $F(O) - n + 1$, whence total degree $F(O)+1$, as expected.

\begin{figure}[h!]
	\centering
	\includegraphics{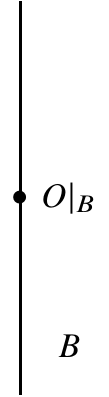} \hspace{2cm} \includegraphics{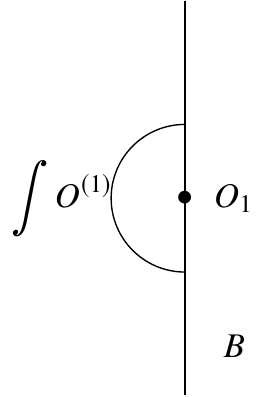} \hspace{2cm} \includegraphics{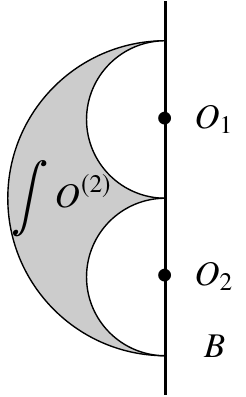}
	\caption{Illustrations of the first three operations mapping a bulk local operator $O$ to a Hochschild cochain $hc_O$. Left: the 0-cochain $hc^0_O$ comes from the boundary value $O|_B$, which is an integral over the $0$-chain $\mathcal{B}_0(\epsilon)$. Middle: the 1-cochain $hc^1_O$ applied to the boundary local operator $O_1$ comes from integrating the first descendant $O^{(1)}$ over the 1-chain $\mathcal{B}_1(\epsilon)$. Right: the 2-cochain $hc^2_O$ applied to the boundary operators $O_1, O_2$ comes from integrating the second descendant $O^{(2)}$ over the 2-chain $\mathcal{B}_2(\epsilon)$.}
	\label{fig:hochschild}
\end{figure}

Finally, we construct the cochains $hc^n_O$ that realize a (co)chain map from bulk local operators to the Hochschild cochain complex. Perhaps unsurprisingly, the $n=0$ map corresponds to bringing the bulk operator to the boundary $B$, i.e. it sends $1 \in \CC$ to $O|_B \in A$:
\be
hc^0_O(1) = O|_B\,.
\ee
It immediately follows that $hc^0_{QO}(1) = (QO)|_B = Q_\partial (O|_B) = (Q_\partial^* hc^0_O)(1)$, as desired. We think of the point $t=0$ on the boundary as the $0$-chain $\mathcal{B}_0(\epsilon)$ in the configuration space $\mathcal{C}_{\mathbb{H}^2}(1)$ of one point (open ball) on half-space $\mathbb{H}^2$. The case for higher $n$ is more interesting. For $n = 1$, we have to use $O$ to construct a map $hc^1_O$ that sends boundary local operators to boundary local operators. For $O_1 \in A$ such a boundary local operator, we consider
\be
hc^1_O(O_1) = \int_{\mathcal{B}_1(\epsilon)} (O O_1)^{(1)} = \int_{\mathcal{B}_1(\epsilon)} O^{(1)} O_1\,,
\ee
where $\mathcal{B}_1(\epsilon)$ is the contour depicted on the middle of Figure \ref{fig:hochschild}, thought of as a 1-chain in the configuration space $\mathcal{C}_{\mathbb{H}^2}(1;1)$ of one bulk point (open ball) and one boundary point (open half-ball) on the half-space $\mathbb{H}^2$. A straightforward application of the descent equation and Stokes' theorem implies that
\be
\begin{aligned}
	hc^1_{QO}(O_1) & = Q_\partial hc^1_O(O_1) - (-1)^{F(O)-1} hc^1_O(Q_\partial O_1) + (-1)^{F(O) F(O_1)} O_1 O|_B - O|_B O_1\\
	& = (Q^*_\partial hc^1_O)(O_1) + (\delta hc^0_O)(O_1)\,,\\
\end{aligned}
\ee
as desired. Note that if $O$ is $Q$-closed and $O_1$ is $Q_\partial$-closed, it follows that $hc^1_O(O_1)$ can be viewed as a homotopy between the two possible ways to collide $O$ with the boundary:
\be
Q_\partial hc^1_O(O_1) = O|_B O_1 - (-1)^{F(O) F(O_1)} O_1 O|_B\,.
\ee

Similarly, for $n = 2$ we can build a map $hc^2_O$ that requires two elements of $A$:
\be
hc^2_O(O_1, O_2) = \int_{\mathcal{B}_2(\epsilon)} (O O_1 O_2)^{(2)} = \int_{\mathcal{B}_2(\epsilon)} O^{(2)} O_1 O_2\,,
\ee
where $\mathcal{B}_2(\epsilon)$ is the region depicted on the right of Figure \ref{fig:hochschild}, thought of as a 2-chain in the configuration space $\mathcal{C}_{\mathbb{H}^2}(1;2)$ of one bulk point (open ball) and two boundary points (open half-balls). Again, a straightforward application of the descent equations and Stokes' theorem implies
\be
hc^2_{QO}(O_1, O_2) = (Q_\partial^* hc^2_O)(O_1, O_2) + (\delta hc^1_O)(O_1, O_2)\,.
\ee
It follows that if $O$ is $Q$-closed and $O_1,O_2$ are $Q_\partial$-closed, then $hc^2_O(O_1, O_2)$ serves as a homotopy between moving $O|_B$ through $O_1 O_2$ and moving $O|_B$ first through $O_1$ and then through $O_2$:
\be
Q_\partial hc^2_O(O_1, O_2) = hc^1_O(O_1 O_2) - \big(hc^1_O(O_1) O_2 + (-1)^{(F(O)-1)F(O_1)} O_1 hc^1_O(O_2)\big)\,.
\ee

\begin{figure}[h!]
	\centering
	\includegraphics{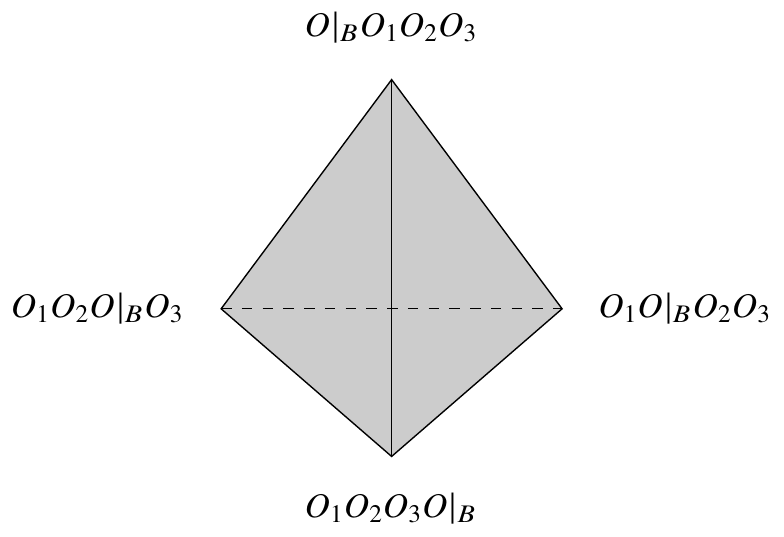}
	\caption{The tetrahedron $\Delta^3$ with its vertices decorated by the locations of the bulk local operator $O|_B$ in the product of the boundary local operators $O_1, O_2, O_3$. The edges of the tetrahedron correspond to a homotopies between these positions, e.g. the dashed edge corresponds to the homotopy $O_1 hc^1_O(O_2) O_3$ between $O_1 O|_B O_2 O_3$ and $O_1 O_2 O|_B O_3$. The faces of the tetrahedron correspond to homotopies between these homotopies, e.g. the rear face corresponds to the homotopy $hc^2_O(O_1, O_2) O_3$ between $O|_B (O_1 O_2) O_3 \to (O_1 O_2) O|_B O_3$ and $O|_B O_1 O_2 O_3 \to O_1 O|_B O_2 O_3 \to O_1 O_2 O|_B O_3$.}
	\label{fig:polytopesHH}
\end{figure}

\begin{figure}[h!]
	\centering
	\includegraphics{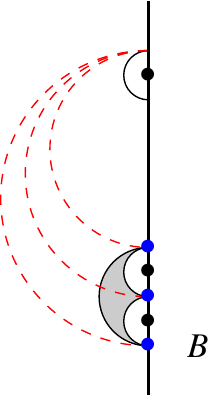} \hspace{1cm} \includegraphics{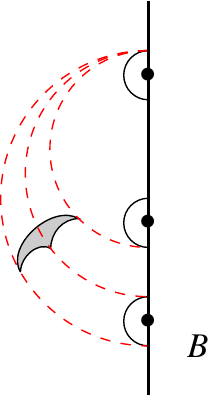} \hspace{1cm} \includegraphics{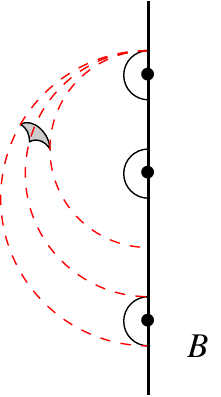} \hspace{1cm} \includegraphics{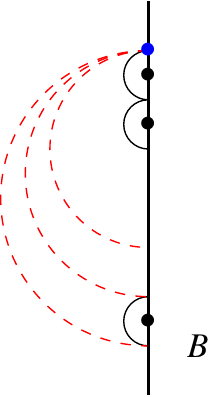}
	\caption{The 3-chain $\mathcal{B}_3(\epsilon)$ in the configuration space $\mathcal{C}_{\mathbb{H}^2}(1;3)$ of one bulk open ball and three boundary open half-balls. The blue vertices are the vertices of the tetrahedron in Figure \ref{fig:polytopesHH}; the gray shaded regions are horizontal slices of the tetrahedron, with the leftmost region corresponding to the bottom face of the tetrahedron; and the dashed red edges are the vertical edges of the tetrahedron.}
	\label{fig:HH3chain}
\end{figure}

We can systematize this process. Just as with the $A_\infty$ operations described in Section \ref{sec:Ainfty}, we construct a polytope whose various faces encode the above homotopies. At Hochschild degree $n$, we get an $n$-dimensional polytope with $n+1$ vertices corresponding to possible places to collide $O$ with the boundary, e.g. $O|_B O_1 \ldots O_n$ or $O_1 O|_B \ldots O_n$ and so on. The 1-dimensional faces, a.k.a. edges, represent homotopies between placements of $O|_B$ and connect vertices that differ by moving $O|_B$ across a single operator (or a single product of operators) and, more generally, the $k$-dimensional faces represent higher homotopies between the various ways to move $O|_B$ through $k$ boundary operators (or $k$ separate products of operators). We saw above that the $n=0$ is a point, $n=1$ is an interval, and $n=2$ a triangle; in general, the corresponding polytope is the \textit{$n$-simplex} $\Delta^n$. We illustrate the case the 3-simplex $\Delta^3$, a.k.a. the tetrahedron, in Figure \ref{fig:polytopesHH} and the corresponding 3-cycle in the configuration space $\mathcal{C}_{\mathbb{H}^2}(1;3)$ in Figure \ref{fig:HH3chain}.

In this fashion, we can construct a multilinear map with $n$ inputs by integrating over a region $\mathcal{B}_n(\epsilon)$ in the configuration space $\mathcal{C}_{\mathbb{H}}(1;n)$
\be
hc^n_O(O_1, \ldots, O_n) = \int_{\mathcal{B}_n(\epsilon)} (O O_1 \ldots O_n)^{(n)}\,.
\ee
From the boundary structure of the $n$-simplex $\Delta^n$ and the descent equations, it immediately follows that $hc^n_O$ can be identified with a $n$ cochain with internal degree $F(O)-n$. Thus, the assignment
\be
hc: O \mapsto \sum\limits_n hc^n_O
\ee
yields a (co)chain map from local operators (with differential $Q$) to the Hochschild cochain complex $HC^\bullet(A)$ of the DGA $(A, Q_\partial)$ (with differential $Q_\partial^* + \delta$)
\be
hc_{QO} = Q_\partial^* hc_O + \delta hc_O\,.
\ee

\subsubsection{Example: $B$-twisted free chiral multiplet}
As an example computation, consider the case of $B$-twisted free chiral multiplet. In this context, the category of boundary conditions is generated by the $B$-type Neumann boundary condition. (It is also illustrative to rederive the following result using a Dirichlet boundary condition!\footnote{We remark that these two basic boundary conditions, each of which generates the category of boundary conditions in the TQFT of the $B$-twisted free chiral multiplet, provide a simple instance of \textit{Koszul duality}. In particular, two ``transverse'' generating objects $B_1, B_2$ in a category of boundary conditions, i.e. generators such that $\textrm{Hom}(B_1, B_2) = \mathbb{C}$, support Koszul dual boundary operator algebras \cite{Dim17, PW}. For a richer example with applications to mirror symmetry, see e.g. \cite{Aganagic:2021ubp}, which employs a version of Koszul duality on the derived Fukaya-Seidel category. The latter can be viewed as being generated by the boundary conditions corresponding to either left or right-thimbles. \label{fn:Koszul}}) The algebra of local operators is simply the polynomial algebra $\CC[\phi]$, so we are interested in the Hochschild cohomology $HH^\bullet(\CC[\phi])$.

As mentioned above, the zeroth Hochschild cohomology group is the center $HH^0(A) \cong Z(A)$; since $\CC[\phi]$ is commutative, we conclude the Hochschild cohomology group is all of $\CC[\phi]$:
\be
HH^0(\CC[\phi]) \cong \CC[\phi] \cong H^{0}(\CC).
\ee
Similarly, the first Hochschild cohomology group corresponds to \textit{outer derivations} of the algebra $HH^1(A) \cong \textrm{OutDer}(A)$, i.e. derivations of $A$ (maps $f: A\to A$ with $f(ab) = f(a)b \pm a f(b)$) modulo inner derivations of $A$ (maps given by commutators $a \to [a,b]$ for $b \in A$). Since $\CC[\phi]$ has no inner derivations, and derivations of $\CC[\phi]$ take the form $\phi^k \partial_\phi$, it follows that $HH^1(\CC[\phi])$ can be identified as
\be
HH^1(\CC[\phi]) \cong \CC[\phi] \partial_\phi.
\ee
The remaining Hochschild cohomology groups vanish, so we conclude
\be
HH^\bullet(\CC[\phi]) \cong \CC[\phi] \oplus \CC[\phi] \partial_\phi  \cong H^{(0,\bullet)}(\CC, \mbox{$\bigwedge$}^\bullet T^{(1,0)}\CC)\,,
\ee
precisely matching the vector space of local operators in a $B$-twisted chiral, c.f. Section \ref{sec:CRdolbeault}.

The product structure on Hochschild cohomology is given by the \textit{cup product} $\alpha \cup \beta$. Explicitly, if $\alpha$ is a $p$-cochain and $\beta$ is a $(n-p)$-cochain, we have
\be
\alpha \cup \beta(O_1, \ldots, O_n) = (-1)^{F(\alpha)(F(O_1) + \ldots + F(O_p))} \alpha(O_1, \ldots, O_p) \beta(O_{p+1}, \ldots, O_n).
\ee
For example, we have $\phi^k \cup \phi^l = \phi^{k+l}$, $\phi^k \cup \phi^l \partial_\phi = \phi^{k+l} \partial_\phi = \phi^k \partial_\phi \cup \phi^l$. More interestingly, we have
\be
\partial_\phi \cup \partial_\phi(\phi^k, \phi^l) = kl \phi^{k+l-2}.
\ee
Note that this 2-cochain is $\delta$-exact
\be
\partial_\phi \cup \partial_\phi = -\delta (\partial_\phi{}^2)\,, \qquad \partial_\phi{}^2(\phi^k) = k(k-1) \phi^{k-2}\,,
\ee
so that $\partial_\phi \cup \partial_\phi$ vanishes at the level of cohomology, as expected. Thus, we not only recover the vector space of local operators, but also its product structure.

One recent appearance of these considerations may be found in \cite{Costello:2015xsa, Costello:2019jsy}, which proves that open-closed (type IIB and type I) topological strings have a unique quantization. The authors employ cohomological arguments to prove the cancellation of anomalies in open-closed string couplings. In this context, the Hochschild cohomology (more precisely, the cyclic cohomology; see footnote \ref{fn:cyclic}) that classifies consistent deformations of the open-string algebra on a(n infinite) stack of twisted D-branes reproduces the space of closed-string fields arising from BCOV theory \cite{BCOV94} (the twisted closed string sector). The open-string algebra is readily obtained using the holomorphic Chern-Simons theory description of the D-brane worldvolume theory. One may write down universal deformations of the open-string theory via couplings with BCOV theory; cf. our discussions of deformations and descent in Section \ref{sec:deformations}. Koszul duality, which provides complementary descriptions of a category of boundary conditions (see footnote \ref{fn:Koszul}), also plays a role in these contexts \cite{CL16}: Koszul dual algebras possess the same Hochschild cohomology, so one may use Koszul duality to pass to a different, and potentially more useful, description of the open-string sector.

\subsubsection{Higher dimensional considerations}
\label{sec:higherdim}
Much of the structure present in $d=2$ applies to more generally to $d \geq 2$-dimensional topological theories, c.f. \cite[Section 2]{BBBDN18}. Collision of local operators induces an associative and commutative product on $Q$-cohomology classes; the secondary product described in Section \ref{sec:sec-prod} upgrades this commutative algebra to a ($(1-d)$-shifted) Poisson algebra. Considering these operations at the chain-level would imply that the various properties of this algebra are only satisfied up to suitable homotopies encoded by descendants. This notion is formalized that of an \textit{$E_d$ or $d$-disk algebra} \cite{L09}: there is a product operation for every choice of insertion points and data relating homotopic configurations; for $d=1$, an $E_1$ algebra is equivalent to an $A_\infty$ algebra. 

\begin{figure}[h!]
	\centering
	\includegraphics{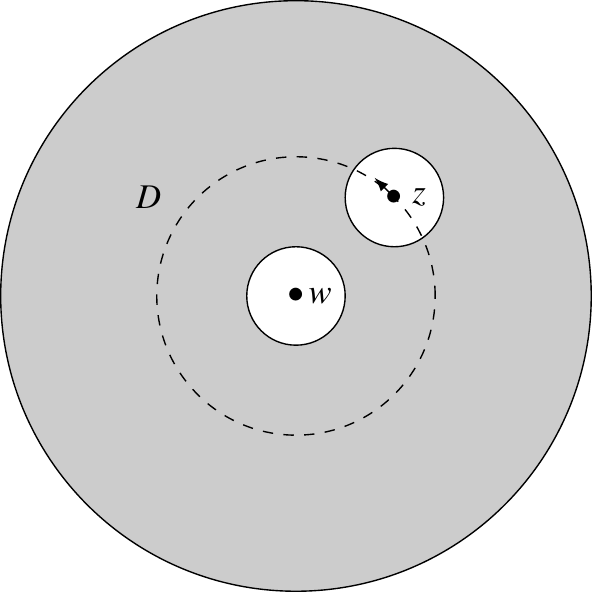}
	\caption{A slice of the 3-cycle $\mathcal{L}_3$ in the configuration space $\mathcal{C}_{\RR^2}(3)$ used to construct the ternary product $\ell_3$. The local operator $O_3$ is placed at a point $w$, the 1-form valued local operator $O_2^{(1)}$ is placed at $z$, and the 2-form valued local operator $O_1^{(2)}$ at $u$. The variable $u$ is integrated over the 2-dimensional region $D$ and $z$ is integrated along the dotted contour encircling $w$.}
	\label{fig:l3int}
\end{figure}

If we focus on the secondary product, i.e. on the \textit{Lie part} of this $E_d$ algebra, there is essentially%
\footnote{There is also the graded skew-symmetry of the bracket, but it is less interesting. See \cite[Section 3.2.2]{BBBDN18}.} %
only one relation to be considered: the (shifted) Jacobi identity
\be
\{O_1,\{O_2,O_3\}\} = \{\{O_1,O_2\},O_3\} + (-1)^{(F(O_1)+1)(F(O_2)+1)}\{O_2, \{O_1,O_3\}\}\,.
\ee
The above chain level considerations are encoded in the notion of a (shifted) \textit{homotopy Lie algebra} or \textit{$L_\infty$ algebra}.%
\footnote{The recent works \cite{GMW115, GMW215} obtain the secondary product in massive Landau-Ginzburg theories in the A-twist. The algebraic structure, including the secondary product, gets extended to a full $L_{\infty}$ structure. In fact, the complete bulk-boundary system enjoys a larger homotopy algebraic structure, incorporating an $A_{\infty}$ algebra associated to the algebra of local boundary operators, called an $LA_{\infty}$-algebra, and this is believed to be further enhanced to a complete $E_2$-algebra structure. We thank G. Moore and A. Khan for correspondences on this point. (There is also an $A_{\infty}$ structure in these theories describing the fusion of one-dimensional topological lines in two-dimensions, associated to the $(\infty, 2)$-categorical structure of 2d TQFTs.) Homotopy algebras in bulk-boundary systems have also been considered in the context of open-closed string field theory, e.g. \cite{KSopenclosed}.}
For example, in $d = 2$, we can define the following ternary operation $\ell_3$: given three points (open balls) on $\RR^2$, we consider the region illustrated in Figure \ref{fig:l3int}. The resulting 3-chain $\mathcal{L}_3$ in the configuration space $\mathcal{C}_{\RR^2}(3)$ is a solid torus with two solid tori removed. Given $Q$-closed local operators $O_1, O_2, O_3$ we consider the resulting product
\be
\ell_3(O_1, O_2, O_3) = \int_{\mathcal{L}_3} (O_1 O_2 O_3)^{(3)} = \int_{\mathcal{L}_3} O_1^{(2)} O_2^{(1)} O_3\,.
\ee
A straight-forward computation shows that
\be
Q \ell_3(O_1, O_2, O_3) = \{O_1, \{O_2, O_3\}\} - \{\{O_1, O_2\}, O_3\} - (-1)^{(F(O_1)+1)(F(O_2)+1)} \{O_2, \{O_1, O_3\}\},
\ee
whence $\ell_3(O_1, O_2, O_3)$ is a homotopy trivializing the (shifted) Jacobi identity. Similarly, the higher $n$-ary $L_\infty$ products can be realized by integrating over cycles in the configuration space of $n$ points on $\RR^2$.

It is also be possible to recover the algebra of bulk local operators from sufficient knowledge of the theory's boundary conditions. Indeed, this is a baby instance of the \textit{cobordism hypothesis} of Baez-Dolan \cite{BD95}, which essentially says a $d$-dimensional TQFT is entirely determined (up to homotopy) by its $(d-1)$-category of boundary conditions; see also \cite{L09}.

One situation where it is possible to get quite far with much less occurs when a 3d TQFT admits a (sufficiently large) holomorphic boundary condition furnishing a \textit{vertex operator algebra} (VOA). The prototypical example is the classic relation between Chern-Simons theory and the boundary Wess-Zumino-Witten (WZW) current algebra \cite{W89}, but also applies to the $A$- and $B$-twisted $\mathcal{N}=4$ theories described in Section \ref{sec:3dAB} \cite{CG19}. In such a situation, the algebra of local operators of the bulk TQFT can be realized as the \textit{space of conformal blocks} of the boundary VOA (on a sphere), also called \textit{chiral homology} in the derived setting \cite{FG12}; there is a single conformal block of the WZW current algebra, corresponding to the lack of local operators in the bulk, and typically infinitely many for the VOAs arising in twisted 3d $\mathcal{N}=4$ theories \cite{CCG19}, to account for (typically) infinite-dimensional algebra of holomorphic functions on the Higgs and Coulomb branches.

\subsubsection{A word on higher algebras \& deformations}\label{sec:deformations}

There is much to say about higher algebras, and we have only scratched the surface. One prominent line of thought, that we unfortunately mention only briefly, is the deep relationship in mathematics between formal deformation, or moduli, problems, and homotopy Lie and associative algebras. 

The relationship arises as follows. Consider solutions to the Maurer-Cartan (MC) equation of a DG-Lie algebra $(A, d_A)$, i.e. a graded Lie algebra $(A = \bigoplus A^j, [-,-])$ with a differential $d_A: A^j \to A^{j+1}$ that is nilpotent $d_A{}^2 = 0$:
\begin{equation}
	d_A x + {1 \over 2}[x, x] = 0, \ \ x \in A.
\end{equation} 
This equation can be viewed as a requirement for the differential in the presence of a first-order deformation, $d_A + \left[x , \cdot \right]$, to remain nilpotent. The solution space to this equation is equivalent to the space of (perturbative)  deformations of the DG-Lie algebra. There is similarly a version of the MC equation for a DG-associative algebra, where the quadratic term is given by the associative product. More generally, one can write a MC equation for homotopy algebras, where the quadratic term is replaced by a sum over all higher $n$-ary operations, $n \geq 2$.  

Maurer-Cartan equations arise quite generally in physics. At first blush, they are familiar from Chern-Simons or BF-like theories, as the equations of motion. For example, string field theorists may recognize the MC equations as equations of motion in (open-)closed string field theory. In that context, solutions to the equations of motion describe closed string backgrounds, which have the interpretation as consistent deformations around a given geometric background (i.e. vacuum solution). 

More generally, if one formulates any gauge theory in the BV formalism, the resulting first-order form of the action leads naturally to a(n $L_{\infty}$-type) MC equation \cite{Movshev:2003ib}. The MC equation arising from the (classical) BV master equation admits a deformation theory interpretation in terms of an underlying formal manifold equipped with the BV-differential; $L_{\infty}$ algebras are, quite generally, baked into perturbative quantum field theory and string theory! This point of view also dovetails beautifully with modern formulations of quantum field theories as factorization algebras, following Costello and Gwilliam \cite{CG16}, to which we refer for details.

We have also seen that descendants in a twisted QFT could be added to the action as deformations. There, too, the MC equation for (homotopy) algebras is lurking behind the scenes. In that context, the deforming element $x$ should be viewed as the operator whose descendant deforms the action, and $d_A$ is given by the twisted BRST differential. $A$ is the operator algebra of the twisted QFT. For example, in the A and B-twists we have been spending most of our time on, these are the corresponding topological chiral and twisted chiral rings. The fact that $x$ is a MC element is equivalent to the statement that the deformation term is $d_A$-invariant (physically, gauge-invariant) to all orders in perturbation theory. This can be shown by a straightforward path integral manipulation; see \cite{CP20, PW} for the computation in the context of DG-associative algebras and \cite{GO} for the A$_{\infty}$ expression. We stress that, here, the MC equation is not a classical equation of motion, but describes gauge-invariance at the quantum level to all orders in perturbation theory, often resulting in rich deformations of higher algebraic structures. 

One may also wish to study perturbative deformations arising from coupling two uncoupled sectors together, such as a (twisted) bulk QFT with some local order-type defect, or open-closed (topological) string theories. This leads to the appearance of a notion in homological algebra called \textit{Koszul duality}, which captures (BV)-BRST invariance of the resulting deformation to all orders in perturbation theory. See \cite{PW} for a recent review and for original references. Because Koszul duality appears universally when studying perturbative deformations (couplings) of two operator algebras, its appearance in open-closed string theories has applications to holography, viewed as a type of open-closed string duality \cite{CP20, CP22}. These connections are still under active study and development.

\subsection{Holomorphic and mixed topological-holomorphic descent}
\label{sec:HTdescent}
We have been focusing on topological descent in topologically twisted theories. In our earlier discussion of twisting, we noticed that in general one may instead have holomorphic dependence in some or all of the twisted directions. There is a holomorphic version of the descent procedure, and the resulting enrichments of algebraic structure in twisted theories is an area of active research. As mentioned above, the algebraic structure underlying descent and collision in a topological theory is an $E_d$ algebra. The corresponding structure for holomorphic or mixed holomorphic-topological is not well understood.

The structure of twists across various dimensions implies that a minimal twist in even dimensions will be fully holomorphic ($n= 0$) and in odd dimensions will have a single topological direction ($n = 1$) \cite{ESW20,ESW21}. Somewhat more generally, the twisting supercharge $Q$ will trivialize, say, $n$ real translations and $m$ anti-holomorphic translations, where spacetime has dimension $d = n + 2m$. If we work in local real coordinates $x^\mu$ for $\mu = 1, ..., n$ and complex coordinates $z^a,\bar{z}^{\bar{a}}$ for $a, \bar{a} = 1, ..., m$, we expect to find fermionic operators $Q_\mu, Q_{\bar{a}}$, that commute with the momenta and anti-commute amongst themselves, such that
\be
\{Q, Q_\mu\} = i P_\mu \qquad \{Q, Q_{\bar{a}}\} = i P_{\bar{a}}.
\ee
Just as in purely topological descent, we can use these to construct mixed holomorphic-topological descendants $O^{(k)}$ of a local operator $O$ satisfying a mixed holomorphic-topological descent equation
\be
Q O^{(k)} = d' O^{(k-1)} + (QO)^{(k)}
\ee
where, locally, we have $d' = d x^\mu \partial_\mu + d z^{\bar{a}} \partial_{\bar{a}} = d_{\RR^n} + \bar{\partial}_{\CC^m}$ is the sum of the de Rham differential on $\RR^n$ and the Dolbeault differential on $\CC^m$. Just as in the fully topological setting, $O^{(k)}$ contains explicit information about the $x^\mu$ and $\bar{z}^{\bar{a}}$ dependence of $Q$-closed local operators. In particular, at the level of $Q$-cohomology, it follows that correlation functions of $Q$-closed local operators are topological along $\RR^n$ and holomorphic along $\CC^m$.

\subsubsection{Holomorphic twists in 2d}
\label{sec:2dholo}
The simplest theories that exhibit holomorphic descent are holomorphically twisted of 2d $\mathcal{N} = (0,2)$ or $\mathcal{N} = (2,2)$ theories \cite{W98, K05, W07}; the latter were historically called \textit{half-twisted} theories.\footnote{Although it is beyond the scope of these notes, there has been much fascinating recent work on novel dualities of 2d $\mathcal{N}=(0,2)$ theories \cite{Gadde:2013lxa, Gadde:2014wma} and applications to the geometry of four-manifolds \cite{Gadde:2013sca, Dedushenko:2017tdw, Dimofte:2019buf}.} In this setting, there are two supercharges $Q_+, \bar{Q}_+$, with non-trivial bracket $\{\bar{Q}_+, Q_+\} = -2 i \partial_{\bar z}$ so that e.g. $Q = \bar{Q}_+$ is a holomorphic supercharge with $Q_{\bar{z}} = \tfrac{i}{2} Q_+$ trivializing $\partial_{\bar{z}}$. The fact that correlation functions of $Q$-closed local operators only depend holomorphically on their position implies that collision gives local operators the structure of a \textit{vertex algebra} or \textit{chiral algebra}.

A local operator $O$ in a holomorphically twisted 2d $\mathcal{N} = (0,2)$ or $\mathcal{N} = (2,2)$ theory only has a single descendant $O^{(1)} = -d\bar{z} [Q_{\bar{z}}, O]$ satisfying $Q O^{(1)} = \bar{\partial} O$. Note that the $Q$-variation of the 2-form valued local operator $dz O^{(1)}$ is the exterior derivative of a 1-form local operator when $QO = 0$: $Q (dz O^{(1)}) = d (dz O)$. It is this descendant that ensures contour integrals of ($dz$ times) $Q$-closed local operators only depend on the homology class of the integration contour: if $D$ is some compact, connected region on $\CC$
\be
Q \int_{D} dz O^{(1)} = \int_{D} d(dz O) =\oint_{\partial D} dz O.
\ee 
Of course, this invariance under deforming an integration contour is expected from the holomorphic nature of $Q$-cohomology. 

If we have two local operators $O_1$ and $O_2$ at points $z$ and $w$, respectively, we can use this deformation invariance to define a secondary product%
\footnote{We call this a ``secondary product'' but it is closer in spirit to the (graded) commutator in 1d topological theories. In particular, the bracket is degree 0 and is directly tied to the primary product; this bracket merely extracts the simple pole in the OPE. The $\lambda$-bracket described below can be interpreted as the Fourier transform of the OPE, c.f. \cite[Chapter 2]{K97}.} %
by, say, integrating $dz O_1$ along a small circle centered at $w$. More generally, we can define their \textit{$\lambda$-bracket} $\{-{}_\lambda-\}$, which is a local operator at $w$, by integrating against $dz e^{\lambda (z-w)}$:
\be
\{O_1 {}_\lambda O_2\}(w) = \oint_{S^1} (dz e^{\lambda (z-w)} O_1(z)) O_2(w).
\ee
This is one of the simplest instance of the sphere algebras present in any holomorphic field theory, c.f. \cite{CG16, GW21}. 

In an honest holomorphic theory, e.g. a chiral 2d CFT, the $\lambda$-bracket satisfies various properties and the resulting structure is sometimes called a \textit{Lie conformal algebra}; together with the normal-ordered product $:O_1 O_2:$, the resulting structure is equivalent to the usual notion of a vertex algebra \cite{BK03}. We will mostly focus on the Lie part of this structure, i.e. those properties only involving the $\lambda$-bracket, to keep the discussion brief.

In a twisted theory, the defining relations of the Lie conformal algebra structure may be automatic or they may need to be imposed cohomologically. For example, the action of the holomorphic derivative $\partial_w$ is given by:
\be
\partial_w \{O_1 {}_\lambda O_2\}(w) = \lambda \{O_1 {}_\lambda O_2\}(w) + \{O_1 {}_\lambda \partial O_2\}(w),
\ee
where $\partial O_2(w) = \partial_w O_2(w)$. Similarly, we have
\be
\{\partial O_1 {}_\lambda O_2\}(w) = - \lambda \{O_1 {}_\lambda O_2\}(w).
\ee
These properties (together called the \textit{sesquilinearity} of the $\lambda$-bracket) are immediate from the definition of the $\lambda$-bracket and aren't imposed cohomologically. 

More interestingly, the $\lambda$-bracket has a version of graded skew-symmetry
\be
\{O_1 {}_\lambda O_2\} = -(-1)^{F(O_1)F(O_2)} \{O_2 {}_{-\lambda-\partial} O_1\}\,.
\ee
This is usually verified as follows:
\be
\begin{aligned}
	\{O_1 {}_\lambda O_2\}(0) & = \oint dz e^{\lambda z} O_1(z) O_2(0) = \oint dz e^{\lambda z} \big(e^{z \partial + \bar{z} \bar{\partial}}O_1(0)\big) \big(e^{z \partial + \bar{z} \bar{\partial}}O_2(-z)\big)\\
	& = (-1)^{F(O_1) F(O_2)} \oint dz e^{z(\lambda + \partial)} \big(O_2(-z) O_1(0)\big)\\
	& = -(-1)^{F(O_1) F(O_2)} \{O_2 {}_{-\lambda-\partial} O_1\}(0)\,.\\
\end{aligned}
\ee 
Note that the holomorphic nature of the theory is used to remove the anti-holomorphic part of the translation $e^{\bar{z} \bar{\partial}}$. When this holomorphy is imposed cohomologically, we see that this skew-symmetry relation must be imposed cohomologically. The essential step to constructing such a homotopy lies in dealing with the anti-holomorphic translation; in particular, we need to construct an operator $H_{\bar{z}}(O(0))$ of cohomological degree $F(O)-1$ such that:
\be
O(z)(\equiv e^{z \partial + \bar{z}\bar{\partial}} O(0)) = e^{z \partial} O(0) + Q H_{\bar{z}}(O(0))\,.
\ee
It is straightforward to construct such a homotopy using the descent supercharge $Q_{\bar{z}}$, e.g.
\be
\begin{aligned}
	H_{\bar{z}}(O(0)) & = e^{z \partial} \int\limits_{0}^{\bar{z}} d\bar{z}' e^{\bar{z}' \bar{\partial}} O(0) = \sum\limits_{n \geq 0, \bar{n} \geq 1} \frac{z^n \bar{z}^{\bar{n}}}{n! \bar{n}!} \partial^n \bar{\partial}^{\bar{n}-1} Q_{\bar{z}} O(0)\,,\\
	\rightsquigarrow & Q H_{\bar{z}}(O(0)) = \sum\limits_{n \geq 0, \bar{n} \geq 1} \frac{z^n \bar{z}^{\bar{n}}}{n! \bar{n}!} \partial^n \bar{\partial}^{\bar{n}}O(0) = O(z) - e^{z \partial} O(0)\,.
\end{aligned}
\ee
Since we are working with a single complex dimension, there are no higher homotopies for translations -- these would require spacetime $(0,p)$ forms with $p>1$. We can construct the homotopy for the (graded) skew-symmetry by integrating this translation homotopy over the configuration space of 2-point on the complex plane $\mathcal{C}_\CC(2)$.%
\footnote{We use $\CC$ instead of $\RR^2$ to emphasize the complex nature on this configuration space, i.e. that it has a natural complex structure inherited from $\CC$.} %
For example, we denote
\be
H(O_1, O_2)(0) = \oint dz e^{\lambda z} H_{\bar{z}}(O_1(0) O_2(-z))\,;
\ee
if $O_1$ and $O_2$ are $Q$-closed, then
\be
Q H(O_1, O_2)(0) = \{O_1 {}_\lambda O_2\}(0) + (-1)^{F(O_1)F(O_2)} \{O_2 {}_{-\lambda-\partial} O_1\}(0)\,.
\ee

%Similarly, the $\lambda$-bracket is a (twisted) derivation of the OPE: if $O_2$ is placed at a point $y$ and $O_3$ at $w$ with $y-w$ small, we have 
%\be
%	\{O_1 {}_\lambda O_2(y) O_3\}(w) = e^{\lambda(y-w)} \{O_1{}_\lambda O_2\}(y) O_3(w) + (-1)^{F(O_1)F(O_2)} O_2(y) \{O_1{}_\lambda O_3\}(w)
%\ee
%Just as with the graded symmetry, this relation is imposed cohomologically with by integrating $(O_1 O_2 O_3)^{(2)}$ over the chain $D$ given on the right of Figure \ref{fig:PVhomotopies}: if $O_1$, $O_2$, and $O_3$ are $Q$-closed, it follows that
%\be
%	Q \big(\star_D(O_1, O_2, O_3)\big) = e^{\lambda (y-w)}\{O_1 {}_\lambda O_2\} O_3 + (-1)^{F(O_1)F(O_2)} O_2 \{O_1 {}_{\lambda} O_3\}\,.
%\ee

Even more interestingly, there is the $\lambda$-Jacobi identity: for any local operators $O_1, O_2, O_3$ and $\lambda, \mu \in \CC$ the bracket satisfies
\be
\{O_1 {}_\lambda \{O_2{}_{\mu} O_3\}\} = \{\{O_1 {}_\lambda O_2\}{}_{\lambda + \mu} O_3\} + (-1)^{F(O_1)F(O_2)} \{O_2 {}_\mu \{O_1{}_{\lambda} O_3\}\}.
\ee
Just as with the graded skew symmetry of the bracket, this relation is generally imposed cohomologically. We can construct the desired homotopy by integrating over the same region $\mathcal{L}_3$ (see Figure \ref{fig:l3int}) used for the ternary $L_\infty$ operation $\ell_3$ in the topological setting
\be
\{O_1 {}_\lambda O_2 {}_\mu O_3\}:= \int_{\mathcal{L}_3} (du e^{\lambda (u-w)} O_1^{(1)})  (dz e^{\mu (z-w)} O_2) O_3.
\ee
Since this operation involves a single descendant, it follows that this is a degree $-1$ operation. A straightforward application of the holomorphic descent equation and Stokes' theorem implies that the $Q$-variation of $\{O_1 {}_\lambda O_2 {}_\mu O_3\}$ for $Q$-closed $O_i$ is exactly the $\lambda$-Jacobiator
\be
Q\{O_1 {}_\lambda O_2 {}_\mu O_3\} = \{O_1 {}_\lambda \{O_2{}_{\mu} O_3\}\} - \{\{O_1 {}_\lambda O_2\}{}_{\lambda + \mu} O_3\} - (-1)^{F(O_1)F(O_2)} \{O_2 {}_\mu \{O_1{}_{\lambda} O_3\}\},
\ee
where the first term on the right-hand side comes from the exterior boundary of $D$, the second term comes from the boundary of $D$ surrounding $w$, and the third term comes from the boundary of $D$ surrounding $z$. Unsurprisingly, there is a whole tower of higher $n$-ary brackets that serve as homotopies between higher Jacobi identities in complete analogy with an $L_\infty$ algebra; we call the resulting structure an \textit{$L_{\infty}$ conformal algebra}.%
\footnote{The recent paper \cite{WLW22} introduces the related notion of an $n$-Lie conformal algebra. This structure corresponds to the $L_\infty$ conformal algebra we describe, where only the $n$-ary bracket $\{O_1 {}_{\lambda_1} \ldots {}_{\lambda_{n-1}} O_n\}$ is non-vanishing.} %
Once combined with the normal ordered product, and the various homotopies realizing the $Q$-exactness of the defining relations, this $L_\infty$ conformal algebra structure should extend to a \textit{homotopy vertex algebra} in analogy with the homotopy-free setting.%
\footnote{It would be interesting to explore this homotopies in more detail and compare the resulting structure to the notion of \textit{homotopy chiral algebra} as developed by Francis-Gaitsgory \cite{FG12}.}

\subsubsection{Holomorphic-topological twists in 3d}
\label{sec:3dmixed}
Descent in mixed holomorphic-topological theories is an interesting admixture of descent in topological and holomorphic theories. Many simple examples come from 3d theories with (at least) $\mathcal{N}=2$ supersymmetry. Consider the 3d $\mathcal{N}=2$ superalgebra:
\be
\left\lbrace Q_+, \bar{Q}_+ \right\rbrace = -2 i \partial_{\bar{z}} \qquad  \left\lbrace Q_-, \bar{Q}_- \right\rbrace = 2 i \partial_{z} \qquad \left\lbrace Q_-, \bar{Q}_+ \right\rbrace = \left\lbrace Q_+, \bar{Q}_- \right\rbrace = i \partial_{t}.
\ee
It follows that, e.g., $Q:= \bar{Q}_+$ is a holomorphic-topological supercharge with derivatives $\partial_t, \partial_{\bar{z}}$ are $Q$-exact via $Q_t = -i Q_-$ and $Q_{\bar{z}} = \tfrac{i}{2} Q_+$.

Just as in 2d, collision gives $Q$-closed local operators the structure of vertex algebra. Importantly, the OPE of two $Q$-closed local operators is necessarily non-singular (up to $Q$-exact terms): these OPEs may only be singular in the limit as the insertion points coincide but are simultaneously independent of the separation in $t$ up to $Q$-exact terms, whence they are non-singular in $Q$-cohomology. The resulting vertex algebra is said to be \textit{commutative}.%
\footnote{The reason for the name comes from the fact that non-singular OPEs implies that the Lie algebra of modes is abelian, i.e. (graded) commutative.} %
This is a general feature of mixed holomorphic-topological theories. 

Twisted 3d $\mathcal{N}=2$ theories also admit a $\lambda$-bracket, now given by integrating $dz e^{\lambda(z-w)} O_1^{(1)}(t,z,\bar{z})$ over an $S^2$ surrounding $O_2(s, w,\bar{w})$:
\be
\{O_1 {}_\lambda O_2\}(s,w,\bar{w}) = \int_{S^2} (dz e^{\lambda (z-w)} O_1^{(1)}(t, z,\bar{z})) O_2(s,w,\bar{w}).
\ee
Since this operation involves 1 descendant, the $\lambda$-bracket now has degree $-1$. Together with the (commutative) OPE, the resulting structure can be called a \textit{$(-1)$-shifted Poisson vertex algebra}. Thus, we conclude that the $Q$-cohomology of local operators in a holomorphically twisted 3d theory is naturally have this structure \cite{CDG20, OY19}.

As with the $A_\infty$ and $L_\infty$ algebras appearing in topological 1d and 2d theories, respectively, working at the chain level should yield yet more refined information about the collision and descent of local operators in general mixed holomorphic-topological theory. For example, the $(-1)$-shifted $\lambda$-bracket should satisfy a $(-1)$-shifted version of the $\lambda$-Jacobi identity that is imposed cohomologically by the $Q$-variation of integrated descendant as in 2d. Similarly, descent in the topological $t$ direction should allow for an $A_\infty$ enrichment of this $\lambda$-bracket. See \cite[Section 2.4]{CDG20} for more details about what such a structure, which they call an ``$A_\infty$-chiral algebra,'' would entail.

It is also worth mentioning that there is an analog of the Hochschild cohomology construction described at the end of Section \ref{sec:branecat} that applies to this mixed 3d setting. In particular, the algebra of bulk local operators in a holomorphic-topological twist should be able to be recovered from a (sufficiently large) boundary condition. Any such boundary condition admits a (homotopy) vertex algebra of local operators, and the algebra of bulk local operators can be extracted from its (derived) center, c.f. \cite[Section 2.4]{CDG20}. Recent work of Zeng \cite{Z21} has used this construction to describe aspects of the mode algebra of the commutative vertex algebra of local operators in twisted 3d $\mathcal{N}=2$ abelian gauge theories, which receives interesting non-perturbative corrections due to monopole operators.

\subsubsection{Homotopy actions of vertex algebras and Poisson vertex algebras}
It is natural to expect that instances of vertex algebras and (shifted) Poisson vertex algebras appearing in twisted supersymmetric field theories and string theories have chain-level lifts involving the homotopy algebras described above. It would be gratifying to understand in detail how the homotopical enrichments of these vertex algebras transfer across dualities. There are many beautiful examples of vertex algebras, such as those accessible from twisted theories, transferring across duality maps. One classic example from string theory arises in the duality between the heterotic string on a four-torus $T^4$ and type IIA on a K3 surface. BPS states in the heterotic string are related to vertex operators on the worldsheet such that the supersymmetric right-movers are in their ground states; in the half-twist, one retains a holomorphic vertex algebra structure, essentially from the action of the left-movers on BPS states. The heterotic BPS states get mapped in type IIA to various configurations of wrapped D-branes on K3. More precisely, the BPS states are represented by differential forms in the cohomology of the moduli space of semi-stable coherent sheaves on K3; see \cite{HM} for details. The algebra action, which is obvious on the heterotic side, acts geometrically on these instanton moduli spaces by \textit{correspondences}, in a manner first elucidated by Nakajima \cite{nakajima1994instantons, nakajima1997heisenberg} and Grojnowski \cite{grojnowski1995instantons}.

There are many beautiful recent examples of vertex algebras which enjoy actions on various moduli spaces of interest to mathematicians and physicists. For example, the work of Gaiotto-Rapcak identified a large family of ``corner VOAs'' $Y_{(L,M,N)}[\Psi]$ at interfaces of $\tfrac{1}{2}$-BPS boundary conditions in twisted $\mathcal{N}=4$ super Yang-Mills induced from junctions of elementary 5-branes in IIB string theory \cite{GR19}. String duality relates this setup to configurations of wrapped D-branes in type IIA string theory on Calabi-Yau threefolds, and these VOAs are also expected to describe the algebra of local operators on a certain stack of M5 branes in an Omega-deformed M-theory \cite{GR22}. The corner VOAs act \cite{Rapcak:2018nsl} on the equivariant cohomology of moduli spaces of spiked instantons discovered by Nekrasov \cite{Nekrasov:2015wsu, Nekrasov:2016gud, Nekrasov:2016qym, Nekrasov:2016ydq, Nekrasov:2017rqy, Nekrasov:2017gzb}; in turn, this may be viewed as a generalization of the action of the \textit{W-algebra} $W_N^\Psi = Y_{(0,0,N)}[\Psi]$ on (the equivariant cohomology of) $U(N)$ instanton moduli spaces \cite{SV13, MO12} predicted by the AGT conjecture \cite{AGT10}. BPS algebras from configurations of D0 and D2-branes wrapping four-manifolds (i.e., divisors in the Calabi-Yau threefolds) have also been studied from the perspective of a 4d-2d correspondence \cite{Dedushenko:2017tdw, Gadde:2013sca, Feigin:2018bkf}. Vertex algebras arising from holomorphic-topological twists of 3d $\mathcal{N}=2$ theories with boundary \cite{Gadde:2013sca, DGP18, CDG20} are related to the aforementioned string and M-theoretic constructions \cite{Gukov:2017kmk, Gadde:2013sca, Gadde:2014wma, Dimofte:2019buf} and have been a recent source of many beautiful insights in their own right, e.g. \cite{Gukov:2018iiq, Cheng:2018vpl, Cheng:2022rqr}. 

This line of reasoning raises the (currently unanswered) questions: what is the chain-level lift of these (corner and boundary) VOAs, and does the resulting (homotopy) VOA have an action on (a suitable chain-level realization of) the corresponding moduli space homologies?%
\footnote{These developments also touch on numerous deep structures in physics and mathematics, such as algebras of BPS states and cohomological Hall algebras, Donaldson-Thomas theory in enumerative geometry, (mock and quantum) modularity in analytic number theory, the topology of low-dimensional manifolds, and more, that go far beyond the scope of these notes. The reader is encouraged to explore the cited texts, and references therein.}

\section{Concluding remarks}
This concludes our whirlwind tour of the mathematics of string dualities. Regrettably, our tour of the subject has been brief, with many important topics omitted. Nonetheless, we hope that the selected topics provide a useful foundation upon which to build up further expertise in this vibrant, multifaceted area.

Let us make brief concluding remarks and recapitulate some important lessons. Cohomology provides a powerful, covariant way to \textit{resolve}, and study, subspaces or quotients of physical and mathematical systems when we have a natural nilpotent differential, as we do in the (BV-)BRST quantization of gauge theories, and in (twists of) supersymmetric theories.  In the context of string theory, twists on the worldsheet produce topological strings, that have been a source of powerful geometric equivalences like mirror symmetry. Both the parent chain complexes and the cohomology groups themselves can carry rich and intricate algebraic structures that relate the operators (or states, which we did not emphasize in these lectures) of the constrained, or simplified, physical system to one another. These algebras are generally higher algebras, of which homotopical algebras and Poisson vertex algebras form two concrete and ubiquitous examples. 

Although we have only alluded to the BV-BRST formalism in a few brief remarks, we encourage the reader to consult \cite{Batalin:1981jr, henneaux1990lectures, Costello:2007ei, C11, henneaux2020quantization} for thorough reviews. There, too, cohomology and homotopical algebras are inescapable, even though there need not be any supersymmetry in sight. There are also long-standing, beautiful mathematical connections to derived algebraic geometry. 

We have also briefly mentioned equivariant cohomology; supersymmetric localization may be viewed as an infinite-dimensional generalization of this notion that can be used to compute a host of supersymmetric observables, including correlation functions of BPS states. BPS states, which may be isolated by the twisting procedure, are also deeply related to the geometry of moduli spaces, governing deformations of physical theories or families of inequivalent vacua. Indeed, such special subspaces of the physical Hilbert space can span the tangent space to the moduli space. Deformation theory, again, connects to higher algebras in a variety of ways, some of which are still partially understood.

Finally, we recall that the web of string dualities is vast and intricate, passing through compactifications, perturbative and nonperturbative dynamics, and many branches of mathematics. Although many strong-weak dualities give us access to various corners of the string theory landscape, our souped-up ``Fourier transforms'' are useless when the coupling $g \sim 1$, as there is no duality frame in which the coupling is weak. We still lack a useful description of string theory away from limits that are perturbative in some duality frame, and we may hope that a deeper understanding of the mathematics of string and M-theory can help get us there.

\section{Acknowledgments}

We thank K. Costello, G. Moore, and B. Williams for helpful comments on a draft of this manuscript, and to M. Duff and W. Gu for comments on an earlier version of the manuscript. We are also grateful to K. Costello, T. Dimofte, B. Williams, and I. Saberi for wonderful collaborations that have informed and clarified our thinking on the topics presented in these notes.  NP is grateful to the organizers of TASI 2021 for the opportunity to give these lectures and the encouragement to contribute these proceedings, and to the students for their engagement and insightful questions. The work of NG and NP is supported by the University of Washington. NP is also supported by the DOE award DE-SC0022347.

\appendix
\section{A brief summary of localization}
\label{sec:localization}
Supersymmetric QFTs often admit exactly calculable quantities that are remarkably robust under deformations of the theory. Often, the ability to compute these quantities stems from the indispensable tool of \textit{supersymmetric localization}, or simply \textit{localization}. For a diverse array of pedagogical lecture notes on localization in QFT, see \cite{P17, PlFY18}. Because of the importance of localization in checking dualities and computing correlation functions of special classes of observables (including in twisted theories), we want to give a sketch of the basic idea, in case it is unfamiliar to the reader.

Numerous quantities computed with localization techniques (partition functions on a variety of (curved) spaces, supersymmetric indices, correlation functions) often have rich translations into mathematical quantities. Most obviously, computing two dual representations of a single partition function and setting them equal to one another, for example, can produce highly nontrivial integral identities. More deeply, supersymmetric localization provides a physical realization/generalization of the classical notion of localization in equivariant cohomology, due to Duistermaat-Heckman \cite{DH82} and Atiyah-Bott \cite{AB84}.

We will introduce this idea first in a geometric language, and then translate the basic idea into the language of supersymmetric field theories. As always in life, we seek a way to compute a path integral, preferably exactly. If the theory is free, the integral is Gaussian and can of course be computed exactly. In a weakly coupled theory, we can hope to expand the path integral in a perturbation series, but even if we could resum the perturbation series around a free theory this is at best an asymptotic series that has vanishing radius of convergence in the coupling constant.\footnote{Note, however, that the subject of resurgence, concerned with deducing the nonperturbative parts of a path integral from its perturbative series, has proven very powerful for many physical and mathematical applications. It has connections with Stokes phenomena, wall-crossing, and much more. For an introduction, see \cite{M12, D14}.} Moreover, when studying dualities, the dual frame of a weakly coupled theory is just about guaranteed to be strongly coupled and not amenable to this sort of analysis. In the presence of supersymmetry, however, bosonic and fermionic contributions to the path integral cancel one another and this cancellation can be used to reduce the computation to the path integral for lower dimensional field theory, or, even better, an integral over a finite-dimensional space of fields, fixed by a suitable supersymmetry transformation: we say that the path integral \textit{localizes to BPS field configurations}.

\subsection{Localization in equivariant cohomology}
As a warm up,  we first recall the idea of localization in equivariant cohomology to compute integrals exactly by exploiting a symmetry on the integration manifold. The idea is to compute an integral defined on a manifold $\mathcal{M}$ with a symmetry group $G$. (Notice already the formal similarity to the problem of path integration in the presence of gauge symmetries!). One could try to compute the integral on $\mathcal{M}/G$ instead, including only a single representative from each orbit. This works nicely when $G$ acts freely on $\mathcal{M}$, but this space can be singular when the $G$-action on $\mathcal{M}$ has fixed loci. Instead of trying to make sense of integrals over the (possibly singular) space $\mathcal{M}/G$, we instead consider integration of $G$-equivariant forms over $\mathcal{M}$, i.e. elements of $G$-equivariant cohomology. Our abridged discussion of localization will largely follow the beautiful introduction \cite{Cremonesi:2013twh}, which the reader is advised to consult for a more careful treatment and details.

Consider for simplicity a $U(1)$ action on a closed, connected, oriented Riemannian manifold $\mathcal{M}$. We assume that the vector field $V$ generating the $U(1)$ action is a \textit{Killing vector field}: $\mathcal{L}_V g_{\mu \nu}=0$. A $U(1)$-equivariant cohomology class $[\omega]$ is represented by a $U(1)$-invariant differential form $\mathcal{L}_V \omega = 0$ (typically with inhomogeneous form degree, and dressed by a polynomial in the equivariant parameter $\sigma$) that is \textit{equivariantly closed} $d_\sigma \omega = 0$, modulo the addition of \textit{equivariantly exact} forms $\omega \sim \omega + d_\sigma \lambda$ for $\mathcal{L}_V \lambda = 0$. These cohomology groups agree with ordinary cohomology when the symmetry acts freely, and otherwise gives a well-defined notion of cohomology for non-free actions. 

Integration of $U(1)$-equivariant forms comes from integrating over the top-form piece, i.e. with form degree equal to the dimension of $\mathcal{M}$, as usual. If one is integrating an equivariantly exact form $\omega = d_\sigma \lambda$, then its top-form component is automatically de Rham-exact $\omega|_{\Omega^n(\mathcal{M})} = d \lambda|_{\Omega^{n-1}(\mathcal{M})}$, and hence its integral vanishes by Stokes theorem:
\be
\int_{\mathcal{M}} \omega = \int_{\mathcal{M}} d \lambda|_{\Omega^{n-1}(\mathcal{M})} = 0,
\ee
thus integrals of equivariantly closed forms over $\mathcal{M}$ only depend on their class in equivariant cohomology. 

The beautiful fact about equivariant integrals is that they \textit{localize}. This means that if we integrate a $U(1)$-equivariant closed form over $\mathcal{M}$, the integral will only receive contributions from the $U(1)$-fixed points on $\mathcal{M}$. Let us quickly a sketch a proof of this statement, which will readily generalize to the case of supersymmetric field theories. 

Since integration only depends on the $U(1)$-equivariant cohomology class of the form, we are free to replace $\omega$ by a more convenient representative. Explicitly, let us deform the integrand by a $d_\sigma$-exact piece:
\be
\omega \mapsto \omega_t := \omega e^{t d_\sigma \lambda}, \ t \in \mathbb{R}, \ \  \mathcal{L}_V \lambda = 0.
\ee
By construction, changing $t$ doesn't change the cohomology class of the integrand since $\omega$ and $\omega_{t}$ are equivariantly cohomologous:
\begin{align}
	\omega_t = \omega\big(1 + t d_\sigma \lambda + \ldots\big) = \omega + d_\sigma\big((-1)^{F(\omega)} t \omega \lambda + \ldots\big)
\end{align}
As a result, since $\mathcal{M}$ has no boundary, the $t$-derivative of the original integral also vanishes. This means we can compute the integral at any convenient value of $t$ and the result is equivalent to the original integral at $t=0$. One can think of $t = {1 \over \hbar}$ as the inverse of some auxiliary quantum parameter, so that $t \rightarrow \infty$ is a semiclassical limit. The $t$-independence tells us that by simply expanding in a one-loop approximation around the saddle points, we can obtain the exact result for the original integral! 

The limit exists if the 0-form part of $d_\sigma \lambda$ is negative semi-definite and has maximum equal to 0. Assuming this is the case, then in the $t \rightarrow \infty$ limit the integral is dominated by the zeroes (i.e. minima) of $-d_\sigma \lambda$. Let us further choose $\lambda$ to be the 1-form $\mu$ dual to the vector field $V$, so that $d_\sigma \mu = d \mu + \sigma |V|^2$. Then we are evaluating the integral:
\begin{align}
	\lim_{t \rightarrow \infty} \int_{\mathcal{M}} \omega e^{t d \mu} e^{\sigma t |V|^2}.
\end{align}
Since the term $e^{t d \mu}$ is an exponential of a 2-form, it can be expanded as a \textit{polynomial} of forms whose degree is no higher than $n/2$, where $n$ is the dimension of the manifold (integrals of higher degree forms will all vanish trivially). The other term $e^{\sigma t |V|^2}$ has an exponential dependence on $t$, which dominates the first term and tends towards a delta function in the $t \rightarrow \infty$ limit, supported on the zero-locus of the Killing vector $V$. 

With this choice of $\lambda$, we have seen that the semiclassical limit is delta-function-supported on the zero-locus of the Killing vector, up to a suitable Jacobian factor measuring first-order fluctuations around the zero-locus of $V$. If the manifold has several Killing vectors with respect to which the integrand is equivariantly closed, we can choose which vector or combination of vectors to localize on. We of course also have the general freedom to choose $\lambda$, subject to the condition that the ``semiclassical'' limit exists.

\subsection{Localization in QFT}

Let's now extend the above argument to quantum field theory. We assume there is a fermionic symmetry generator $Q$ that squares to a bosonic symmetry generator $Q^2 = J$; $Q$ is analogous to the $U(1)$-equivariant exterior derivative $d_\sigma$ and $J$ the Lie derivative $\mathcal{L}_V$ generating the symmetry. In the simplest case, $J = 0$ and $Q$ is simply a nilpotent fermionic symmetry. More generally, $J$ can be some general linear combination of spacetime symmetries, global symmetries, or gauge symmetries. If $Q$ is the BRST charge, the following localization argument produces a gauge-fixing of the path integral; if it is an ordinary supercharge, we will obtain the general prescription for supersymmetric localization. In practice, we are often interested in a differential $Q$ that is a combination of the BRST differential and a supersymmetry generator, as discussed in at the beginning of Section \ref{sec:twistinto}. In analogy with the above, we expect the path integral of a supersymmetric quantum field theory will localize to field configurations that are fixed by the bosonic symmetry $J$. Of course, we are integrating over a noncompact, infinite-dimensional $\mathcal{M}$, and in practice there are various subtleties to consider. We will simply sketch the main idea. 

Our goal is to compute an integral of the form
\be
\langle O \rangle = \int_{\mathcal{F}} \mathcal{D}\phi O e^{-S[\phi]},
\ee
where $\mathcal{F}$ is our space of fields and $O$ is a $Q$-closed, $J$-invariant observable. It can be a local operator, a collection of local operators, or operators of nontrivial spatial extent, like a Wilson line. We assume throughout that $S[\phi]$ is a $Q$-closed and $J$-invariant. We will also assume that $Q,J$ are non-anomalous, so that the measure is also invariant under the action of $Q$. From this, it follows this expectation value vanishes if $O$ is $Q$-exact $O = [Q, O']$ by an infinite-dimensional generalization of Stokes' theorem%
\footnote{This argument tacitly assumes the resulting boundary terms vanish, i.e. that the integrand decays sufficiently fast as we approach infinity in field space. We will assume this in what follows, though there are exceptions to this that can lead to interesting subtleties.}: %
\begin{align}
	\langle [Q, O'] \rangle &= \int_{\mathcal{F}} \mathcal{D}\phi [Q,O'] e^{-S[\phi]}\\
	&= \int_{\mathcal{F}} \mathcal{D}\phi \delta_Q( O' e^{-S[\phi]}) =0.
\end{align} We have used the fact that the action of $Q$ is a symmetry variation $\delta_Q$ which can be moved to act on the entire integrand and become a total derivative in field space. 

We are again in a situation where we want to compute an integral that only depends on the cohomology class of the argument, in this case the $Q$-cohomology class of $O$. Let us run the same kind of localization argument as before. In contrast to the simple geometric examples, the analytics of the integral are much more subtle due to the infinite dimensionality of the field space $\mathcal{F}$. We will assume that the path integral is free from IR divergences; in practice, this often involves turning on an Omega-background or placing a supersymmetric theory on a compact manifold. 

As before, we deform the integral by a $Q$-exact term that will ultimately force the integral to localize to a fixed locus in field space:
\be
\langle O \rangle_t = \int_{\mathcal{F}} \mathcal{D}\phi O e^{-S[\phi] - t [Q, V[\phi]]}.
\ee
$V[\phi]$ is necessarily fermionic, and must also be $J$-invariant to produce a good equivariant cohomology problem. We choose the bosonic part of $[Q, V[\phi]]$ to be positive semidefinite, so that the exponential is bounded from below when $t>0$. The argument for $t$-invariance proceeds identically to the previous case, provided our various assumptions hold. We will use the $t$-independence to take the $t \rightarrow \infty$ limit, in which the integral will be dominated by the saddle points of the deformation term. Again, we have the freedom to choose $V[\phi]$, and this freedom can result in uncovering \textit{dual} representations of the integral, localizing over different-looking degrees of freedom. In the BRST context, we have the freedom to gauge-fix as we choose, with different gauges often producing different insights. Of course, physical observables are ultimately insensitive to the choice of gauge, and BPS observables must similarly give the same answer regardless of localizing term. 

To evaluate this quantity, we expand all the fields around their saddle point values, $\phi = \phi_0 + {1 \over \sqrt{t}} \delta \phi$ so that the action becomes $S[\phi_0] + {1 \over 2}\int \int  {\delta^2  [Q, V[\phi]] \over \delta \phi^2}|_{\phi = \phi_0}(\delta \phi)^2.$ In the strict $t \rightarrow \infty$ limit, this expansion is one-loop exact in the parameter ${1 \over t}$. All that remains is to integrate out fluctuations of the field $\delta \phi$ that are normal to the localization locus $\mathcal{F}_Q$ to obtain one-loop determinants for the bosonic and fermionic fields. The result is therefore
\be
\langle O \rangle = \int_{\mathcal{F}_Q} \mathcal{D}\phi_0 O|_{\phi = \phi_0} e^{-S[\phi_0]} \left(\text{SDet} {\delta^2 [Q, V[\phi_0]] \over \delta \phi_0^2} \right)^{-1}.
\ee

The path integral has equivariantly localized to a lower-dimensional integral, over the BPS sublocus $\mathcal{F}_Q$ of field space. The original classical action, evaluated on BPS field configurations, is corrected by a one-loop determinant encapsulating field fluctuations normal to this locus. In specific contexts, this localization may even reduce the computation to an ordinary integral. The freedom in choosing localization schemes is a powerful way to obtain different dual representations of various observables. Moreover, the freedom to choose the fermionic symmetry $Q$ enables the computation of a vast array of supersymmetric quantities in the original theory. In the context of (topologically)-twisted field theories, one can employ the twisting supercharge in a localization procedure to compute \textit{supersymmetric indices}, which often have interesting mathematical expressions.

In the context of the 2d $\mathcal{N}=(2,2)$ theories emphasized in these notes, one may readily compute, for example, their Witten indices and elliptic genera via localization \cite{Benini:2013nda, Benini:2013xpa}. Partition functions of these theories on two-spheres have also recently been computed by localization \cite{Gomis:2012wy}, and used to directly extract Gromov-Witten invariants in the A-model without recourse to mirror symmetry \cite{Jockers:2012dk}. Further, as sketched in the main text, correlation functions of topological rings may be computed with these methods \cite{W98, HKKPTVVZ03}. Since these results are well-discussed elsewhere, and do not play a role in the remainder of these proceedings, we content ourselves with being telegraphic and referring the reader to the cited texts.  

%\subsection{2d $\mathcal{N}=(2,2)$ theories}
%
%\subsection{The elliptic genus}

\section{$A$-type Quantum Mechanics and Morse Theory}
\label{sec:morseintro}

In this appendix, we review some of the essential features of a 1d $\mathcal{N}=2$ theory, called \textit{$A$-type quantum mechanics}, and its relation to Morse theory. As described in Section \ref{sec:Abranes}, this theory is one of the starting points for studying topological $A$-branes in the $A$ twist of 2d $\mathcal{N}=(2,2)$ theories; in fact, the analysis of \cite{GMW115} uses lessons from this perspective to grapple with algebraic properties of the entire $A$-twisted 2d theory, e.g. the $LA_\infty$ structure encoding colliding of bulk and boundary local operators. The material presented in this appendix is adapted from \cite[Section 10]{HKKPTVVZ03} and \cite[Section 10]{GMW115}; we also recommend the original work \cite{WittenMorse} relating $A$-type quantum mechanics to Morse theory.

\subsection{$\mathcal{N}=2$ superspace}
\label{sec:superspace}
A convenient language to describe $\mathcal{N}=2$ quantum mechanical theories is with $\mathcal{N}=2$ \textit{superspace} $\RR^{1|2}$. We introduce coordinates on Euclidean $\RR^{1|2}$ given by $t, \theta, \bar{\theta}$. The derivative with respect to $t$ is denoted $\partial_t$. Derivatives with respect to $\theta$ and $\bar{\theta}$ are denoted $\partial_\theta$ and $\partial_{\bar{\theta}}$. They satisfy
\be
\{\partial_\theta, \theta\} = \{\partial_{\bar{\theta}}, \bar{\theta}\} = [\partial_t, t] = 1,
\ee
with all other (anti-)commutators vanishing. We use the convention that complex conjugation of Grassman odd variables exchanges order: $(\chi \xi)^* = \bar{\xi} \bar{\chi}$.

The generators of superspace translations are given by the Hamiltonian $H = - \partial_t$%
\footnote{In Lorentzian signature, we identify the Hamiltonian with the vector field $H = -i \partial_{x^0}$; the Wick rotation to Euclidean signature $t = i x^0$ implies $H = - \partial_t$.}
and the supercharges
\be
Q := \partial_\theta + \bar{\theta} \partial_t \hspace{1cm} \bar{Q} := -\partial_{\bar{\theta}} -  \theta \partial_t.
\ee
We will also need the chiral and anti-chiral superderivatives
\be
D := \partial_\theta -  \bar{\theta} \partial_t \hspace{1cm} \bar{D} := -\partial_{\bar{\theta}} +  \theta \partial_t .
\ee
Together, these operators satisfy the following relations:
\be
\{\bar{Q}, Q\} = - 2 \partial_t = 2H \qquad \{\bar{D}, D\} = 2 \partial_t = - 2H,
\ee
with all other (anti-)commutators vanishing. There is a $U(1)_R$ $R$-symmetry that acts on $\theta$ with charge $-1$ and $\bar{\theta}$ with degree $1$; this implies that $Q$ has $R$-charge $1$ and $\bar{Q}$ has $R$-charge $-1$.

A function on superspace, or section of a more general bundle over superspace, is called a \textit{superfield}. A general superfield $S$ can be Taylor expanded in terms of the odd variable $\theta, \bar{\theta}$; since these odd variables are nilpotent, this expansion is a finite sum:
\be
S = a + \theta \alpha - \bar{\theta}\, \bar{\beta} + \theta \bar{\theta} b.
\ee
Additionally, the superfield can either be bosonic, which implies $a,b$ are bosonic and $\alpha, \beta$ are fermionic, or fermionic, which implies the $a,b$ are fermionic and $\alpha, \beta$ are bosonic. 

The $\mathcal{N}=2$ supersymmetry algebra acts on a superfield by the vector fields $Q, \bar{Q}$, and this induces an action on the component fields. For example, we have
\be
\begin{aligned}
	Q S & = \alpha + \bar{\theta} (b + \partial_t a) + \theta \bar{\theta} (-\partial_t \alpha) \\
	& :=  Q a - \theta Q \alpha + \bar{\theta} Q \bar{\beta} + \theta \bar{\theta} Q b
\end{aligned}
\ee
so that $Q a = \alpha$, $Q \alpha = 0$, $Q \bar{\beta} = b + \partial_t a$, and $Qb = -\partial_t \alpha$. The additional signs appearing in the second line come from passing the fermionic symmetry $Q$ through $\theta$ and $\bar{\theta}$.

A general superfield $S$ yields a reducible representation of the supersymmetry algebra, and we may impose some conditions on $S$ to yield an \textit{irreducible} superfield. In the following, we will make use of the (bosonic) \textit{real superfield} $X$ satisfying $X^* = X$:
\be
X = x + \theta \xi - \bar{\theta} \bar{\xi} + \theta \bar{\theta} f.
\ee
The action of the $\mathcal{N}=2$ supersymmetry algebra on these fields is thus given by
\be
\begin{aligned}
	Q x & = \xi \qquad & Q \xi & = 0 \qquad & Q \bar{\xi} & = \partial_t x + f \qquad & Q f & = \partial_t \bar{\xi}\\
	\bar{Q} x & = \bar{\xi} \qquad & \bar{Q} \xi & = \partial_t x - f \qquad & \bar{Q} \bar{\xi} & = 0 \qquad & \bar{Q} f & = -\partial_t \xi\\
\end{aligned}
\ee
It is natural to give the superfield $X$ $U(1)_R$ charge $0$, whence $x, f$ have charge 0, $\xi$ has charge 1 and $\bar{\xi}$ has charge $-1$. We will ultimately be interested in the cohomological grading $C = -R$, so that the twisting supercharge $\bar{Q}$ has cohomological degree $1$.

\subsection{$A$-type quantum mechanics}
\label{sec:SQMA}
We can write down manifestly supersymmetric actions by integrating superfields over superspace. We suppose that the map $X$ takes values in a Riemannian manifold $M$ with metric $g_{ab}(X)$; the natural kinetic term for our $\mathcal{N}=2$ sigma-model is as follows:
\be
\begin{aligned}
	S_{kin} & = \tfrac{1}{2}\int dt d^2\theta \bigg( g_{ab}(X) DX^a \bar{D}X^b\bigg)\\
	& = \int dt \bigg(\tfrac{1}{2} g_{ab} \partial_t x^a \partial_t x^b - g_{ab} \bar{\xi}^a \partial_t \xi^b - \tfrac{1}{2} R_{abcd} \bar{\xi}^a \xi^b\bar{\xi}^c \xi^d\\
	& \qquad - \tfrac{1}{2} g_{ab}(f^a + \Gamma^a{}_{cd} \bar{\xi}^c \xi^d)(f^b + \Gamma^b{}_{ef}\bar{\xi}^e \xi^f) \bigg)\\
\end{aligned}
\ee
Since the only natural chiral superfield around is $\bar{D}X$, we are essentially restricted to considering interactions coming from integrals over all of superspace. We choose a smooth, real function $h: M \to \RR$ (also called the \textit{superpotential}) and integrate it over superspace
\be
S_h = -\int dt d^2\theta\, h(X) = \int dt \bigg(-f^a \partial_a h(x) + \partial_a \partial_b h(x) \bar{\xi}^a \xi^b\bigg)
\ee
from which we obtain the full action as the sum of these two terms
\be
S = S_{kin} + S_h = \int dt d^2 \theta \bigg(\tfrac{1}{2} g_{ab}(X) DX^a \bar{D}X^b - h(X)\bigg).
\ee
It follows that the $f^a$ are auxiliary fields whose equations of motion specialize them to the values $f^a = - g^{ab}\partial_b h - \Gamma^a{}_{bc}\bar{\xi}^b \xi^c$.

It is important to note that the bosonic potential energy, after integrating out $f$, is given by $\sim |\partial h|^2 = g_{ab} (\partial_a h)(\partial_b h)$, thus the ground states will localize around the \textit{critical points of $h$}, i.e. locations where $\partial_a h = 0$ for every $a$. In a neighborhood of such a critical point $p$, the fermions $\bar{\xi}^a, \xi^a$ have a mass matrix depends on the \textit{Hessian of $h$} $\sim \partial_a \partial_b h(p)$. We will restrict to superpotentials $h$ where there are no massless fermions%
\footnote{This constraint implies that, to lowest order in perturbation theory, there will be a single ground state for each critical point. If there were a massless fermion, both the fermionic vacuum $|0\rangle$ (annihilated by, e.g., $\xi$) and $\bar{\xi}|0\rangle$ are zero-energy states.} %
in the neighborhood of any critical point, i.e. that the Hessian $\sim \partial_a \partial_b h(p)$ is non-degenerate at each critical point $p$. Such an $h$ is called a \textit{Morse function}.

\subsection{BPS states and cohomology}
\label{sec:BPScoho}
Before quantizing the above theory, we pause to remark about some structural properties of Hilbert spaces in $\mathcal{N}=2$ theories. Our aim is to identify what distinguishes ground states in such theories. Very generally, we assume that there are fermionic symmetries $Q, \bar{Q} = Q^\dagger$, i.e. $[H, Q] = [H,\bar{Q}] = 0$, such that $[Q,\bar{Q}] = Q \bar{Q} + \bar{Q} Q = 2H$, where $[-,-]$ is the \textit{graded commutator}, for $H$ the Hamiltonian of the quantum theory. One immediate structural point is that states necessarily have non-negative energies due to a \textit{unitarity bound}: let $|\psi\rangle$ be an energy $E$ eigenstate with unit norm $\langle \psi| \psi \rangle = 1$, it follows that
\be
E = E \langle \psi | \psi \rangle = \langle \psi | H | \psi \rangle = \tfrac{1}{2} \langle\psi|[Q, \bar{Q}]|\psi \rangle = \tfrac{1}{2}\langle Q\psi| Q\psi\rangle + \tfrac{1}{2}\langle \bar{Q}\psi| \bar{Q}\psi\rangle \geq 0\,,
\ee
where $|Q \psi\rangle = Q |\psi\rangle$ and $|\bar{Q} \psi\rangle = \bar{Q} |\psi\rangle$. In particular, the state $|\psi\rangle$ is a ground state if and only if it has vanishing energy $E = 0$, which is true if and only if $Q |\psi\rangle = 0$ and $\bar{Q} |\psi\rangle = 0$. We call $|\psi\rangle$ \textit{$Q$-closed} if $Q |\psi\rangle = 0$, and similarly \textit{$\bar{Q}$-closed} if $\bar{Q} |\psi\rangle = 0$; thus, ground states are simultaneously $Q$-closed and $\bar{Q}$-closed. These states are often called \textit{BPS states}.

It turns out that we can encode ground states solely in terms of a single supercharge, e.g., $\bar{Q}$ and the corresponding $\bar{Q}$-closed states. Note that $\bar{Q}^2 = \tfrac{1}{2} [\bar{Q}, \bar{Q}] = 0$ implies that if $|\psi\rangle$ is $\bar{Q}$-closed, so too is $|\psi\rangle + \bar{Q} |\psi'\rangle$ for any $\psi'$. We call states of the form $\bar{Q} |\psi'\rangle$ \textit{$\bar{Q}$-exact}; $\bar{Q}$-exact states are necessarily $\bar{Q}$-closed. We then form the $\bar{Q}$-cohomology $H(\mathcal{H}, \bar{Q})$ as the space of $\bar{Q}$-closed states, modulo $\bar{Q}$-exact states:
\be
H(\mathcal{H}, \bar{Q}) := \textrm{Ker}\bar{Q} / \textrm{Im} \bar{Q} = \{|\psi\rangle \in \mathcal{H} | \bar{Q} |\psi\rangle = 0\}/\big(|\psi\rangle \sim |\psi\rangle + \bar{Q} |\psi'\rangle\big).
\ee
We now show that we can identify ground states with $\bar{Q}$-cohomology classes.

First, since $\bar{Q}$ commutes with $H$, we can decompose $\mathcal{H}$ into $H$-eigenspaces, where $\mathcal{H}^E$ is the eigenspace of energy $E$:
\be
H(\mathcal{H}, \bar{Q}) = \bigoplus_{E \geq 0} H(\mathcal{H}^E, \bar{Q}|_{\mathcal{H}^E}).
\ee
Suppose $|\psi\rangle$ is a $\bar{Q}$-closed state of energy $E > 0$. We can construct a state $|\psi'\rangle$ that realizes the $\bar{Q}$-exactness of $|\psi\rangle$ as follows:
\be
|\psi'\rangle = \tfrac{1}{2E} Q |\psi\rangle \Rightarrow \bar{Q}|\psi'\rangle = \tfrac{1}{2E} [\bar{Q}, Q]|\psi\rangle = |\psi\rangle
\ee
Thus every $\bar{Q}$-closed state of positive energy is necessarily $\bar{Q}$-exact, i.e. $H(\mathcal{H}^E, \bar{Q}|_{\mathcal{H}^E}) = 0$. On the other hand, by the above unitarity bound, a state of energy $0$ is necessarily $\bar{Q}$-closed, $\bar{Q}|_{\mathcal{H}^0} = 0$. In particular, any $\bar{Q}$-exact 0-energy vector is necessarily the $0$ vector: $H(\mathcal{H}^0, \bar{Q}|_{\mathcal{H}^0}) = \mathcal{H}^0$. As claimed, we see that the ground states are identified with $\bar{Q}$-cohomology classes:
\be
\mathcal{H}^0 \cong H(\mathcal{H}, \bar{Q}).
\ee

We will often be able to put more structure on the theory. For example, we will have a conserved charge $R$, $[R,H] = 0$, such that $Q$ has charge $1$ and $\bar{Q}$ has charge $-1$. If the charges of $R$ are integral, we can interpret this as the generator of a $U(1)_R$ $R$-symmetry of the theory; we can then give Hilbert space $\mathcal{H}$ a $\ZZ$-grading by (the negative of) their $R$-charge:
\be
\mathcal{H} = \bigoplus \mathcal{H}^n\,, \qquad R |\psi\rangle = i n|\psi\rangle, |\psi\rangle \in \mathcal{H}^n.
\ee
It follows that $\bar{Q}, Q$ act on this $\ZZ$-graded vector as follows:
\be
\includegraphics{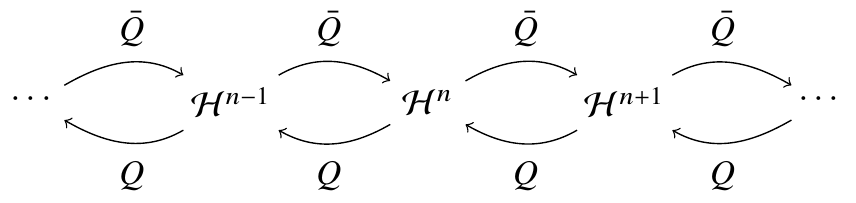}
\ee
We call the data of a $\ZZ$-graded vector space $\mathcal{H} = \bigoplus_n \mathcal{H}^n$ with square-zero operator of degree 1 $\bar{Q}: \mathcal{H}^n \to \mathcal{H}^{n+1}$ a \textit{$\ZZ$-graded (cochain) complex} or simply a \textit{complex}.

We will assume that the fermionic parity of a state $|\psi\rangle$ agrees with $R$-charge $n$ mod $2$, so that $\mathcal{H}^n$ is purely bosonic for even $n$ and purely fermionic if $n$ is odd. We can still compute the $\bar{Q}$-cohomology to obtain the space of ground states, but we can restrict our attention to a fixed degree:
\be
H(\mathcal{H}, \bar{Q}) = \bigoplus_n H^n(\mathcal{H}, \bar{Q})\,, \qquad H^n(\mathcal{H}, \bar{Q}) := \textrm{Ker}\bar{Q}|_{\mathcal{H}^n} / \textrm{Im} \bar{Q}|_{\mathcal{H}^{n-1}}\,.
\ee
The cohomology group $H^n(\mathcal{H}, \bar{Q})$ is merely the ground states of charge $n$.

\subsection{$A$-type quantum mechanics and Morse cohomology}
Let's determine the space of ground states in the $A$-type quantum mechanical theory of Section \ref{sec:SQMA} as the cohomology of the supercharge $\bar{Q}$. First, in a local patch on the target space, there is the vacuum $|0\rangle$ annihilated by all of the $\xi^a$%
\footnote{Somewhat more precisely, the state $|0\rangle$ should be identified with the zero-energy state of lowest $R$-charge, c.f. \cite[Section 10]{GMW115}. We note that globalizing the vacuum state requires a choice of orientation on $M$. Roughly speaking, the Clifford vacuum vector $|0\rangle$ can be identified with the algebra element $\xi^1 \ldots \xi^N$, with the action by the Clifford algebra $\xi^a,\bar{\xi}^a$ given by multiplication on the left. Since $\xi^a$ transforms as a tangent vector, such a global choice corresponds to a trivialization of the tangent bundle, i.e. an orientation. Thankfully, the Riemannian metric gives us a natural orientation.} %
and independent of the bosonic coordinates $x^a$. We can then identify states in the Hilbert space $\mathcal{H}$ with differential forms on the target space $\mathcal{H} \cong \Omega^\bullet (M)$%
\footnote{It is natural to take the functions $\omega_{a...b}$ to be complex valued, so we are really considering $\Omega^\bullet(M, \CC)$.}; %
a $p$-form $\omega \in \Omega^p(M)$ is identified as
\be
\omega = \omega_{a...b}(x) dx^a ... dx^b \leftrightarrow |\omega\rangle = \omega_{a...b}(x) \bar{\xi}^a \ldots \bar{\xi}^b |0\rangle
\ee
If the target space $M$ has dimension $d$, we will give a $p$-form state $R$-charge $n = p-\tfrac{d}{2}$.%
\footnote{We will mostly work with symplectic target spaces, so that $\tfrac{d}{2}$ is an integer and we have integral $R$-charges. On odd manifolds, this still yields a $\ZZ + \tfrac{1}{2}$-grading but this grading isn't necessarily correlated with fermion parity.} %
There is then a natural inner product on these states given by integration:
\be
\langle \omega_1| \omega_2 \rangle = \int \omega_1 \wedge \star \omega_2^*
\ee
where $\star$ is the Hodge operator sending $p$-forms to $d-p$ forms via the formula
\be\label{eq:hodgestar}
(\star \omega)_{a \ldots b} = \tfrac{1}{p!}\sqrt{g}\epsilon_{c \ldots d a \ldots b} g^{cc'} \ldots g^{d d'} \omega_{c' \ldots d'}.
\ee
We can thus identify the operation $\omega \to \star \omega^*$ as Hermitian conjugation on the space of (complex-valued) differential forms.

In terms of operators acting on differential forms, we similarly identify functions of $x^a, \bar{\xi}^a$ with differential forms acting via multiplication; then the conjugate momenta are identified as derivatives thereof: $\partial_t x^a \leftrightarrow g^{ab} \partial_{b}$, $\xi^a \leftrightarrow  g^{ab} \partial_{\bar{\xi}^b} \sim g^{ab} \iota_{\partial_b}$, where $\iota_V$ denotes the operation of contracting with a vector field $V$. From this, it follows that $\bar{Q}$ can be identified as
\be
\bar{Q}  \leftrightarrow dx^a\big(\partial_a + \partial_a h(x)\big)) = d + dh \wedge\,,
\ee
where $d = dx^a \partial_a$ is the exterior derivative or de Rham differential, and $dh \wedge$ denotes wedging with the 1-form $dh$. We immediately conclude that the space of supersymmetric ground states are identified as cohomology of the \textit{Morse complex}:
\be
\mathcal{H}_0 \cong H(\Omega^\bullet(M), d + dh \wedge)
\ee

It is worth noting that, at least for compact $M$, the complex $(\Omega^\bullet(M), d + dh \wedge)$ is quasi-isomorphic to the de Rham complex $(\Omega^\bullet(M), d)$, i.e. the complex with $h = 0$: this follows from noting that the differentials are conjugate $d+ dh = e^{-h} d e^h$ so that we can relate the complexes by simply multiplying forms by $e^h$. This conjugation is well-defined so long as $h$ doesn't go to $\pm \infty$; if $M$ is compact any $h$ is necessarily bounded. More generally, we find that we can relate differentials $d + dh$ and $d + dh'$ so long as $h-h'$ is sufficiently small, e.g. grows at the same rate as $h$ or $h'$ at infinity in field space. This feature results from the localization procedure described in Appendix \ref{sec:localization}; see \cite{GMW115} or \cite{HKKPTVVZ03} for more details. So long as this is valid, we can conclude that the dimensions of the graded components are topological invariants of the target space $M$, known as the \textit{Betti numbers} $b_p(M) = \dim H^p(\Omega^p(M), d)$.

\subsection{The Morse-Smale-Witten complex}
\label{sec:MSW}

We now end this appendix by using the independence of $h$ to obtain a model for the cohomology of the Morse complex by way of perturbation theory. This is the main tool utilized in \cite{GMW115} to understand infinite dimensional generalizations of the present problem. The idea is as follows: we use the independence on the superpotential to send $h \to \lambda h$ and take the limit $\lambda \to \infty$. As described in Section \ref{sec:SQMA}, the theory will localize to the critical points $p$ of the superpotential $\partial_a h(p) = 0$.

To first order in perturbation theory, the resulting quantum system has a ground state $|p\rangle$ for each critical point (in terms of differential forms, this is roughly a distributional form with support at $p$). The $R$-charge of such a perturbative ground state is determined by considering the simplest case of a supersymmetric harmonic oscillator -- in a small neighborhood of a critical point $p$, the superpotential takes the form $h(p) + \tfrac{1}{2}\partial_a \partial_b h (x^a - p^a) (x^b - p^b) + \ldots$. Moreover, by diagonalizing the Hessian, it suffices to consider the case of $d = 1$ with $h = \tfrac{1}{2} m x^2$. Thus, we seek (square-integrable) states of the form $|\psi\rangle = (a + b dx)|0\rangle$ with $\bar{Q} |\psi\rangle = (m x a + \partial_x a)dx|0\rangle = 0$, thus $a(x) = \# e^{-m x^2}$. When $m > 0$, this represents a square-integrable $0$-energy state $\# e^{-mx^2} |0\rangle$ of form degree $0$. On the other hand, when $m < 0$ this state is not square-integrable and instead the ground state is represented by $\# e^{m x^2} dx |0\rangle$ of form degree $1$. Applying this reasoning more generally, we see that the $R$-charge of perturbative ground state $|p\rangle$ depends on the number of negative eigenvalues of the Hessian $\partial_a \partial_b h(p)$, also called the \textit{Morse index} $\mu(p)$. More precisely, we give such a state $R$-charge $\mu(p) - \tfrac{d}{2}$.

For a general Morse function, the number of critical points of Morse index $p$ is far greater than the Betti number number $b_p(M)$, indicating that some of the perturbative ground states $|p\rangle$ are lifted at higher orders in perturbation theory. As described in, e.g., \cite[Section 10]{GMW115} this lifting is controlled by the transition amplitude $\langle q | \bar{Q} | p\rangle$; namely, $|p\rangle$ may no longer be a $\bar{Q}$-closed state to all orders in perturbation theory, and we must measure the overlaps with the other perturbative groundstates to measure this failure. Importantly, since the cohomological degree of $\bar{Q}$ is 1 (it has $R$-charge $-1$), for this amplitude to be non-zero, the Morse indices of these critical points must have $\mu(q) = \mu(p) + 1$.

We are thus interested in computing the path integral of our quantum mechanical theory with boundary conditions $x \to p$ at $-\infty$ and $x \to q$ at $\infty$ for critical points $p,q$ whose Morse indices satisfy $\mu(q) = \mu(p) + 1$. By utilizing supersymmetric localization, we can reduce this computation to a sum over 1-loop computations around the $\bar{Q}$-fixed locus (as sketched in Appendix \ref{sec:localization}). In the present context, the $\bar{Q}$-fixed points are solutions to the gradient flow equations $\partial_t x^a + \partial_a h = 0$; if we think of $h$ as the ``height'' of $x$, these are paths of steepest descent. The contribution to the full path integral from such a trajectory $\gamma$ is simply a sign $\pm 1$ (after rescaling the states $|p \rangle \to e^{-h(p)} |p \rangle$) given by a ratio of regularized determinants, see, e.g., \cite[Section 10.4]{GMW115} for a detailed presentation:
\be
\langle q | \bar{Q} | p \rangle = \sum_{\gamma: p \to q} \frac{\det'(L)}{\det'(L^\dagger L)} = \sum_{\gamma: p \to q} (-1)^{|\gamma|} =: m_{qp}\,,
\ee
where $L$ is the linearized Dirac operator along the curve $\gamma$.%
\footnote{Heuristically, this sign can be determined as follows: we compare the orientation of $\gamma$ induced by the of orientation $M$ and the orientation of $\gamma$ pointing towards the ``future''. If these orientations agree, the path contributes $1$, otherwise it contributes $-1$.} %
The integer $m_{qp}$ represents a (signed!) count of the gradient flows $p \to q$, up to overall translation invariance. In more detail, the index of the Dirac operator $L$ implies that the moduli space of gradient flows from $p \to q$, at least for generic superpotential $h$, has dimension $\mu(q) - \mu(p)$. In the present situation, the moduli space is 1-dimensional. Since we can always shift the time parameter, we find that the \textit{reduced moduli space} of such flows, i.e. the quotient of the full moduli space by these translations, is 0-dimensional, i.e. a collection of points. All together, we conclude that the space of supersymmetric ground states can be identified with the cohomology of the \textit{Morse-Smale-Witten complex}:
\be
\mathcal{H}_{MSW} = \bigoplus |p\rangle\,, \qquad  \bar{Q} |p\rangle = \sum_{q| \mu(q) = \mu(p) + 1} m_{qp} |q \rangle.
\ee

This last claim requires some justification -- at the very least, we should justify why the Morse-Smale-Witten complex is actually a complex, i.e. why $\bar{Q}^2 = 0$. Explicitly, we have
\be
\bar{Q}^2 |p\rangle = \sum_{r| \mu(r) = \mu(q) + 1} \sum_{q| \mu(q) = \mu(p) + 1} m_{rq} m_{qp} |r \rangle\,,
\ee
so we must check that $\sum_q m_{rq} m_{qp}$ vanishes for each critical point $r$ with $\mu(r) = \mu(p)+2$, where we sum over intermediate critical points $q$ with $\mu(q) = \mu(p) + 1 = \mu(r) - 1$. This sum can only be non-zero if there exists critical point $q$ with gradient flows $\gamma_{qp}: p \to q$ and $\gamma_{rq}:q \to r$, so let us suppose that this is the case. As described in \cite[Section 10.6]{GMW115}, the composed path $\gamma_{rq} + \gamma_{qp}$ can be approximated arbitrarily well by an gradient flow from $p \to r$; moreover, since the expected dimension of the moduli space of such trajectories is $\mu(r) - \mu(p) = 2$, it belongs to a 2-dimensional family of such gradient flows. Once we quotient by overall translations of the time parameter, we find a 1-dimensional reduced moduli space -- this is some connected component of the full moduli space of flows $p \to r$.

Once again, index theory implies that a given component of the reduced moduli space of trajectories $p \to q$ is necessarily a smooth, 1-dimensional manifold \textit{without} boundary. There are two choices: either the moduli space is compact, hence a copy of $S^1$, or non-compact, hence a copy of $\RR$. For the present component, we know it cannot be compact -- the composed path $\gamma_{rq} + \gamma_{qp}$ is a limit point of this moduli space. In particular, there must be a \textit{second} limiting trajectory (corresponding to the other limit point on $\RR$), again corresponding to a composed trajectory $\gamma_{rq'} + \gamma_{q'p}$, with $q'$ a (possibly different) fixed point with $\mu(q') = \mu(p) + 1.$ Moreover, $\gamma{rq'}+\gamma_{q'p}$ necessarily has the opposite (relative) orientation to $\gamma_{rq} + \gamma_{qp}$. Thus, the contribution of the composed trajectory $\gamma_{rq} + \gamma_{qp}$ to $\bar{Q}^2 |p\rangle$ is necessarily canceled by a second composed trajectory $\gamma_{rq'} + \gamma_{q'p}$.

\section{Landau-Ginzburg $B$-models and Matrix Factorizations}
\label{sec:MFintro}

In this appendix, we review aspects of $B$-twisted Landau-Ginzburg models and the origin of matrix factorizations in the context of boundary conditions as proposed by Kontsevich \cite[Section 7]{KapustinLi}. We start by reviewing some aspects of 1d $\mathcal{N}=2$ \textit{$B$-type quantum mechanics} in Section \ref{sec:SQMB}. In Section \ref{sec:LGbdy} we rewrite a 2d Landau-Ginzburg model in terms of a 1d $\mathcal{N}=2$ algebra containing the $B$-twist supercharge $\bar{Q}_B = \bar{Q}_+ + \bar{Q}_-$%
\footnote{For this appendix alone, we will call the $B$-twist supercharge $\bar{Q}_B$ to facilitate comparison between our discussion of $\mathcal{N}=2$ quantum mechanics and the reduction of $B$-twisted 2d $\mathcal{N}=(2,2)$.} %
and describe $B$-branes by introducing boundary Fermi multiplets factoring the boundary superpotential. Finally, in Section \ref{sec:bdyMF} we describe how the category of matrix factorizations arise as a useful description of junctions of $B$-branes.

\subsection{$B$-type quantum mechanics}
\label{sec:SQMB}
We start as in Section \ref{sec:SQMA} with superspace $\RR^{1|2}$ with coordinates $t, \theta, \bar{\theta}$; we use the same conventions for the supersymmetry generators $Q, \bar{Q}$ and the superderivatives $D, \bar{D}$ as in Section \ref{sec:superspace}. The first type of irreducible superfield we will consider is a \textit{chiral superfield} $\Phi$, which is bosonic and satisfies $\bar{D} \Phi = 0$. Chiral superfields have the following component expansion:
\be
\label{eq:chiral}
\Phi^n = \phi^n + \theta \psi^n + \theta \bar{\theta} \big(-\partial_t \phi^n\big).
\ee
If $\Phi^n$, and therefore $\phi^n$, has $R$-charge $r_n$, then $\psi^n$ has $R$-charge $r_n+1$. The second type of irreducible superfield we consider is fermionic and called a \textit{Fermi superfield} $\Gamma$. The most general form of Fermi multiplet depends on a choice of holomorphic function $E^a(\phi)$ of the chiral superfields, called the \textit{$E$-term}, and satisfies a modified chirality constraint: $\bar{D} \Gamma^a - E^a(\Phi) = 0$. This constraint implies that $\Gamma^a$ can be expanded as
\be
\label{eq:fermi}
\Gamma^a = \gamma^a + \theta g^a - \bar{\theta} (E^a(\phi)) + \theta \bar{\theta} \big( -\partial_t \gamma^a + \psi^n \partial_n E^a(\phi) \big).
\ee
where $\partial_n = \partial_{\phi^n}$. Such a Fermi multiplet is called an \textit{$E$-type} Fermi multiplet. To preserve $U(1)_R$ $R$-symmetry, we assume that $E^a$ has homogeneous degree $r_a + 1$, where $r_a$ is the $R$-charge of $\Gamma^a$.

From the above, the action of the supercharge $Q$ on the components of these superfields, and their complex conjugates, is as follow:
\be
\label{eq:Q1d}
\begin{aligned}
	Q \phi^n & = \psi^n & Q \psi^n & = 0 & Q \gamma^a & = g^a &  Q g^a & = 0\\
	Q \bar{\phi}^{\bar{n}} & = 0 & Q \bar{\psi}^{\bar{n}} & = 2 \partial_t \bar{\phi}^{\bar{n}} & Q \bar{\gamma}^{\bar{a}} & = -\bar{E}^{\bar{a}} & Q \bar{g}^{\bar{a}} & = -2 \big(\partial_t \bar{\gamma}^{\bar{a}} + \tfrac{1}{2}\bar{\psi}^{\bar{n}} \partial_{\bar{n}} \bar{E}^{\bar{a}}\big)\\
\end{aligned}
\ee
where $\partial_{\bar{n}} = \partial_{\bar{\phi}^{\bar{n}}}$. Similarly, the action of $\bar{Q}$ is:
\be
\label{eq:barQ1d}
\begin{aligned}
	\bar{Q} \phi^n & = 0 & \bar{Q} \psi^n & = 2 \partial_t \phi^n & \bar{Q} \gamma^a & = E^a &  \bar{Q} g^a & = 2\big( \partial_t \gamma^a - \tfrac{1}{2} \psi^n \partial_n E^a\big)\\
	\bar{Q} \bar{\phi}^{\bar{n}} & = \bar{\psi}^{\bar{n}} & \bar{Q} \bar{\psi}^{\bar{n}} & = 0 & \bar{Q} \bar{\gamma}^{\bar{a}} & = -\bar{g}^{\bar{a}} & \bar{Q} \bar{g}^{\bar{a}} & = 0\\
\end{aligned}
\ee

We can write a simple action of chiral multiplets and Fermi multiplets that is invariant under the transformations in Eq. \eqref{eq:Q1d} and Eq. \eqref{eq:barQ1d} as an integral over superspace. Each Fermi multiplet has an associated $E$-term, and we additionally introduce a second holomorphic function $J_a(\phi)$ the \textit{$J$-term}. To preserve $U(1)_R$, we need $J_a$ to be homogeneous of degree $1 - r_a$. The action of the theory can be expressed as an integral over superspace as
\be
\label{eq:1dphysaction}
\begin{aligned}
	S & = \tfrac{1}{4}\int d t d^2\theta \big(\delta_{n \bar{n}}(D \Phi^n)(\bar{D} \bar{\Phi}^{\bar{n}}) - \delta_{a \bar{a}}\Gamma^a\bar{\Gamma}^{\bar{a}}\big) - \tfrac{1}{4}\int d t d \theta \Gamma^a J_a(\Phi)\big|_{\bar{\theta} = 0} + \textrm{c.c.}\\
	&  = \int dt \bigg[\big|\partial_t \phi^n\big|^2 - \tfrac{1}{2} \bar{\psi}_n \partial_t \psi^n - \tfrac{1}{2} \bar{\gamma}_a \partial_t \gamma^a +  \tfrac{1}{4}E^a \bar{E}_a - \tfrac{1}{4}g^a \bar{g}_a - \tfrac{1}{4}g^a J_a(\phi) - \tfrac{1}{4} \bar{g}_a \bar{J}^a(\bar{\phi})\\
	& \qquad \qquad + \tfrac{1}{4} \big(\bar{\gamma}_a \psi^n \partial_n E^a+ \gamma^a \psi^n  \partial_n J_a - \gamma^a \bar{\psi}_n \bar{\partial}^n \bar{E}_a  - \bar{\gamma}_a \bar{\psi}_n \bar{\partial}^n \bar{J}^a\big)\bigg],
\end{aligned}
\ee
where we have used the metrics $\delta_{\bar{n} n}$ and $\delta_{\bar{a} a}$ to remove all barred indices, and $\bar{\partial}^n = \delta^{n \bar{n}} \partial_{\bar{\phi}^{\bar{n}}}$. As with 2d $\mathcal{N}=(2,2)$, chiral multiplets, we could replace the kinetic term $\delta_{\bar{n} n} (\bar{D}\bar{\Phi}^{\bar{a}})(D \Phi^a)$ by $\bar{D} D K(\Phi, \bar{\Phi})$ for $K(\phi, \bar{\phi})$ a more general K\"{a}hler potential or, equivalently, $\delta_{\bar{n} n}$ with a more general K{\"a}hler metric $G_{\bar{n} n}(\Phi, \bar{\Phi})$; we could similarly introduce a non-trivial Hermitian metric $H_{\bar{a} a}(\Phi, \bar{\Phi})$ for the Fermi multiplets. Note that the bosons $g, \bar{g}$ coming from the Fermi multiplets are auxiliary fields whose equations of motion specialize them to $g^a = -\bar{J}^a$ and $\bar{g}_a = - J_a$. Once we integrate out these auxiliary fields, we see that the action $S$ and the supersymmetry transformations are invariant under the simultaneous exchange of the fermions $\gamma \leftrightarrow \bar{\gamma}$ and the $E$- and $J$-terms $E \leftrightarrow J$, a 1d analog of \textit{fermionic $T$-duality} in 2d $\mathcal{N}=(0,2)$ theories, c.f. \cite[Appendix A]{DGP18}.

For the action in Eq. \eqref{eq:1dphysaction} to be supersymmetric, i.e. so that the presented actions of $Q, \bar{Q}$ are actual symmetries if and only if $E^a J_a$ is constant. The kinetic terms are invariant for the same reason as in Section \ref{sec:SQMA}, so the trouble comes from the $J$-term. For example:
\be
\begin{aligned}
	Q \int dt d\theta \Gamma^a J_a(\Phi)\big|_{\bar{\theta}=0} &= 0\\
	\bar{Q} \int dt d\theta \Gamma^a J_a(\Phi)\big|_{\bar{\theta}=0} &= \int dt \big(2\partial_t(\gamma^a J_a) - \psi^n \partial_n(E^a J_a) \big)\\
\end{aligned}
\ee
The variation under $Q$ is trivially invariant, but the variation under $\bar{Q}$ is nearly a total derivative -- it is if we require $\partial_n(E^a J_a) = 0$, i.e. $E^a J_a$ is constant. If we want to preserve the $U(1)_R$ $R$-symmetry, this constant must vanish, leading to the famous relation $E^a J_a = 0$. We will see that we can relax this condition once we place these theories on the boundary of a 2d theory: the boundary terms from the variation of the bulk action can be used to cancel the variation of the boundary action.

It's fairly straight-forward to quantize the above classical theory. The Hilbert space of the theory can be identified with polynomials in the bosons $\phi^n, \bar{\phi}_n$ and half of fermions, e.g. $\bar{\psi}_a$ and $\bar{\gamma}_n$, $\mathcal{H} \simeq \CC[\phi^n, \bar{\phi}_n, \bar{\psi}_a, \bar{\gamma}_n]$.%
\footnote{More precisely, to get square-normalizable states we should introduce a real mass parameter $m$ for a $U(1)$ flavor symmetry, c.f. \cite[Section 2]{BF18}. This real mass parameter induces an exponential suppression $\sim e^{|\phi|^2}$ on top of the polynomial dependence on $\phi, \bar{\phi}$ described here.} %
The supercharges $Q, \bar{Q}$ are the represented as the differential operators
\be
Q = \partial_{\bar{\psi}_n} \partial_n - \bar{J}^a \bar{\gamma}_a - \bar{E}_a \partial_{\bar{\gamma}_a} \qquad \bar{Q} = \bar{\psi}_n \bar{\partial}^a + E^a \bar{\gamma}_a + J_a \partial_{\bar{\gamma}_a}.
\ee
The $\bar{Q}$-cohomology of $\mathcal{H}$ has a natural algebraic interpretation. The first term removes $\bar{\phi}_n$ and $\bar{\psi}_n$ from cohomology, and we interpret the remainder $\CC[\phi^n, \bar{\gamma}_a]$ as a $\ZZ$-graded $\CC[\phi^n]$-module $\mathcal{E}$, where the $\ZZ$-grading comes from the $U(1)_R$ $R$-symmetry. The second term in $\bar{Q}$ corresponds to a differential $\delta$ on $\mathcal{E}$, thus the $\bar{Q}$-cohomology of $\mathcal{H}$ can be identified with the cohomology of this complex of $\CC[\phi]$-modules: $H^\bullet(\mathcal{H}) \simeq H^\bullet(\mathcal{E}, \delta)$.%
\footnote{This can also be stated in terms of complex differential geometry as follows. We identify the fermion $\bar{\psi}_n$ with the differential form $d \bar{\phi}_n$, therefore the first term of $\bar{Q}$ is the Dolbeault differential $\bar{\partial} = d\bar{\phi}_n \bar{\partial}^n$ on the target space. We then identify the polynomials in fermion $\bar{\gamma}^{\bar{n}}$ as section of a $\ZZ$-graded vector bundle $E$ and the remainder of $\bar{Q}$ turns this into a complex of vector bundles $\mathcal{E}$. Putting this together, we find that that the $\bar{Q}$-cohomology of $\mathcal{H}$ is simply the zeroth Dolbeault cohomology group with values in $\mathcal{E}$.}

\subsection{Boundary conditions in $B$-twisted Landau-Ginzburg models}
\label{sec:LGbdy}
One of most efficient ways to describe (1/2-BPS) boundary conditions of $B$-twisted 2d $\mathcal{N}=(2,2)$ theories is to choose an 1d $\mathcal{N}=2$ subalgebra of the 2d $\mathcal{N}=(2,2)$ supersymmetry algebra that contains the $B$-twist supercharge $\bar{Q}_B$ and then write the 2d superspace action as an integral over ``space'' $s$ and 1d $\mathcal{N}=2$ superspace. For the $B$-twist supercharge $\bar{Q}_B = \bar{Q}_+ + \bar{Q}_-$, the conjugate supercharge is $Q_B = Q_+ + Q_-$. Together they generate the desired 1d $\mathcal{N}=2$ algebra $\{Q_B, \bar{Q}_B\} = -2 \partial_t$.

With respect to this 1d $\mathcal{N}=2$ subalgebra, a 2d $\mathcal{N}=(2,2)$ chiral multiplet $\Phi^n$ decomposes as a chiral multiplet $\Phi^n$ and an $E$-type Fermi multiplet $\Psi^n$ with $E_{2d}^n = 2i\partial_s \phi^n$. In more detail, we define $\theta = \tfrac{1}{2}(\theta^+ + \theta^-)$ and $\eta = \tfrac{1}{2}(\theta^+ - \theta^-)$, as well as the corresponding derivatives. The 2d chiral conditions on $\Phi^n$ imply that $\Phi_{2d}^n := \Phi^n|_{\eta = \bar{\eta} = 0}$ satisfies the 1d chirality condition $\bar{D} \Phi_{2d}^n = 0$, where $\bar{D} = \bar{D}_+ + \bar{D}_-$. We also find that $\Gamma_{2d}^n := (D'\Phi^n)|_{\eta = \bar{\eta} = 0}$, where $D' = D_+ - D_-,$ is the aforementioned $E$-type Fermi multiplet: 
\be
\overline{D} \Gamma_{2d}^n = i\overline{D} D'\Phi^n|_{\eta = \bar{\eta} = 0} = i\{\overline{D}, D'\}\Phi^a|_{\eta = \bar{\eta} = 0} = 2i\partial_s \Phi^n_{2d}.
\ee
In terms of the component fields, we have
\be
\Phi^n_{2d} = \phi^n + \theta \psi^n + \theta \bar{\theta} \big(-\partial_t \phi^n\big) \qquad \Gamma_{2d}^n = \gamma^n + \theta g^n - \bar{\theta}\big(2i\partial_s \phi^n\big) + \theta \bar{\theta} \big(-i \partial_t\gamma^n + 2i\partial_s \psi^n \big)
\ee
where $\psi^n = \psi^n_+ + \psi^n_-$, $\gamma^n = (\psi^n_+ - \psi^n_-)$, and $g^n = 2F^n$. The action of $Q_B, \bar{Q}_B$ can be read off directly from the Eq. \eqref{eq:Q1d} and Eq. \eqref{eq:barQ1d}:
\be
\label{eq:QB}
\begin{aligned}
	Q_B \phi^n & = \psi^n & Q_B \psi^n & = 0 & Q_B \gamma^n & = g^n &  Q_B g^n & = 0\\
	Q_B \bar{\phi}_n & = 0 & Q_B \bar{\psi}_n & = 2 \partial_t \bar{\phi}_n & Q_B \bar{\gamma}_n & = 2i\partial_s\bar{\phi}_n & Q_B \bar{g}_n & = -2 (\partial_t \bar{\gamma}_n - i\partial_s\bar{\psi}_n)\\
\end{aligned}
\ee
\be
\label{eq:barQB}
\begin{aligned}
	\bar{Q}_B \phi^n & = 0 & \bar{Q}_B \psi^n & = 2 \partial_t \phi^n & \bar{Q}_B \gamma^n & = 2i \partial_s \phi^n &  \bar{Q}_B g^n & = 2(\partial_t \gamma^n - i\partial_s \psi^n)\\
	\bar{Q}_B \bar{\phi}_n & = \bar{\psi}_n & \bar{Q}_B \bar{\psi}_n & = 0 & \bar{Q}_B \bar{\gamma}_n & = -\bar{g}_n & \bar{Q}_B \bar{g}_n & = 0\\	
\end{aligned}
\ee

By performing the integral over $\eta$, the superpotential term $\int d^2\theta W(\Phi)$ becomes a $J$-type superpotential with $J_{2d,n} = \partial_n W(\phi)$. As mentioned above, we assume that the 2d chiral multiplets have $U(1)_R$ charges $r_a$ so that $W$ has $R$-charge 2. We conclude that the 2d action can be expressed as follows:
\be
\begin{aligned}
	S_{2d} & = \tfrac{1}{4}\int dt d^2\theta \bigg(\int ds (D \Phi_{2d}^n)(\bar{D} \bar{\Phi}_{2d,n}) - \Gamma_{2d}^n \bar{\Gamma}_{2d,n}\bigg)\\
	& \qquad + \tfrac{1}{4}\int dt d\theta \bigg(\int ds \Gamma_{2d}^n J_{2d,n}\bigg)\bigg|_{\bar{\theta}=0} + \textrm{c.c.}
\end{aligned}
\ee
Note that, when there is no spatial boundary, we find that $\int ds E_{2d}^n J_{2d,n} = 2i\int ds \partial_s W(\phi) = 0$, as expected. On the other hand, if we consider a half-spacetime with $s \leq 0$ we find that 
\be
\int ds E_{2d}^n J_{2d,n} = 2i W(\phi|),
\ee
where $\phi|$ is the value of $\phi$ on the boundary $s = 0$. When $W(\phi|)$ is non-vanishing, we are forced to introduce boundary degrees of freedom if we wish to make construct a boundary condition compatible with the $B$-twist; this is sometimes called the \textit{Warner problem} \cite{W95}. The standard solution to this problem is to factor $W$: we introduce boundary Fermi multiplets $\Gamma_{1d}^a$ with $E$- and $J$-type superpotentials $E_{1d}^a(\phi|), J_{1d,a}(\phi|)$ such that $E_{1d}^a(\phi|) \cdot J_{1d,a}(\phi|) = -2i W(\phi|)$:
\be
S_{1d+2d} = S_{2d} + \tfrac{1}{4}\int dt d^2\theta \Gamma_{1d}^a \bar{\Gamma}_{1d,a} - \tfrac{1}{4}\int dt d\theta \Gamma_{1d}^a J_{1d,a}(\Phi|)\bigg|_{\bar{\theta}=0} + \textrm{c.c.}
\ee
We see that the total $E$- and $J$-terms satisfy $E_{1d}^a J_{1d,a} + \int ds E_{2d}^n J_{2d,n} = 0$ so that the combined action $S_{1d+2d}$ is invariant under the 1d $\mathcal{N}=2$ superalgebra generated by $Q_B, \bar{Q}_B$.

\subsection{Matrix factorizations from $B$-branes}
\label{sec:bdyMF}

Now that we have the classical data required to define 1/2-BPS $B$-type boundary conditions, also called \textit{$B$-branes}, we can move to local operators bound to such boundary conditions and, more generally, the local operators that can be used to interpolate between two different boundary conditions, i.e. morphisms in the category of $B$-branes. It is conventional to work exclusively with Neumann boundary conditions on the entire 2d chiral multiplet, so that the boundary value $\phi|$ is unconstrained, and work with factorizations of the full superpotential $W(\phi)$. (We will make the substitution $W \to \tfrac{i}{2}W$ to simplify expressions in the following.)

As in the classic work \cite{KapustinLi}, we can view local operators at the junction of two boundary conditions as states on a strip with the two boundary conditions on either side via a state-operator correspondence; c.f. Figure \ref{fig:bdystateoperator}. Instead of performing a detailed analysis of this Hilbert space, we will intuit the result from the perspective on the left of Figure \ref{fig:bdystateoperator} and some algebra.

We start by interpreting the factorization $(E^a,J_a)$ of the superpotential $W$ algebraically. As in Section \ref{sec:SQMB}, we think of the boundary fermions as encoding a module for the boundary value $\phi|$. Now, however, the map $\delta$ no longer acts as a differential -- it squares to $W(\phi|)$! In particular, if we choose a basis for the module $\mathcal{E}$, the endomorphism $\delta$ is a matrix of polynomials in $\phi^n|$ that squares to $2W$ times the identity matrix. We call the data of such a $\CC[\phi^n|]$-module, or more generally a \textit{coherent sheaf}, $\mathcal{E}$ with a $R$-charge/degree 1 endomorphism $\delta_{\mathcal{E}}: \mathcal{E} \to \mathcal{E}$ such that $\delta_{\mathcal{E}}^2 = W(\phi|) \textrm{id}_{\mathcal{E}}$ a \textit{matrix factorization of $W$}.

A local operator at the junction between two boundary conditions necessarily commutes with multiplication by $\phi^n|$: we are free to pull an insertion of $\phi^n|$ into the bulk and back to the boundary at an arbitrary value of $t$. Thus, a local operator at the junction of two boundary conditions  encoded by matrix factorizations $(\mathcal{E},\delta_\mathcal{E})$ and $(\mathcal{F},\delta_\mathcal{F})$ should be a map of these $\CC[\phi^n|]$-modules $f: \mathcal{E} \to \mathcal{F}$; the physical local operators are going to be those module maps $f$ that are $\bar{Q}_B$-closed, modulo those that are $\bar{Q}_B$-exact. Although $\delta_{\mathcal{E}}$ and $\delta_{\mathcal{F}}$ are not differentials, they do induce a differential $\delta$, identified with the action of $\bar{Q}_B$, on this space of maps between matrix factorizations: if $f$ has $R$-charge/degree $r$, then the degree $r+1$ map $\delta f$ is given by $\delta f := \delta_{\mathcal{F}} f - (-1)^r f \delta_{\mathcal{E}}$ so that
\be
\delta^2 f = \delta_{\mathcal{F}} \delta f - (-1)^{r+1} \delta f \delta_{\mathcal{E}} = W f - f W = 0.
\ee
We thereby define the (DG-)category of matrix factorizations $MF(W)$ as the category whose objects are matrix factorizations of $W$ and whose morphisms from  $(\mathcal{E},\delta_\mathcal{E})$ to $(\mathcal{F},\delta_\mathcal{F})$ are the (DG)-vector space of $\CC[\phi^n|]$-module morphisms $\textrm{Hom}_{\CC[\phi^n|]-\textrm{mod}}(\mathcal{E}, \mathcal{F})$ with differential $\delta f := \delta_{\mathcal{F}} f - (-1)^r f \delta_{\mathcal{E}}$.

To make the above somewhat more explicit, let's briefly describe the space of local operators on a simple boundary condition, i.e. the endomorphism space in a simple matrix factorization. Consider a matrix factorization $(\mathcal{E}, \delta_{\mathcal{E}})$ with underlying $\CC[\phi^n|]$-modules $\mathcal{E} = \CC[\phi^n, \bar{\gamma}_a]$ and endomorphism $\delta_{\mathcal{E}} = E^a \bar{\gamma}_a + J_{a}\partial_{\bar{\gamma}_a}$ as above. The space of $\CC[\phi^n|]$-module maps $\mathcal{E} \to \mathcal{E}$, viewed as a $\CC[\phi^n|]$-module itself, is generated by the degree $r_a$, fermionic maps $\partial_{\bar{\gamma}_a}$ and the degree $-r_a$, fermionic maps $\bar{\gamma}_a$ (i.e. multiply by $\bar{\gamma}_a$), i.e. this space of $\CC[\phi^n|]$-module maps is simply $\CC[\phi^n|, \bar{\gamma}_a, \partial_{\bar{\gamma}_a}]$. The differential $\delta$ on this algebra is induced by graded-commutator with $\delta_{\mathcal{E}}$; on these generators, it is given by
\be
\delta \phi^n| = 0 \qquad \delta \bar{\gamma}_a = J_a \qquad \delta \partial_{\bar{\gamma}_a} = E^a.
\ee
Modulo replacing $\phi^n| \rightsquigarrow \phi^n$, this is exactly the description given in Section \ref{sec:Ainf-xyz}.

\bibliographystyle{JHEP}
\bibliography{ref}

\end{document}